\documentclass[3p, 12pt, sort&compress]{elsarticle}
\pdfoutput=1
\journal{Nuclear Physics B}
\usepackage{graphicx,psfrag,color}
\usepackage{bm}
\usepackage{mathbbol,verbatim, amsmath, amssymb}
\usepackage{slashed}
\usepackage{graphics}
\usepackage{color,ulem}

\usepackage{tikz}
\usetikzlibrary{trees}
\usetikzlibrary{decorations.pathmorphing}
\usetikzlibrary{decorations.markings}
\usetikzlibrary{decorations, decorations.markings, decorations.pathmorphing, arrows, graphs, shapes.geometric, snakes}
\usetikzlibrary{arrows}

\tikzset{
    photon/.style={decorate, decoration={snake}, draw=red},
    dark/.style={draw=gray, postaction={decorate},
        decoration={markings,mark=at position .55 with {\arrow[draw=gray]{>}}}},
antidark/.style={draw=gray, postaction={decorate},
        decoration={markings,mark=at position .55 with {\arrow[draw=gray]{<}}}},
electron/.style={draw=violet, postaction={decorate},
        decoration={markings,mark=at position .55 with {\arrow[draw=violet]{>}}}},
quark/.style={draw=blue, postaction={decorate},
        decoration={markings,mark=at position .55 with {\arrow[draw=blue]{>}}}},
antiquark/.style={draw=blue, postaction={decorate},
        decoration={markings,mark=at position .55 with {\arrow[draw=blue]{<}}}},
        gluon/.style={decorate, draw=or,
        decoration={coil,amplitude=2pt, segment length=3pt}},
  ZZ/.style={decorate, decoration={snake,amplitude=1.5pt, segment length=5pt}, draw=greeen},
left,
  }

\definecolor{greeen}{rgb}{0.03,0.84,0.13}
\definecolor{test}{rgb}{0.03,0.74,0.33}
\definecolor{viol}{rgb}{0.44,0,0.94}
\definecolor{or}{rgb}{0.95,0.65,0}

\begin{document}
\begin{frontmatter}
\begin{flushright}
\begin{footnotesize}
UMD-PP-017-21, ULB-TH/17-05
\end{footnotesize}
\end{flushright}

\vspace{-0.5cm}

\title{{\bf Long Lived Light Scalars  as Probe of Low Scale Seesaw Models}}

\author[a]{P. S. Bhupal Dev}
\ead{bdev@wustl.edu}
\author[b]{Rabindra N. Mohapatra}
\ead{rmohapat@umd.edu}
\author[c]{Yongchao Zhang}
\ead{yongchao.zhang@ulb.ac.be}
\address[a]{~Department of Physics and McDonnell Center for the Space Sciences,  \\
Washington University, St. Louis, MO 63130, USA}
\address[b]{~Maryland Center for Fundamental Physics, Department of Physics, \\
University of Maryland, College Park, MD 20742, USA}
\address[c]{~Service de Physique Th\'{e}orique, Universit\'{e} Libre de Bruxelles, \\
Boulevard du Triomphe, CP225, 1050 Brussels, Belgium}


\begin{abstract}
We point out that in generic TeV scale  seesaw models for neutrino masses with local $B-L$ symmetry breaking, there is a phenomenologically allowed range of parameters where the Higgs field responsible for $B-L$ symmetry breaking leaves a physical real scalar field with mass around GeV scale. This particle (denoted here by $H_3$) is weakly mixed with the Standard Model Higgs field ($h$) with mixing $\theta_1\lesssim m_{H_3}/m_h$, barring fine-tuned cancellation. In the specific case when the $B-L$ symmetry is embedded into the TeV scale left-right seesaw scenario, we show that the bounds on the  $h-H_3$  mixing $\theta_1$  become further strengthened due to low energy flavor constraints, thus forcing the light $H_3$ to be long lived, with displaced vertex signals at the LHC. The property of left-right TeV scale seesaw models are such that they make the $H_3$ decay to two photons as the dominant mode. This is in contrast with a generic light scalar that mixes with the SM Higgs boson, which could also have leptonic and hadronic decay modes with comparable or larger strength. We discuss the production of this new scalar field at the LHC and show that  it leads to testable displaced vertex signals of collimated photon jets, which is a new distinguishing feature of the left-right seesaw model. We also study a simpler version of the model where the $SU(2)_R$ breaking scale is much higher than the ${\cal O}$(TeV) $U(1)_{B-L}$ breaking scale, in which case the production and decay of $H_3$ proceed differently, but its long lifetime feature is still preserved for  a large range of parameters. Thus, the search for such long-lived light scalar particles  provides a new way to probe TeV scale seesaw models for neutrino masses at colliders.
\end{abstract}

\medskip


\end{frontmatter}


\makeatletter
\def\appendixname{Appendix}
\renewcommand\@makefntext[1]
{\leftskip=0em\hskip1em\@makefnmark\space #1}
\makeatother

\tableofcontents

\newpage

\section{Introduction}\label{sec:intro}
Seesaw mechanism seems to provide a very simple and elegant  way to understand the smallness of neutrino masses~\cite{seesaw1, seesaw2, seesaw3, seesaw4, seesaw5} . Two key ingredients of this mechanism are: (i) addition of the right-handed neutrinos (RHN) to the Standard Model (SM),  and (ii) a large Majorana mass for the RHNs which  breaks the accidental $B-L$ symmetry of the SM. There exist a large class of well-motivated ultraviolet (UV)-complete seesaw models which necessarily employ local $B-L$ symmetry, e.g. the TeV-scale left-right (LR) symmetric model~\cite{LR1, LR2, LR3}. It will therefore be an important step to find experimental evidence for the $B-L$ symmetry and its breaking. If the local $B-L$ is broken by a Higgs field that carries this quantum number, there will be a remnant neutral scalar field, denoted here by $H_3$ (for reasons explained below), analogous to the Higgs boson $h$ of the SM. So looking for signatures of  $H_3$ can provide invaluable clues to the nature of the new physics associated with neutrino mass generation. Clearly, such a search is realistic only if the $B-L$ symmetry breaking scale is within the multi-TeV range.

The mass and couplings of the new Higgs field are still unrestricted to a large extent, mainly because it communicates to the SM sector only through its mixing with the SM Higgs and via the heavy gauge boson interactions. The mass range heavier than the SM Higgs boson has been discussed earlier~\cite{Gunion:1986im, Deshpande:1990ip, Dev:2016dja, Fischer:2016rsh}. The mass range near $m_h$ would generically lead to large $h-H_3$ mixing, which is disfavored by the LHC Higgs data~\cite{Khachatryan:2016vau}. So we will focus here on the more interesting regime with $m_{H_3}\ll m_h$ in which case, the $H_3-h$ mixing angle $\theta_1\lesssim m_{H_3}/m_h\simeq  8\times 10^{-3}(m_{H_3}/{\rm GeV})$ from considerations of fine tuning.
Smaller values of $\theta_1$ can be analyzed as part of the allowed parameter range of generic $B-L$ models. However, as we show here, this range is naturally dictated to us from low energy flavor constraints, once the $B-L$ symmetry is embedded into the minimal LR model. This leads to the $H_3$ particle being necessarily long-lived, with interesting displaced vertex signals at the Large Hadron Collider (LHC). The important point is that even though a generic light scalar in a BSM theory that mixes with the SM Higgs can have leptonic, hadronic as well as photonic decay modes, the specific property of LR seesaw models makes the two photon decay mode exclusively dominant, as recently pointed out by us~\cite{DMZ}. In this paper, we elaborate on the details of this scenario, including an in-depth discussion of all relevant high and low-energy constraints on the model parameter space, the production and decay of the new Higgs boson at the LHC and future colliders, as well as the experimental prospects for observing the displaced vertex collimated diphoton signal.  We would like to emphasize the complementarity of the new collider signal discussed here with existing and future low-energy probes of light neutral sectors at the intensity frontier.

We will discuss two classes of UV-complete theories, based on (i) the full LR gauge group $SU(2)_L\times SU(2)_{R}\times U(1)_{B-L}$, and (ii) a simpler $B-L$ model with $SU(2)_L\times U(1)_{I_{3R}}\times U(1)_{B-L}$ local symmetry. The bulk of this paper deals with the class (i), where we consider a LR symmetric model where parity is broken at a much higher scale than the $SU(2)_R$ breaking. As a result, the $SU(2)_L$ Higgs triplet of the familiar left-right model is pushed to a much higher scale and also in neutrino mass formula, type I seesaw dominates. The heavy-light neutrino mixing is a separate parameter in the model and is small i.e. $\sim \sqrt{{m_\nu}/{m_N}}\lesssim 10^{-5.5}$ in the kinematic region where $m_{H_3}$ is less than 10 GeV and much less than the RH neutrino masses (assumed to be of order of 1 TeV), so its contribution to the $H_3$ phenomenology can be ignored.

 The class (ii) scenario that we discuss can be thought of as an ``effective" theory of the LR model, where the $SU(2)_R$ symmetry breaking scale is much higher than the $U(1)_{B-L}$-breaking which is assumed to be at the TeV-scale to be accessible experimentally.  Nevertheless, our results for the $U(1)_{B-L}$ case are applicable to a wider class of $Z'$-models, which have an associated Higgs boson that mixes with SM Higgs field.

The rest of the paper is organized as follows: in Section~\ref{sec:LR}, we briefly review the minimal LR seesaw model and set up our notation. In Section~\ref{sec:LLP}, we present the arguments for the light $B-L$ breaking Higgs  being weakly mixed with the SM Higgs $h$ and the heavy CP-even Higgs $H_1$, and study its decay lifetime and branching ratios. In Section~\ref{sec:cosmo}, we derive a lower limit on the light scalar $H_3$ mass from cosmological considerations at the nucleosynthesis epoch.  In Section~\ref{sec:lab}, we present an in-depth analysis of all available  laboratory constraints on the light scalar parameter space.   In Section~\ref{sec:collider}, we discuss the production of $H_3$ at hadron colliders and its detection prospects at the LHC, as well as in future colliders. In Section~\ref{sec:U1}, we analyze the LLP prospects of a simpler $U(1)_{B-L}$ model. Our conclusions are given in Section~\ref{sec:con}. In~\ref{app:decay}, we collect the two-body partial decay widths of $H_3$ in the LR model, and in~\ref{app:Zdecay}, the partial decay width of the  loop-induced process $Z\to H_3\gamma$. Finally, in~\ref{app:light_RHN}, we study the LLP sensitivity of light RHNs in the LR model at the LHC.

\section{Minimal left-right seesaw model}\label{sec:LR}
For completeness and to set up our notation, we briefly review the minimal LR model~\cite{LR1, LR2, LR3} in this section, based on the gauge group $SU(3)_c \times SU(2)_L\times SU(2)_R\times U(1)_{B-L}$ with parity broken at a higher scale than $SU(2)_R$. The quarks and leptons are assigned to the following irreducible representations:
\begin{align}
& Q_{L,i} \ = \ \left(\begin{array}{c}u_L\\d_L \end{array}\right)_i : \: \left({ \bf 3}, {\bf 2}, {\bf 1}, \frac{1}{3}\right), \qquad \qquad
Q_{R,i} \ = \ \left(\begin{array}{c}u_R\\d_R \end{array}\right)_i : \: \left({ \bf 3}, {\bf 1}, {\bf 2}, \frac{1}{3}\right), \\
& \psi_{L,i} \ = \  \left(\begin{array}{c}\nu_L \\ e_L \end{array}\right)_i : \: \left({ \bf 1}, {\bf 2}, {\bf 1}, -1 \right), \qquad \qquad
\psi_{R,i} \ = \ \left(\begin{array}{c} N_R \\ e_R \end{array}\right)_i : \: \left({ \bf 1}, {\bf 1}, {\bf 2}, -1 \right),
\label{lrSM}
\end{align}
where $i=1,2,3$ represents the family index, and the subscripts $L,R$ denote  the left- and right-handed chiral projection operators $P_{L,R} = (1\mp \gamma_5)/2$, respectively. There are different ways to break the $SU(2)_R\times U(1)_{B-L}$ gauge symmetry and to understand the small neutrino masses in this model, depending on the choice of the Higgs fields. One of the simplest choices is to introduce $SU(2)_L\times SU(2)_R$ bi-doublet and $SU(2)_R$ triplet scalars
\begin{eqnarray}
\Phi \ = \ \left(\begin{array}{cc}\phi^0_1 & \phi^+_2\\\phi^-_1 & \phi^0_2\end{array}\right) : ({\bf 1}, {\bf 2}, {\bf 2}, 0)\, , \qquad
\Delta_R \ = \ \left(\begin{array}{cc}\Delta^+_R/\sqrt{2} & \Delta^{++}_R\\\Delta^0_R & -\Delta^+_R/\sqrt{2}\end{array}\right) : ({\bf 1}, {\bf 1}, {\bf 3}, 2) \, ,
\label{eq:scalar}
\end{eqnarray}
respectively, with the neutral components of the above fields acquiring non-zero vacuum expectation values (VEV):
\begin{align}
\langle \Delta^0_R \rangle \ = \ v_R, \qquad \langle \phi^0_1 \rangle \ = \ \kappa, \qquad \langle \phi^0_2 \rangle \ = \ \kappa',
\end{align}
with the electroweak (EW) VEV given by $v_{\rm EW} = \sqrt{\kappa^2 + \kappa^{\prime \, 2}} \simeq 174$ GeV. We assume the $v_R$ scale to be in the multi-TeV range for phenomenological purposes. This is the version of the LR seesaw that results when parity symmetry is broken at a much higher scale than $SU(2)_R$~\cite{CMP} which, as remarked earlier, removes the $\Delta_L({\bf 1},{\bf 3},{\bf 1}, 2)$ Higgs field from the effective low energy theory.  This makes it simpler to analyze and present our main results, although our conclusions remain unchanged in the fully parity symmetric version being a TeV scale theory.

The fermion masses can be understood from the Yukawa Lagrangian:
\begin{eqnarray}
\label{eqn:Lyukawa}
{\cal L}_Y \ & = & \
h_{q,ij}\overline{Q}_{L, i}\Phi Q_{R, j}
+ \tilde{h}_{q,ij}\overline{Q}_{L, i}\widetilde{\Phi} Q_{R,j}
+ h_{\ell, ij}\overline{\psi}_{L,i}\Phi \psi_{R, j}
+ \widetilde{h}_{\ell, ij}\overline{\psi}_{L, i}\widetilde{\Phi} \psi_{R, j} \nonumber \\
&& 
+ f_{ij} \psi_{R,i}^{\sf T} C i\sigma_2 \Delta_R \psi_{R,j}  ~+~ {\rm H.c.} \, ,
\end{eqnarray}
where $\widetilde{\Phi}=i\sigma_2\Phi^*$ (with $\sigma_2$ being the second Pauli matrix) and $C$ stands for charge conjugation. The standard type-I seesaw mass matrix for neutrinos follows from the leptonic part of the above couplings after symmetry breaking. The triplet VEV $v_R$ breaks lepton number and provides the Majorana mass for the heavy RHNs which goes into the seesaw matrix~\cite{seesaw2}.

\section{Light scalar in TeV-scale LR seesaw} \label{sec:LLP}
In this section, we analyze the TeV-scale LR seesaw parameter space for a light scalar with mass in the (sub) GeV range and its couplings to determine the decay branching ratios and lifetime.

 The most general scalar potential for this model reads as follows:
\begin{eqnarray}
\label{eqn:potential}
\mathcal{V} & \ = \ & - \mu_1^2 \: {\rm Tr} (\Phi^{\dag} \Phi) - \mu_2^2
\left[ {\rm Tr} (\widetilde{\Phi} \Phi^{\dag}) + {\rm Tr} (\widetilde{\Phi}^{\dag} \Phi) \right]
- \mu_3^2 \:  {\rm Tr} (\Delta_R
\Delta_R^{\dag}) \nonumber
\\
&&+ \lambda_1 \left[ {\rm Tr} (\Phi^{\dag} \Phi) \right]^2 + \lambda_2 \left\{ \left[
{\rm Tr} (\widetilde{\Phi} \Phi^{\dag}) \right]^2 + \left[ {\rm Tr}
(\widetilde{\Phi}^{\dag} \Phi) \right]^2 \right\} \nonumber \\
&&+ \lambda_3 \: {\rm Tr} (\widetilde{\Phi} \Phi^{\dag}) {\rm Tr} (\widetilde{\Phi}^{\dag} \Phi) +
\lambda_4 \: {\rm Tr} (\Phi^{\dag} \Phi) \left[ {\rm Tr} (\widetilde{\Phi} \Phi^{\dag}) + {\rm Tr}
(\widetilde{\Phi}^{\dag} \Phi) \right]  \\
&& + \rho_1  \left[ {\rm
Tr} (\Delta_R \Delta_R^{\dag}) \right]^2 
+ \rho_2 \: {\rm Tr} (\Delta_R
\Delta_R) {\rm Tr} (\Delta_R^{\dag} \Delta_R^{\dag}) \nonumber
\\
&&+ \alpha_1 \: {\rm Tr} (\Phi^{\dag} \Phi) {\rm Tr} (\Delta_R \Delta_R^{\dag})
+ \left[  \alpha_2 e^{i \delta_2}  {\rm Tr} (\widetilde{\Phi}^{\dag} \Phi) {\rm Tr} (\Delta_R
\Delta_R^{\dag}) + {\rm H.c.} \right]
+ \alpha_3 \: {\rm
Tr}(\Phi^{\dag} \Phi \Delta_R \Delta_R^{\dag}) \,.  \nonumber
\end{eqnarray}
All the 12 parameters $\mu^2_{1,2,3}$, $\lambda_{1,2,3,4}$, $\rho_{1,2}$, $\alpha_{1,2,3}$ are chosen to be real, with an appropriate redefinition of the fields, and the only CP-violating phase $\delta_2$ is associated with the coupling $\alpha_2$, as explicitly shown in Eq.~\eqref{eqn:potential}. There are 14 scalar degrees of freedom, six of which are Goldstone modes eaten by the $W^\pm, W_R^\pm, Z, Z_R$ gauge bosons, thus leaving 8 physical Higgs bosons, denoted by $h, H_1, A_1, H_3, H_1^\pm$ and $H_2^{\pm\pm}$. The full scalar potential for this class of LR seesaw models was analyzed in detail in Refs.~\cite{Deshpande:1990ip,Dev:2016dja, Maiezza:2016ybz} and here we present only the results relevant for our purpose.

We work in the natural parameter space, where the following parameters are small:
\begin{eqnarray}
\xi \equiv \kappa'/\kappa \ = \ \frac{m_b}{m_t} \ \ll \ 1 \,, \qquad
\epsilon \equiv \kappa'/v_R \ll v_{\rm EW}/v_R \ \ll \ 1\,, 
\label{eq:small}
\end{eqnarray}
and also assume that there is no CP-violation in the scalar sector, i.e. the CP-violating phase $\delta_2$ is zero.
There are 6 degrees of freedom in the neutral scalar sector: three dominantly CP-even and the other three CP-odd. This includes the two Goldstone modes eaten by the neutral $Z$ and $Z_R$ gauge bosons. In the CP-even sector of neutral scalars, after applying the minimization conditions with respect to the three VEVs $\kappa$, $\kappa'$, $v_{R}$ and the phase $\delta_2$ associated with the VEV $\kappa'$, we obtain the squared mass matrices in the basis of $\{ \phi_1^{0 \, \rm Re} ,\, \phi_2^{0 \, \rm Re} ,\, \Delta_R^{0 \, \rm Re} \}$ at the zeroth, first and second order of the small parameters $\xi$ and $\epsilon$ defined in Eq.~\eqref{eq:small} (setting the phase $\delta_2$ to be zero), respectively,
\begin{align}
\label{eqn:scalarmatrix2}
\left(\mathcal{M}_0^2\right)^{(0)} & \ = \ v_R^2 \left(
\begin{array}{ccc}
 0 & 0 & 0 \\
 0 & \alpha _3 & 0 \\
 0 & 0 & 4 \rho _1
\end{array}
\right) \,, \\
\left(\mathcal{M}_0^2\right)^{(1)} & \ = \ v_R^2 \left(
\begin{array}{ccc}
 0 & -\alpha _3\xi   & 2 \alpha _1\epsilon  \\
 -\alpha _3\xi   & 0 & 4 \alpha _2\epsilon  \\
 2 \alpha _1\epsilon  & 4 \alpha _2\epsilon  &0
\end{array}
\right) \,, \\
\left(\mathcal{M}_0^2\right)^{(2)} & \ = \ v_R^2 \left(
\begin{array}{ccc}
 4 \lambda _1 \epsilon ^2+ \alpha _3 \xi ^2& 4 \lambda _4 \epsilon ^2 & 4\alpha _2 \epsilon  \xi  \\
 4 \lambda _4\epsilon ^2 & 4 \left(2 \lambda _2+\lambda _3\right) \epsilon ^2+\alpha _3\xi ^2 & 2
   \left(\alpha _1+\alpha _3\right)\epsilon  \xi \\
 4 \alpha _2\epsilon  \xi  & 2 \left(\alpha _1+\alpha _3\right)\epsilon  \xi & 0
\end{array}
\right).
\end{align}
In the limit of $\xi \ll 1$, one of the scalars, predominantly from $\phi_2^{0 \, \rm Re}$ is heavy, with mass at the $v_R$ scale $\sqrt{\alpha_3} v_R$ and decouples from the lower energy SM sector, thus leading to a reduced scalar mass matrix involving only  $\phi_1^{0 \, \rm Re}$ and $\Delta_R^{0 \, \rm Re}$.  If the quartic coupling $\rho_1$ is large, say of order one, then $\Delta_R^{0 \, \rm Re}$ (denoted here by $H_3$) will also be heavy, close to the $v_R$ scale, and its mixing with the SM Higgs boson $h$ will be given by $\theta_1\simeq  \frac{\alpha_1 v_{\rm EW}}{\rho_1 v_R}$. However, there exists another possibility, mainly motivated by the lack of any direct experimental constraints on its mass, as well as supported by arguments based on radiative stability (see Section~\ref{sec:radiative}), where $H_3$ could also be very light, and in particular, lighter than the SM Higgs boson. This happens when $\rho_1 \ll 1$, in which case, the sub-matrix in the basis of $\{ \phi_1^{0 \, \rm Re} ,\, \Delta_R^{0 \, \rm Re} \}$ becomes
\begin{align}
\label{eqn:scalarmatrix2}
\mathcal{M}_0^2 & \ \simeq \  \left(
\begin{array}{cc}
4 \lambda_1 \epsilon^2 & 2\alpha_1 \epsilon \\
2\alpha_1 \epsilon & 4\rho_1
\end{array}
\right) v_R^2 \, .
\end{align}
In the limit of $m^2_{H_3} \ll m^2_h$, Eq.~\eqref{eqn:scalarmatrix2} can be diagonalized in a seesaw-like approximation, and the two scalar masses are respectively given by
\begin{eqnarray}
m_h^2 & \ \simeq \ & 4 \lambda_1 \epsilon^2 v_R^2 \ = \ 4 \lambda_1 v_{\rm EW}^2 \,, \\
\label{eqn:massH3}
m_{H_3}^2 & \ \simeq \ & 4 \rho_1 v_R^2 - \sin^2 \theta_1 \, m_h^2 \,,
\end{eqnarray}
with the mixing angle
\begin{eqnarray}
\label{eqn:theta1}
\sin\theta_1 \ \simeq \ \frac{\alpha_1}{2\lambda_1 \epsilon} \ = \ \frac{\alpha_1}{2\lambda_1} \frac{v_R}{v_{\rm EW}} \,.
\end{eqnarray}
Note that the mixing has an inverted dependence on the VEV ratio $(v_{\rm EW} / v_R)^{-1}$, compared to the case where $m_{H_3}\gg m_h$. This restricts the parameter $\alpha_1$ to be appropriately small in order to ensure that $\sin\theta_1 \leq 1$, as we show below.

\subsection{Parameters for a light scalar} \label{sec:param}

\begin{figure}[!t]
  \centering
  \includegraphics[width=0.48\textwidth]{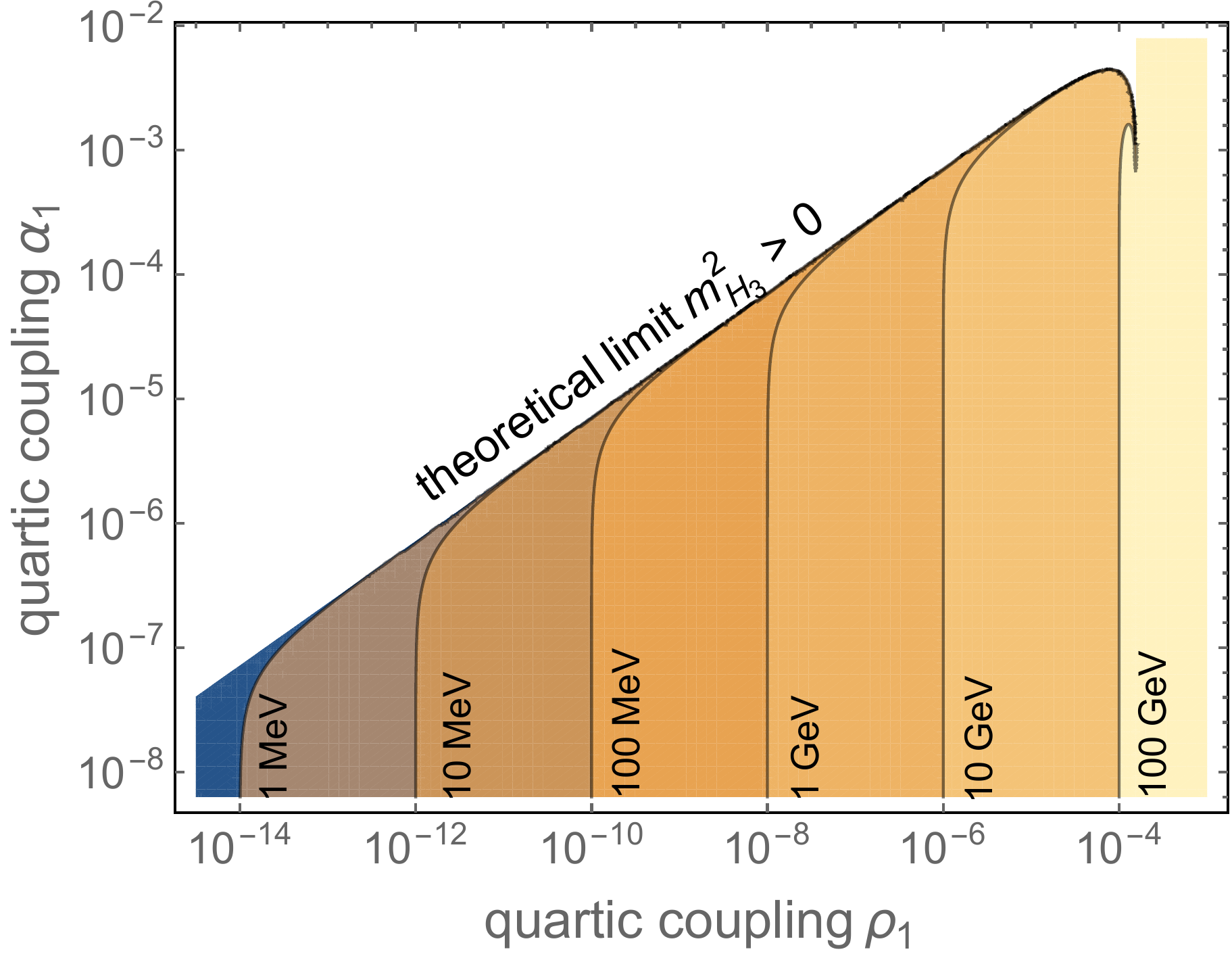}
  \caption{Contours of  $m_{H_3}$ as a function of the quartic coupling parameters $\rho_1$ and $\alpha_1$ defined in the scalar potential~\eqref{eqn:potential}. Here we have set the RH triplet VEV $v_R = 5$ TeV.   }
  \label{fig:mass}
\end{figure}

The question that we address in this subsection is whether there is a range of parameters in the model where $H_3$ could be light, i.e.~in the (sub)-GeV range, which has important phenomenological ramifications. From Eq.~\eqref{eqn:massH3}, it is clear that the lightness of $H_3$ depends on how small the values of $\rho_1$ and $\alpha_1$ can be:
\begin{eqnarray}
m_{H_3}^2 \ \simeq \ \left(4\rho_1-\frac{\alpha_1^2}{\lambda_1}\right)v_R^2 \,.
\label{eq:mH3}
\end{eqnarray}
The exact dependence of $m_{H_3}$ on the parameters $\rho_1$ and $\alpha_1$ is shown in Figure~\ref{fig:mass} for a representative value of $v_R=5$ TeV, which is close to the smallest possible value allowed by the current direct~\cite{Khachatryan:2014dka, Aad:2015xaa, Khachatryan:2016jqo} and indirect~\cite{Bertolini:2014sua} constraints on the $W_R$ mass (with appropriate scaling for $g_R\neq g_L$, where $g_{L,R}$ are the $SU(2)_{L,R}$ gauge couplings). We see that for $\rho_1 \sim 10^{-8}$ and  $\alpha_1 \lesssim 10^{-5}$,
we get $m_{H_3} \sim {\cal O}({\rm GeV})$, which is the region of great interest for long-lived particle (LLP) searches, as we will see below. For a small mixing $\sin\theta_1$, the $H_3$ mass is determined  completely by $\rho_1$ which is required to be very small for light $H_3$. However, when the mixing is sizable, i.e. $\alpha_1$ is comparatively large, there can be large cancellation between the two terms in Eq.~(\ref{eqn:massH3}).

\subsection{Radiative corrections} \label{sec:radiative}
In this subsection, we consider the one-loop corrections to the $H_3$ mass to see whether $m_{H_3}\sim {\cal O}$(GeV) is radiatively stable. The one-loop renormalization group (RG) running of the gauge, quartic and Yukawa couplings in the LR model has been considered in Refs.~\cite{Rothstein:1990qx,Kuchimanchi:2014ota,Chakrabortty:2016wkl}. Due to the large number of parameters in the scalar potential, some of the couplings would easily become non-perturbative before reaching up to the GUT or Planck scale, especially when the RH scale $v_R$ is relatively low~\cite{Maiezza:2016ybz}. There are also stability and perturbativity constraints on some of the quartic couplings~\cite{Maiezza:2016bzp}, but the mass of $H_3$, which is proportional to $\sqrt{\rho_1}$ at the leading order at tree level [cf.~Eq.~\eqref{eq:mH3}], is not limited by these arguments. Moreover, even if the gauge symmetry $SU(2)_R$ is restored at the few-TeV scale, the LR model is far from being a UV-complete theory up to the GUT scale, and more fields have to be introduced at higher energy scales, which hardly leave any imprints at the TeV scale. Our aim here is not to analyze the  ultimate UV-complete theory with parity, rather we remain at the TeV-scale, and using the Coleman-Weinberg effective potential approach~\cite{Coleman:1973jx}, check how the mass of $H_3$ would be affected by interacting with other particles in the TeV-scale LR model at the one-loop level.

Our considerations are similar to what happens to the SM Higgs mass. It is well known that if we neglect the one-loop fermion contributions to the Coleman-Weinberg effective potential, there would be a lower limit of order of 5 GeV on the Higgs boson mass~\cite{linde, weinberg}. However, this bound goes away once the large top-quark Yukawa coupling is included. This approach for the LR models for the case of doublet Higgs but without singlet fermions was carried out in Ref.~\cite{LH}, where a lower bound on the heavy neutral Higgs of the order of 900 GeV was obtained. The crucial difference in our model is the fermion contribution, which enables us to avoid this lower bound on $m_{H_3}$.

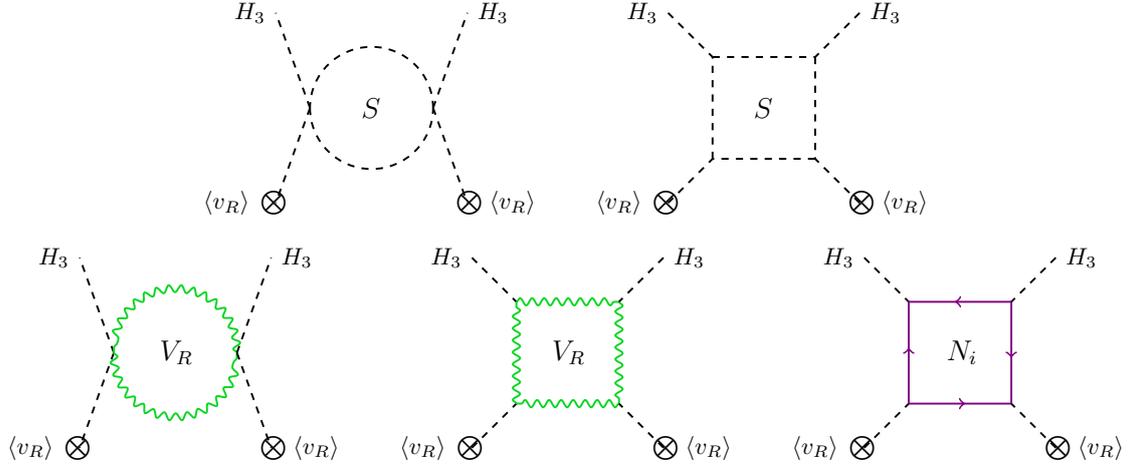
\begin{figure}[t!]
  \def\topdiff{0.25}
  \def\toppos{-1.5}
  \def\vertex{0}
  \def\length{1.4}
  \def\size{0.75}
  \def\topoffset{0.75}

  \centering

  \begin{tabular}{cc}
  \scalebox{0.9}{
  \begin{tikzpicture}[]
  \draw[dashed,thick](-0.9,0) -- (-\length,\length)node[left]{{\footnotesize $H_3$}};
  \draw[dashed,thick](-0.9,0) -- (-\length,-\length)node[left=-9pt]{{\footnotesize $\langle v_R \rangle \; \bigotimes$}};
  \draw[dashed,thick,black] (0.9,0)  arc (0:180 : 0.9 and 0.9) ;
  \draw[dashed,thick,black] (-0.9,0)  arc (180:360 : 0.9 and 0.9) ;
  \draw[dashed,thick](0.9,0) -- (\length,\length)node[right]{{\footnotesize $H_3$}};
  \draw[dashed,thick](0.9,0) -- (\length,-\length)node[right=-9pt]{{\footnotesize $\bigotimes \; \langle v_R \rangle $}};
  \draw (0.275,0) node [rotate=0] {$S$};
  \end{tikzpicture}} &

  \scalebox{0.9}{
  \begin{tikzpicture}[]
  \draw[dashed,thick](-\length,\length)node[left]{{\footnotesize $H_3$}} -- (-\size,\size);
  \draw[dashed,thick](-\length,-\length)node[left=-9pt]{{\footnotesize $\langle v_R \rangle \; \bigotimes$}} -- (-\size,-\size);
  \draw[dashed,thick](-\size,-\size) -- (-\size,\size);
  \draw[dashed,thick](-\size,\size) -- (\size,\size);
  \draw[dashed,thick](\size,\size) -- (\size,-\size);
  \draw[dashed,thick](\size,-\size) -- (-\size,-\size);
  \draw[dashed,thick](\length,\length)node[right]{{\footnotesize $H_3$}} -- (\size,\size);
  \draw[dashed,thick](\length,-\length)node[right=-9pt]{{\footnotesize $\bigotimes \; \langle v_R \rangle $}} -- (\size,-\size);
  \draw (0.275,0) node [rotate=0] {$S$};
  \end{tikzpicture}} 
\end{tabular}
\begin{tabular}{ccc}
  \scalebox{0.9}{
  \begin{tikzpicture}[]
  \draw[dashed,thick](-0.9,0) -- (-\length,\length)node[left]{{\footnotesize $H_3$}};
  \draw[dashed,thick](-0.9,0) -- (-\length,-\length)node[left=-9pt]{{\footnotesize $\langle v_R \rangle \; \bigotimes$}};
  \draw[ZZ,thick] (0.9,0)  arc (0:180 : 0.9 and 0.9) ;
  \draw[ZZ,thick] (-0.9,0)  arc (180:360 : 0.9 and 0.9) ;
  \draw[dashed,thick](0.9,0) -- (\length,\length)node[right]{{\footnotesize $H_3$}};
  \draw[dashed,thick](0.9,0) -- (\length,-\length)node[right=-9pt]{{\footnotesize $\bigotimes \; \langle v_R \rangle $}};
  \draw (0.4,0) node [rotate=0] {$V_R$};
  \end{tikzpicture}} &

  \scalebox{0.9}{
  \begin{tikzpicture}[]
  \draw[dashed,thick](-\length,\length)node[left]{{\footnotesize $H_3$}} -- (-\size,\size);
  \draw[dashed,thick](-\length,-\length)node[left=-9pt]{{\footnotesize $\langle v_R \rangle \; \bigotimes$}} -- (-\size,-\size);
  \draw[ZZ,thick](-\size,-\size) -- (-\size,\size);
  \draw[ZZ,thick](-\size,\size) -- (\size,\size);
  \draw[ZZ,thick](\size,\size) -- (\size,-\size);
  \draw[ZZ,thick](\size,-\size) -- (-\size,-\size);
  \draw[dashed,thick](\length,\length)node[right]{{\footnotesize $H_3$}} -- (\size,\size);
  \draw[dashed,thick](\length,-\length)node[right=-9pt]{{\footnotesize $\bigotimes \; \langle v_R \rangle $}} -- (\size,-\size);
  \draw (0.4,0) node [rotate=0] {$V_R$};
  \end{tikzpicture}} &

  \scalebox{0.9}{
  \begin{tikzpicture}[]
  \draw[dashed,thick](-\length,\length)node[left]{{\footnotesize $H_3$}} -- (-\size,\size);
  \draw[dashed,thick](-\length,-\length)node[left=-9pt]{{\footnotesize $\langle v_R \rangle \; \bigotimes$}} -- (-\size,-\size);
  \draw[electron,thick](-\size,-\size) -- (-\size,\size);
  \draw[electron,thick](\size,\size) -- (-\size,\size);
  \draw[electron,thick](\size,\size) -- (\size,-\size);
  \draw[electron,thick](-\size,-\size) -- (\size,-\size);
  \draw[dashed,thick](\length,\length)node[right]{{\footnotesize $H_3$}} -- (\size,\size);
  \draw[dashed,thick](\length,-\length)node[right=-9pt]{{\footnotesize $\bigotimes \; \langle v_R \rangle $}} -- (\size,-\size);
  \draw (0.4,0) node [rotate=0] {$N_i$};
  \end{tikzpicture}}
  \end{tabular}
  \caption{Dominant loop corrections to $m_{H_3}^2$ from interacting with the heavy scalars $S$ ($H_1$, $A_1$, $H_1^\pm$, $H_2^{\pm\pm}$), the heavy gauge bosons $V_R$ ($W_R$, $Z_R$) and  the heavy RHNs $N_i$.}
  \label{fig:diagrams}
\end{figure}

In the Coleman-Weinberg effective potential, we find that the dominant one-loop corrections to $m_{H_3}^2$ arise from its gauge interaction with the heavy $W_R$, $Z_R$ RH gauge bosons, the Yukawa interaction with the RHNs, and the interactions with the heavy scalars $H_1$, $A_1$, $H_1^\pm$ and $H_2^{\pm\pm}$. All the interactions with the SM fields are suppressed by the small mixing angles. The Feynman diagrams of loop corrections to $m_{H_3}^2$ from the heavy particle loops are collected in Figure~\ref{fig:diagrams}, which sum up to
\begin{eqnarray}
\label{eqn:loop}
\left( m^2_{H_3} \right)^{\rm loop} \ \simeq \
\frac{3}{2\pi^2} \left[ \frac{1}{3} \alpha_3^2 + \frac83 \rho_2^2 - 8 f^4 + \frac{1}{2}g^4_R+(g^2_R+g^2_{BL})^2 \right] v^2_R \,,
\end{eqnarray}
where $g_{BL}$ is the $SU(2)_R$ and $U(1)_{B-L}$ gauge coupling strength, $f$ is the RHN
Yukawa coupling as defined in Eq.~\eqref{eqn:Lyukawa}, and $\alpha_3$ and $\rho_2$ are the scalar quartic couplings defined in Eq.~\eqref{eqn:potential}.
Without any tuning of the scalar, gauge and Yukawa couplings, the loop correction to $m_{H_3}$ is expected to be of order $v_R / 4\pi$. In the above expression, we have kept only the contribution of the scalar coupling $\alpha_3$ since that is the only coupling which is expected to be of order one to satisfy the flavor changing neutral current (FCNC) constraints on bidoublet Higgs mass, as we will discuss below in Section~\ref{sec:lab}. The important point here is that with the minus sign for the $f$ contribution in Eq.~\eqref{eqn:loop}, the bosonic and fermionic contributions can be made to cancel each other keeping $H_3$ mass light even in the presence of radiative corrections. With a tuning of order ${\rm GeV} / \frac{v_R}{4\pi} \sim 10^{-2}$ (with $v_R$ at the TeV-scale) for the parameters in Eq.~(\ref{eqn:loop}), we could easily obtain a light scalar $H_3$ at or below the GeV scale.




\subsection{Couplings of $H_3$}\label{sec:coup}
In order to study the collider phenomenology of light $H_3$, it is essential to delineate its couplings to various particles in the theory.
In the limit of zero CP phase $\delta_2$ in Eq.~\eqref{eqn:potential}, $H_3$ could only mix with the scalars $h$ and $H_1$:
\begin{eqnarray}
\label{eqn:higgsmixing}
\left( \begin{array}{c}
h \\ H_1 \\ H_3 \\
\end{array} \right) \ = \
\left(
\begin{array}{ccccc}
1 - \frac{1}{2}\xi^2 & \xi  & -\sin\theta_1 \\
 -\xi  & 1 -\frac12 \xi^2 & -\sin\theta_2 \\
 \sin\theta_1 & \sin\theta_2 & 1
\end{array} \right)
\left( \begin{array}{c}
\phi_1^{\rm 0 \, Re} \\ \phi_2^{\rm 0 \, Re} \\ \Delta_R^{\rm 0 \, Re} \\
\end{array} \right) \,,
\end{eqnarray}
where $\sin\theta_1$ is the $h-H_3$ mixing already defined in Eq.~\eqref{eqn:theta1} and
\begin{eqnarray}
\label{eqn:theta2}
\sin\theta_2 \ \simeq \ \frac{4 \alpha _2\epsilon}{\alpha _3}
\ = \ \frac{4 \alpha_2}{\alpha _3} \frac{v_{\rm EW}}{v_R} \,.
\end{eqnarray}
is the mixing between $H_3$ and $H_1$. Both $\theta_1$ and $\theta_2$ are expected to be small for a GeV-scale $H_3$. The ``effective'' mixing angles of $H_3$ responsible for the flavor conserving and violating couplings to the SM quarks and charged leptons can be defined as
\begin{eqnarray}
\label{eqn:theta1b}
\sin\tilde\theta_1 & \ \equiv \ & \sin\theta_1 + \xi \sin\theta_2 \,, \\
\label{eqn:theta2b}
\sin\tilde\theta_2 & \ \equiv \ & \sin\theta_2 + \xi \sin\theta_1 \, ,
\end{eqnarray}
which will be used in our subsequent discussion.

\begin{table}[!t]
  \centering
  \caption[]{The couplings of the light scalar $H_3$, up to the leading order in $\epsilon$ and $\xi$. 
}
  \label{tab:coupling}
  \begin{tabular}[t]{lll}
  \hline\hline
  couplings & values \\ \hline
  $H_3 hh$ & $ \frac{1}{\sqrt{2}} \alpha _1 v_R $ \\
  $h H_3 H_3$ & $ - \sqrt2 \alpha _1 v_{\rm EW} $ \\
  $H_3 h H^0_1$ & $ 2 \sqrt{2} \alpha _2 v_R $ \\
  $H_3 H^0_1 H^0_1$ & $ \frac{1}{\sqrt{2}} \alpha _3 v_R $  \\
  $H_3 A^0_1 A^0_1$ & $\frac{1}{\sqrt{2}} \alpha _3  v_R$  \\
  $H_3 H_1^+ H_1^-$ & $\sqrt{2} \alpha _3  v_R$  \\
  $H_3 H_2^{++} H_2^{--}$ & $2 \sqrt{2} \left(\rho _1+2 \rho _2\right) v_R $ \\ \hline
  $H_3 \bar{u} u$ & $\frac{1}{\sqrt2} \widehat{Y}_U
  \sin\tilde\theta_1
  - \frac{1}{\sqrt2} \left( V_L \widehat{Y}_D V_R^\dagger \right) \sin\tilde\theta_2 $\\
  $H_3 \bar{d}d$  & $\frac{1}{\sqrt2} \widehat{Y}_D
  \sin\tilde\theta_1
  -\frac{1}{\sqrt2} \left( V_L^\dagger \widehat{Y}_U V_R \right) \sin\tilde\theta_2$\\
  $H_3 \bar{e}e$  & $\frac{1}{\sqrt2} \widehat{Y}_E
  \sin\tilde\theta_1 -\frac{1}{\sqrt2} Y_{\nu N}
  \sin\tilde\theta_2$ \\
  $H_3 NN$  & $\frac{m_N}{\sqrt2 v_R}$ \\ \hline
  $H_3  W_{}^+ W_{}^{-}$ & $\frac{1}{\sqrt2}g_L^2 \sin\theta_1\, v_{\rm EW} + \sqrt2 g_R^2 \sin^2\zeta_W \, v_R$ \\
  $H_3  W_{}^+ W_R^{-}$ & $\sqrt2 g_R^2 \sin\zeta_W \, v_R$ \\
  $H_3  W_{R}^+ W_R^{-}$ & $\sqrt2  g_R^2 v_R $ \\ \hline
  $H_3  Z_{} Z^{}$ & $\frac{g_L^2 \sin\theta_1\, v_{\rm EW}}{2\sqrt2 \cos^2\theta_w} +

  \frac{\sqrt2 g_R^2 \sin^2\zeta_Z \, v_R}{\cos^2\phi} $ \\
  $H_3  Z_{} Z_{R}$ & $-\frac{g_L g_R \sin\theta_1 \cos\phi \, v_{\rm EW}}{\sqrt2 \cos\theta_w} +
  \frac{2\sqrt2 g_R^2 \sin\zeta_Z \, v_R }{\cos^2\phi} $ \\
  $H_3  Z_{R} Z_R^{}$ & $\frac{\sqrt2 g_R^2 v_R }{\cos^2\phi} $ \\ \hline
  $H_3 H_1^+ W^-$ & $\frac12 g_L (\sin\theta_2 - \sin\theta_1 \xi)$ \\
  $H_3 H_1^+ W_R^-$ & $\frac12 g_R \epsilon$ \\
  $H_3 A_1 Z$ & $-\frac{i g_L (\sin\theta_2 - \sin\theta_1 \xi)}{2\cos\theta_w}$ \\
  $H_3 A_1 Z_R$ & $\frac{i}{2} g_R (\sin\theta_2 - \sin\theta_1 \xi) \cos\phi$ \\
  \hline\hline
  \end{tabular}
\end{table}

For completeness we collect all the coupling of light $H_3$ in the minimal LR models in Table~\ref{tab:coupling},  which is based on the calculation of Ref.~\cite{Dev:2016dja} and up to the leading order in the small parameters $\xi$ and $\epsilon$ defined in Eq.~\eqref{eq:small}. Here the RH gauge mixing $\phi$ is defined as $\tan\phi \equiv g_{BL}/g_R$, and the $W-W_R$ and $Z-Z_R$ mixings are respectively given by
\begin{align}
\tan\zeta_W & \ \simeq \ -\frac{2g_R\xi}{g_L}\left(\frac{m_W}{m_{W_R}}\right)^2, \label{eq:zetaW}\\
\tan\zeta_Z & \ \simeq \ \left[\frac{g_R^2}{g_L^2}-\left(1+\frac{g_R^2}{g_L^2}\right)\sin^2\theta_w\right]^{1/2}\left(\frac{m_Z}{m_{Z_R}}\right)^2, \label{eq:zetaZ}
\end{align}
where $\theta_w$ is the weak mixing angle.

In Table~\ref{tab:coupling}, the couplings to the charged leptons depend on the neutrino sector via the Dirac coupling matrix $Y_{\nu N} = m_D/v_{\rm EW}$, which can be parameterized through the Casas-Ibarra form~\cite{Casas:2001sr}
\begin{eqnarray}
m_D \ = \ i m_N^{1/2} O m_\nu^{1/2}
\label{eq:mD}
\end{eqnarray}
with $O$ an arbitrary complex orthogonal matrix, $m_\nu$  and $m_N$ are the light neutrino and RHN mass matrix, respectively. Without fine-tuning, the Dirac Yukawa couplings $Y_{\nu N}$, and hence, the light-heavy neutrino mixing, are expected to be small for TeV-scale RH neutrinos. So we can safely ignore those couplings involving higher powers of $Y_{\nu N}$, such as $H_3\nu\nu$.

\subsection{Decay lifetime and branching ratios}\label{sec:decay}

Through the mixing with the SM Higgs and the heavy scalar $H_1$ via respectively the mixing angles $\sin\theta_{1}$ and $\sin\theta_{2}$ in Eq.~(\ref{eqn:theta1}) and (\ref{eqn:theta2}), a light $H_3$ could decay at tree level into the SM fermions,\footnote{We assume here that all the three RHNs (typically at the TeV scale) are heavier than the light scalar $H_3$, and therefore, the $H_3$ decay into RHNs is kinematically forbidden. Allowing for the $H_3\to NN$ decay could lead to additional lepton number violating signatures, as discussed in Refs.~\cite{Maiezza:2015lza, Nemevsek:2016enw}. In addition, $H_3$ could in principle decays into the light neutrinos (or one light neutrino plus one heavy neutrino) through the heavy-light neutrino mixing $V_{\nu N}\simeq m_DM_N^{-1}$. However, for the TeV-scale type-I seesaw without any fine-tuning in the seesaw mass matrix, the mixing $V_{\nu N}$ turns out to be $\lesssim 10^{-6}$ [cf.~Eq.~\eqref{eq:mD}]. So the width $\Gamma (H_3 \to \nu\nu) \propto V_{\nu N}^4$ is completely negligible.} through the Yukawa couplings in Table~\ref{tab:coupling}. Note that the couplings to quarks could be flavor-changing, depending on the magnitude of effective mixing angle $\sin\tilde\theta_{2}$, while the lepton flavor violating (LFV) couplings are proportional to the Dirac coupling $Y_{\nu N}$ and $\sin\tilde\theta_{2}$. At the one-loop level, the Yukawa couplings to the SM fermions induce the decay into digluon and diphoton, i.e. $H_3 \to gg, \, \gamma\gamma$, analogous to the SM Higgs. Considering the flavor limits on the mixing angles $\sin\theta_{1,2}$ below in Section~\ref{sec:lab}, the fermion loops for both the $\gamma\gamma$ and $gg$ channels are highly suppressed. However, for the decay $H_3 \to \gamma\gamma$, there are extra contributions from the heavy $W_R^\pm$, $H_1^\pm$ and $H_2^{\pm\pm}$ loops. In the low mass limit $m_{H_3} \ll v_R$, the diphoton channel is sensitive only to the RH scale $\Gamma_{\gamma\gamma} \propto v_R^{-2}$~\cite{Dev:2015vjd} and dominated by the $W_R$ loop for TeV range $v_R$, as the scalar loops are comparatively suppressed by the loop function $5A_0 (0) / A_1(0) = - 5/21$ [cf. Eq.~\eqref{eqn:H3diphoton}]. Similarly, the SM $W$ loop is highly suppressed by the small $W - W_R$ mixing $\sin\zeta_W$ [cf.~Eq.~\eqref{eq:zetaW}].
The dominant two-body tree- and loop-level decay channels of $H_3$ covering all the parameter space of interest for a light scalar $m_{H_3} \lesssim m_h$ are presented in~\ref{app:decay}, while the other decay channels, such as $H_3 \to h^\ast h^\ast \to bbbb$, are suppressed either kinematically or by the multi-particle phase space.




\begin{figure}[!t]
  \centering
  \includegraphics[width=0.7\textwidth]{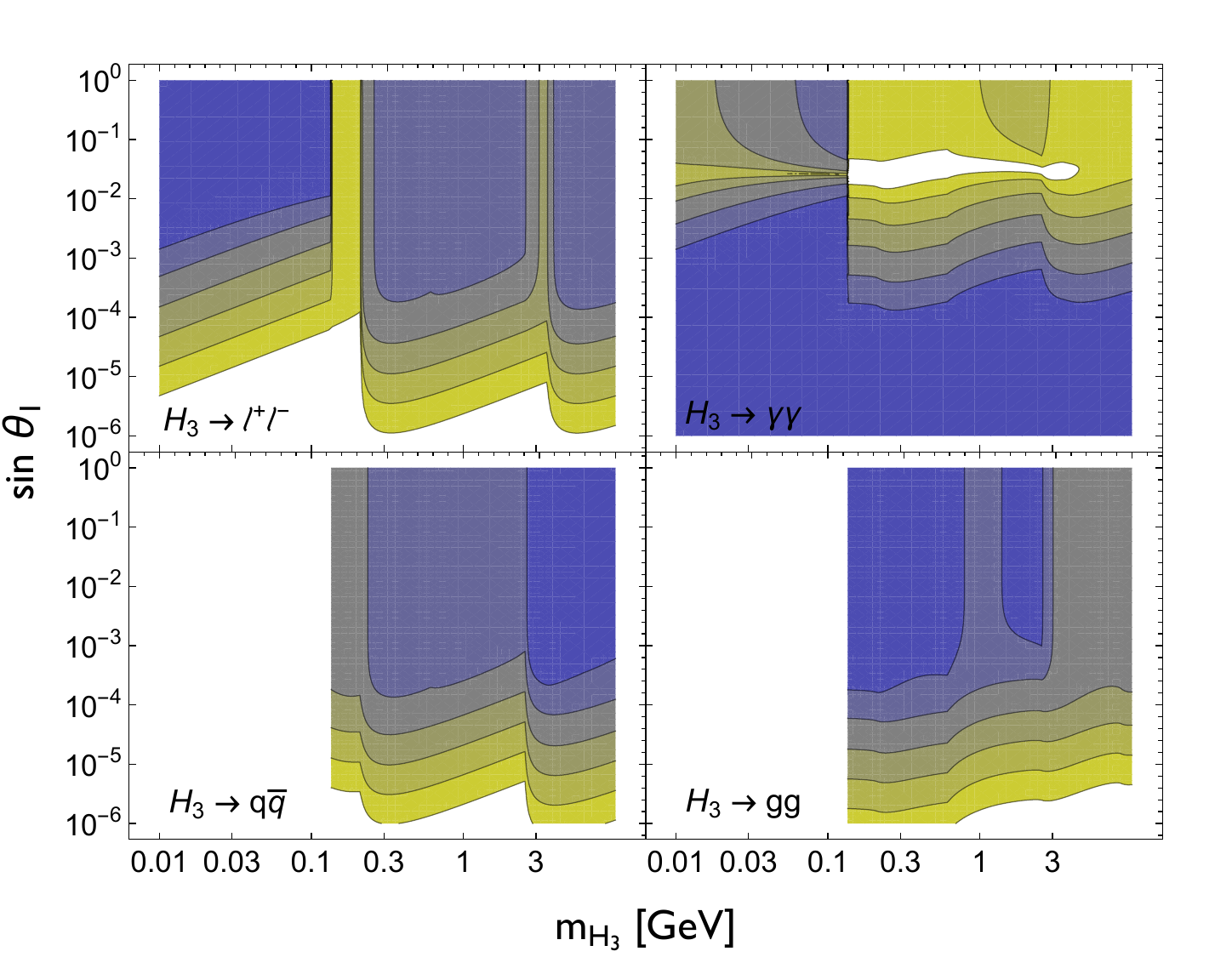}
  \raisebox{0.32\height}{\includegraphics[width=0.07\textwidth]{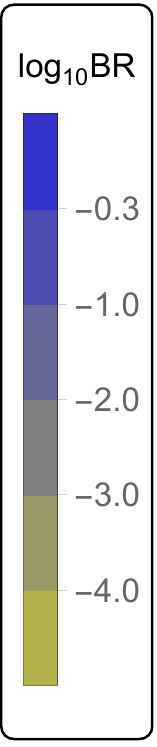}} \\
  \includegraphics[width=0.7\textwidth]{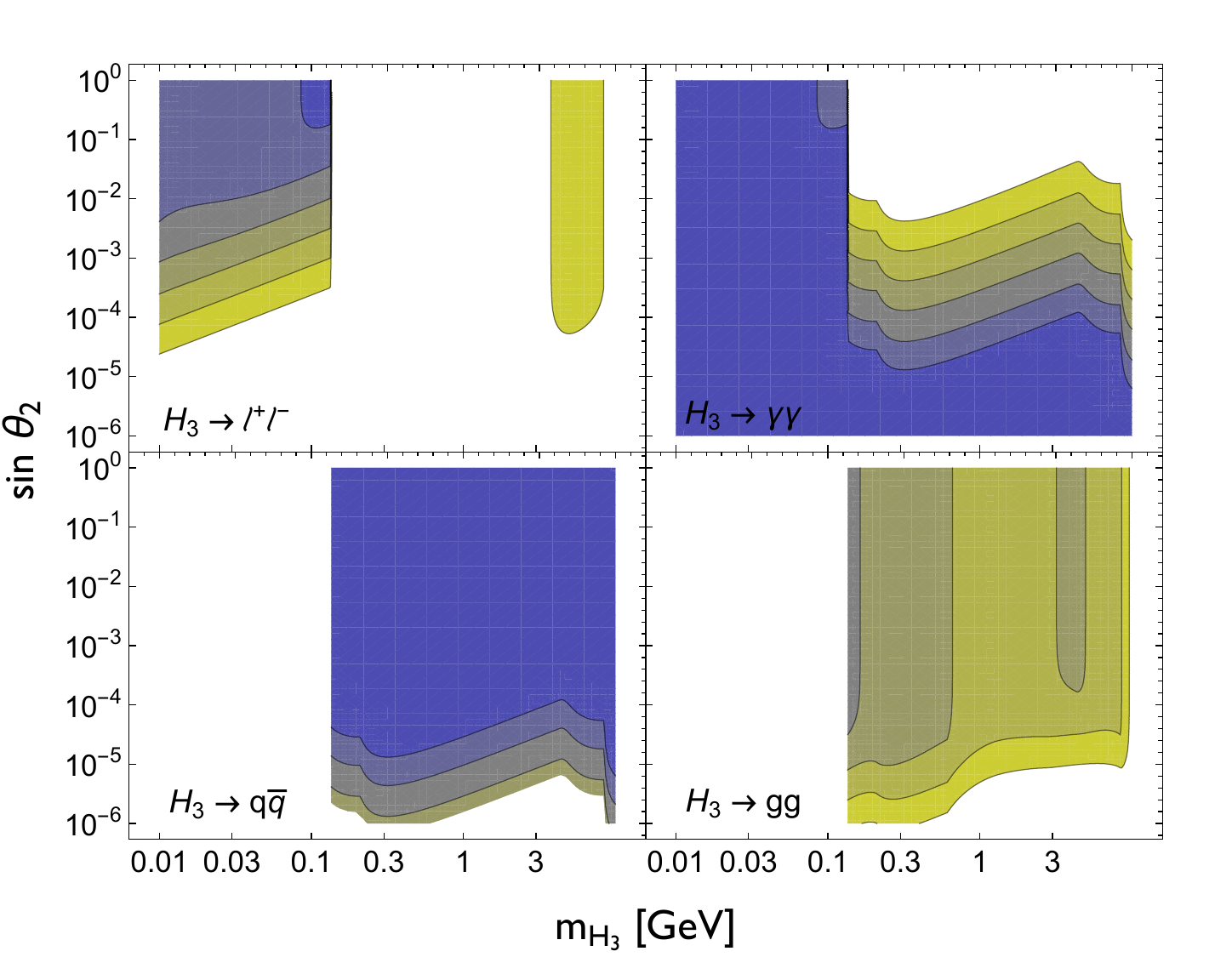}
  \raisebox{0.32\height}{\includegraphics[width=0.07\textwidth]{legend.pdf}}
  \caption{Color-coded branching ratios of $H_3$ as functions of its mass and mixing with $h$ (top) and $H_1$ (bottom) for different decay modes. Here we have set the RH scale $v_R = 5$ TeV and the RHN masses at 1 TeV. }
  \label{fig:BR}
\end{figure}

The decay branching ratios (BR) of $H_3$ to the $qq$, $\ell^+ \ell^-$, $\gamma\gamma$ and $gg$ channels are presented in Figure~\ref{fig:BR} as functions of its mass and mixing with $h$ (top panel) and $H_1$ (bottom panel). For concreteness, we have made the following reasonable assumptions: (i) The RH scale $v_R = 5$ TeV, which is close to the smallest value allowed by the current constraints on $W_R$. (ii) In the minimal LR model, the RH quark mixing matrix $V_R$ is very similar to the CKM matrix $V_L$, up to some additional phases~\cite{Senjanovic:2014pva, Senjanovic:2015yea}. For simplicity, we adopt $V_R = V_L$ in the calculation. Thus with the experimental values of the SM quark masses and CKM mixing, we obtain the numerical values of the $H_3$ couplings to the SM quarks, including the FCNC couplings, arising from its mixing with the heavy scalar $H_1$ [cf.~Table~\ref{tab:coupling}]. (iii) The couplings of $H_3$ to the charged leptons depend on the heavy and light neutrino masses and their mixings via the Yukawa coupling matrix $Y_{\nu N}$. Here we assume the three light neutrino masses are of normal hierarchy with the lightest one to be 0.01 eV, while all the three RHNs are assumed to be degenerate at 1 TeV without any RH lepton mixing, which pushes the couplings $Y_{\nu N}$ to be very small, of order $10^{-7}$ [cf.~Eq.~\eqref{eq:mD}]. Furthermore, when the $H_3$ mass is below the pion mass, its decay to both the quark and gluon channels are kinematically forbidden, as we do not have free hadronic states in Nature lighter than pions. For these hadronic channels, the RG running of the strong coupling constant $\alpha_s$ is taken into consideration, which is important below the EW scale. The flavor violating decays of $H_3$ in both the hadronic and leptonic sectors, e.g. $H_3 \to sb,\, \mu\tau$, are included in making the plots in Figure~\ref{fig:BR}.

From Figure~\ref{fig:BR}, we find that when the mixing angles are sizable, $H_3$ decays mostly into the SM quarks (above the pion mass threshold) and charged leptons (below the pion mass), while when $\sin\theta_{1,2} \lesssim 10^{-4}$, the dominant decay of $H_3$ is into the diphoton channel, which benefits from the heavy gauge boson loops induced by the RH gauge coupling (independent of $\sin\theta_{1,2}$), with a sub-dominant contribution from the heavy scalar loops. With large BR to two photons, the FCNC constraints can be used to set limits on the mixing angles $\sin\theta_{1,2}$ as a function of $m_{H_3}$; see more details in Section~\ref{sec:lab}.

\begin{figure}[!t]
  \centering
  \includegraphics[width=0.48\textwidth]{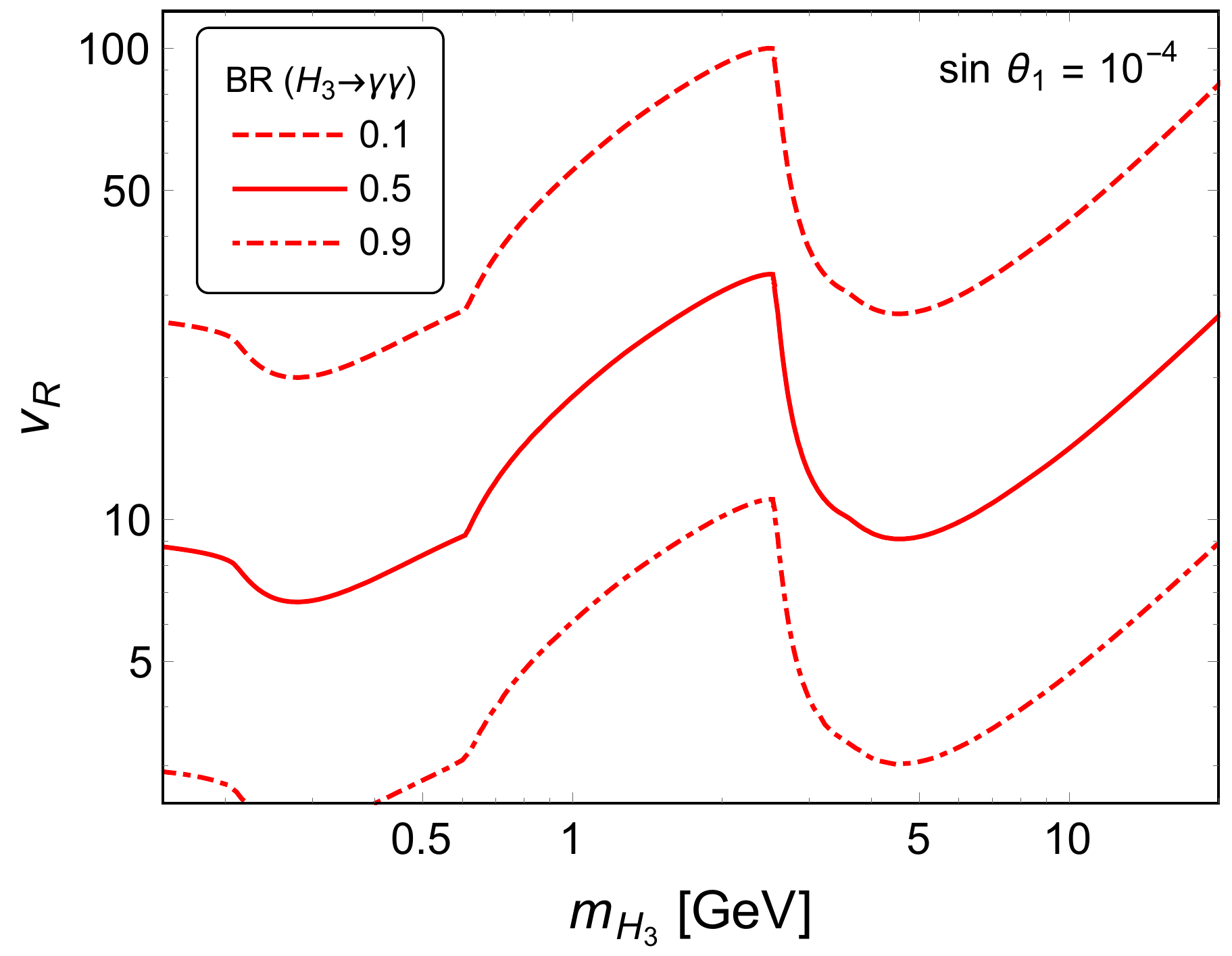}
  \includegraphics[width=0.48\textwidth]{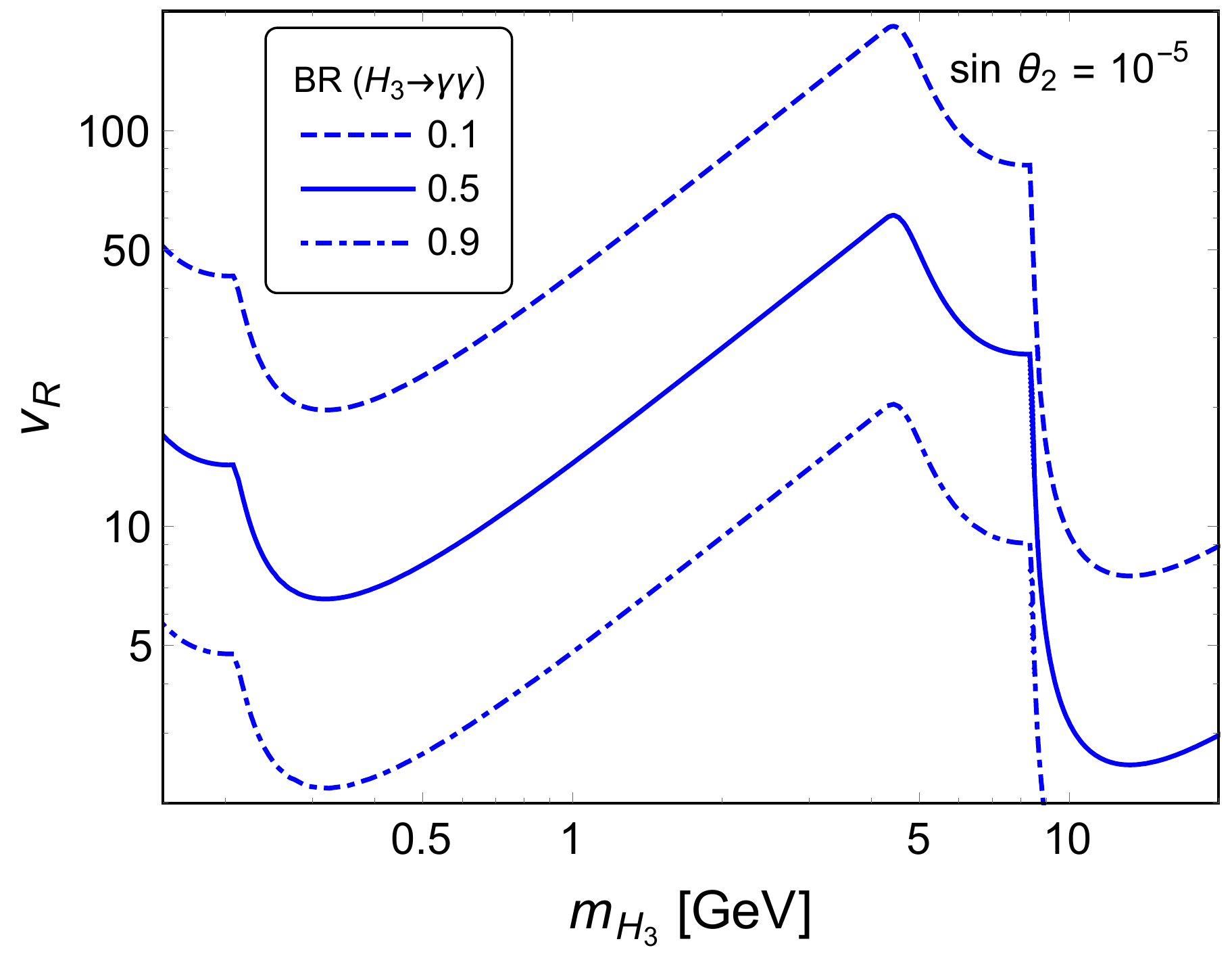}
  \caption{Contours of the ${\rm BR} (H_3 \to \gamma\gamma) = 0.1$, 0.5 and 0.9 in the plane of $m_{H_3} - v_R$ for fixed value of mixing angles $\sin\theta_1 = 10^{-4}$ (left) and $\sin\theta_2 = 10^{-5}$ (right). }
  \label{fig:vR}
\end{figure}

As shown in Figure~\ref{fig:BR}, when the mixing angles $\sin\theta_{1\,(2)}$ are small, e.g. $\lesssim 10^{-4\,(5)}$, all the fermion decay modes are highly suppressed, leaving the diphoton channel as the only dominant mode. When the mixing angles become very small, the fermionic decay modes are completely negligible. In this case, the diphoton channel, being mediated by the heavy gauge and scalar bosons, depends only on the $v_R$ scale, as mentioned above, and therefore, could probe, in principle, up to very high $v_R$ scales, as along as the colliding energy is high enough for a sizable $H_3$ production cross-section through the (off-shell) heavy gauge boson mediation (see Section~\ref{sec:prod}). To be more specific, we show in Figure~\ref{fig:vR} the contours of constant ${\rm BR} (H_3 \to \gamma\gamma) = 0.1$, 0.5 and 0.9, as a function of $v_R$ and the light scalar mass for a fixed value of the small mixing angles $\sin\theta_1 = 10^{-4}$ (left panel) and $\sin\theta_2 = 10^{-5}$ (right panel), as enforced by the meson limits in Section~\ref{sec:lab}, while the other one is set to be zero for simplicity. It is clear that the RH scale could be probed up to tens of TeV through this  diphoton channel. Although such large $v_R$ values might not be directly probed at the LHC, it could be relevant to the searches for LR seesaw at a future 100 TeV collider, such as FCC-hh or SPPC. See Section~\ref{sec:collider} for more realistic LLP searches at hadron colliders.

\begin{figure}[!t]
  \centering
  \includegraphics[width=0.47\textwidth]{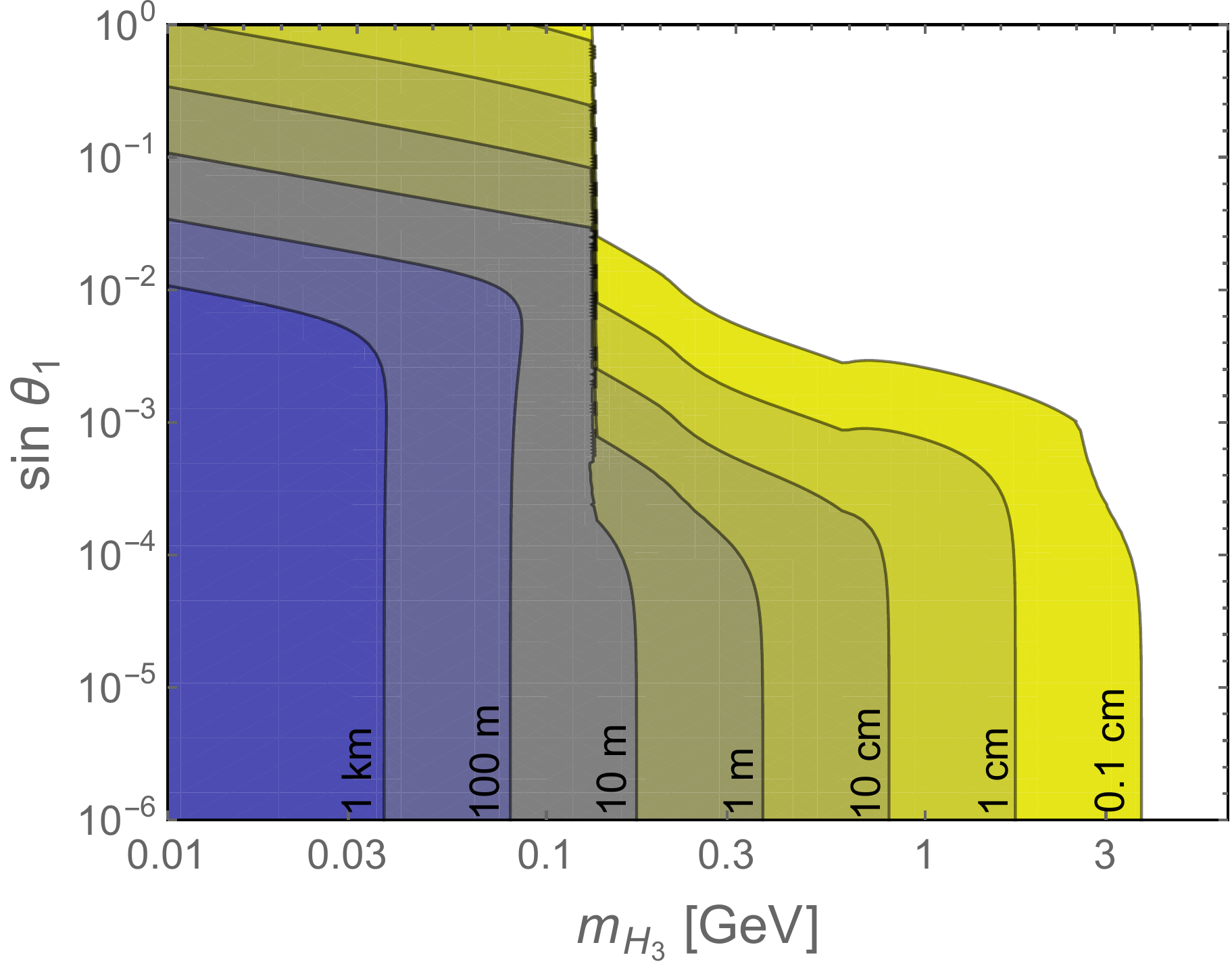}\hspace{0.2cm}
  \includegraphics[width=0.47\textwidth]{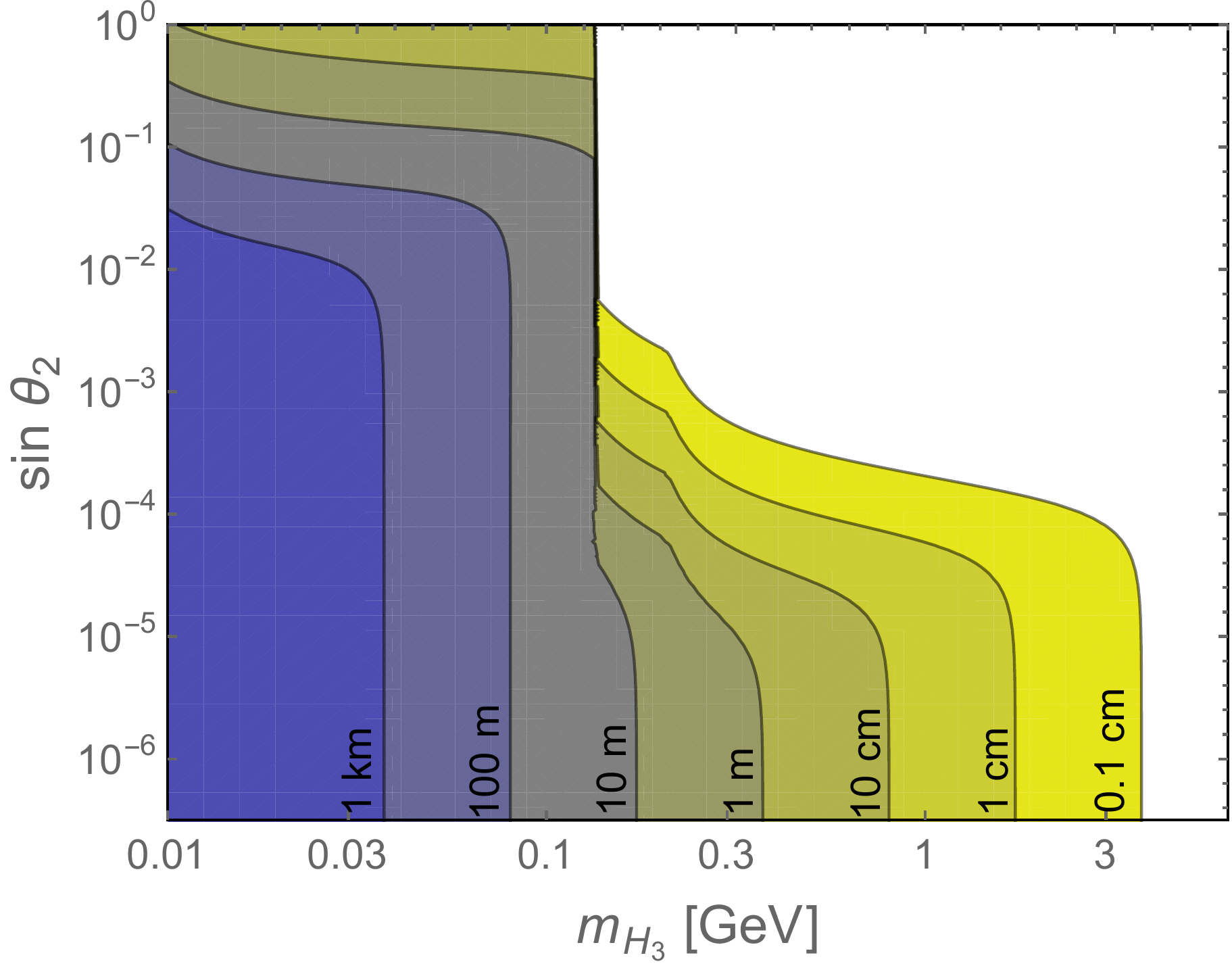}
  \caption{Contours of decay length of $H_3$ at rest as a function of its mass and the mixing angles $\sin\theta_{1,2}$. Here we have set the RH scale $v_R = 5$ TeV and the RHN masses at 1 TeV. }
  \label{fig:lifetime}
\end{figure}

Using the partial decay widths given in~\ref{app:decay}, we calculate the total decay length $L_0$ of $H_3$ at rest, as shown in Figure~\ref{fig:lifetime}, as a function of its mass and mixing with $h$ (left panel) and $H_1$ (right panel). From these lifetime contours, it is obvious that when the scalar $H_3$ mass is in the GeV range, its {\it proper} decay length $L_0 \sim {\cal O}(1)$ cm. When $H_3$ is produced at the LHC, it will be boosted by a Lorentz factor of $E_{H_3} / m_{H_3} \sim {\cal O}(100)$, so the decay length $L$ in the laboratory frame would reach the scale of meters, thus making it a natural LLP candidate and leading to spectacular displaced vertex signals. For even smaller masses, $m_{H_3}\sim {\cal O} (100)$ MeV, the decay length is much longer, of order 100 m. More details of displaced vertex and LLP searches are presented in Section~\ref{sec:collider}.

\section{Cosmological constraints } \label{sec:cosmo}

Light particles can have an impact on the cosmological history of our Universe, depending on their decay properties. In our model, the $H_3$ particle is produced in the early Universe by various processes mediated by heavy scalars and gauge bosons, but at temperatures $T\ll m_{Z'}, m_{W_R}, m_h$, the dominant process that keeps $H_3$ in equilibrium is $\gamma\gamma\to H_3$. For a GeV-scale $H_3$, it stays in equilibrium till below its mass   and decouples at a lower temperature $T_*$, which can be estimated by equating the rate of the process $\gamma\gamma\leftrightarrow H_3$ to the Hubble expansion rate:
\begin{eqnarray}
 \frac{ \alpha^2 T_*^3 x^{3/2} e^{-x}}{1048\pi^3} \ \leq \ \frac{10 T^2_*}{M_{\rm Pl}} \, ,
\label{eq:Tstar}
 \end{eqnarray}
where $x=m_{H_3}/T_* $, $\alpha\equiv e^2/4\pi$ is the fine structure constant and $M_{\rm Pl}$ is the Planck mass. Another relevant parameter is the decay temperature which is determined by the condition $T_d=\sqrt{\Gamma_{H_3} H(T_d)}$, where $\Gamma_{H_3}$ is the total decay rate of $H_3$ (thermal averaged) and $H(T_d)$ is the Hubble expansion rate at temperature $T_d$. Typically, one requires either (i) $T_d\geq \Lambda_{\rm QCD}\sim 150$ MeV, the QCD phase transition temperature, or (ii) $T_*\geq T_{\rm BBN}\sim 1$ MeV, whichever is stronger, so that the primordial synthesis of light elements and the ratio of their abundances is not affected much from their SM predicted values nor the light particle does not contribute like an extra degree of freedom at the epoch of Big Bang Nucleosynthesis (BBN). In our model, $T_d \gg T_*$ till the BBN epoch, and therefore, the condition (ii) is more stringent. Thus, we find that that as long as $T_* \gtrsim 1$ MeV (or the epoch of BBN), the $H_3$ particle will decouple and then decay to photons which will then thermalize with the rest of the cosmic soup (e.g. by Compton scattering). This will simply reset the Hubble temperature and will not affect the cosmological history. On the other hand, if $T_*$ is below the epoch of BBN, the $H_3$ particle is in equilibrium with the thermal soup and will contribute like an extra boson species  and being spin zero will contribute $4/7$ to $\Delta N_{\rm eff}$, 
which is incompatible with the Planck bounds at the 2.5$\sigma$ level~\cite{Ade:2015xua}. Using Eq.~\eqref{eq:Tstar} and setting $T_*\gtrsim 1$ MeV, we therefore obtain a conservative lower bound on $m_{H_3}\gtrsim 20$ MeV, which will be applied to our subsequent discussion.\footnote{$H_3$ masses below the supernovae core temperature of ${\cal O}$(10 MeV) could also be constrained from the observation of SN1987A~\cite{Heurtier:2016otg}.} A more accurate cosmological lower bound on $m_{H_3}$ might be obtained by solving the relevant Boltzmann equations and calculating the temperature rise of the thermal plasma due to energy injection from the $H_3$ decay, but such a detailed analysis is beyond the scope of this work and might be pursued elsewhere.


\section{Laboratory constraints}
\label{sec:lab}

The light scalar $H_3$ mixing with the heavy flavor-changing scalar $H_1$ induces flavor-changing couplings of $H_3$ to the SM quarks and charged leptons [cf.~Table~\ref{tab:coupling}], which are severely constrained by the low-energy flavor data, e.g. the $K^0 - \overline{K}^0$, $B_d - \overline{B}_d$ and $B_s - \overline{B}_s$ meson oscillations and rare $K$ and $B$ meson decays.\footnote{The constraints from $D$ meson sector are much weaker and thus not considered here.} In addition, the $H_3$ couplings are also limited by the SM invisible decay, rare top and $Z$ boson decay, which are either absent or highly suppressed in the SM. In this section, we collect all these laboratory constraints on the $h-H_3$ and $H_1-H_3$ mixing angles $\theta_{1,2}$, as well as their future prospects, which will provide useful guidelines for the collider searches for $H_3$ in the dominant $\gamma\gamma$ channel, as we will discuss in Section~\ref{sec:collider}.

\subsection{$K$ and $B$ meson oscillations}\label{sec:osc}
It should be emphasized that although the flavor-changing couplings of $H_3$ originate from the heavy scalar $H_1$, the masses of $H_1$ and $H_3$ are independent observables (proportional respectively to $\sqrt{\alpha_3}$ and $\sqrt{\rho_1}$ at the leading order), therefore the constraints on $H_3$ derived here from flavor oscillations are different from those on the heavy scalar $H_1$ derived earlier in Ref.~\cite{Zhang:2007da, Bertolini:2014sua}. Furthermore, the constraints on the mixing angles $\sin\theta_{1,2}$ from meson oscillations is sensitive to the mass of $H_3$, especially when $m_{H_3}$ is comparable to the $K$ or $B$ meson masses.

Taking the $K^0 - \overline{K}^0$ mixing as an explicit example, the effective four-fermion interactions mediated by $H_3$ can be cast into linear combinations of the effective dimension-6 operators of form~\cite{Babich:2006bh}
\begin{eqnarray}
\mathcal{O}_2 &\ = \ & [\bar{s} (1-\gamma_5) d] [\bar{s} (1-\gamma_5) d] \,, \\
\widetilde{\mathcal{O}}_2 & \ = \ & [\bar{s} (1+\gamma_5) d] [\bar{s} (1+\gamma_5) d] \,, \\
\mathcal{O}_4 & \ = \ & [\bar{s} (1-\gamma_5) d] [\bar{s} (1+\gamma_5) d] \,.
\end{eqnarray}
Though the flavor-changing couplings of $H_3$ to the SM fermions are from the mixing with $h$ and $H_1^0$ (cf.~the $\sin\tilde\theta_2$ terms of the Yukawa couplings in Table~\ref{tab:coupling}), but here they are not simply multiplied by a factor of $\sin\tilde\theta_2$. This is because the operators of form $\mathcal{O}_2$ and $\widetilde{\mathcal{O}}_2$ are absent in the $H_1$ case, which are canceled by the CP-odd scalar $A_1$ in the mass degenerate limit of $m_{H_1^0} = M_{A_1^0}$. In short, the effective Lagrangian we need is
\begin{eqnarray}
\label{eqn:Leff}
\mathcal{L}_{H_3} & \ = \ & \frac{G_F}{4\sqrt2}
\frac{\sin^2 \tilde{\theta}_2}{m_K^2 - m_{H_3}^2 + i m_{H_3} \Gamma_{H_3}} 
\left[
\left( \sum_i m_i \lambda_i^{RL} \right)^2 \mathcal{O}_2 +
\left( \sum_i m_i \lambda_i^{LR} \right)^2 \widetilde{\mathcal{O}}_2 \right. \nonumber \\
&& \hspace{5cm} \left. + 2
\left( \sum_i m_i \lambda_i^{LR} \right)
\left( \sum_i m_i \lambda_i^{RL} \right) \mathcal{O}_4
\right] \,,
\end{eqnarray}
where $G_F$ is the Fermi constant, $m_i = \{ m_u, m_c, m_t \}$ the running up-type quark masses, $\lambda_i^{LR} = V_{L,\, i2}^\ast V_{R,\, i1}$ and $\lambda_i^{RL} = V_{R,\, i2}^\ast V_{L,\, i1}$ the left- and right-handed quark mixing matrix elements. For simplicity we have assumed that $V_R = V_L$ which is a good approximation in the minimal LR model, up to some additional CP violating phases in the RH matrix~\cite{Senjanovic:2014pva, Senjanovic:2015yea}. As for the heavy scalars $H_1$ and $A_1$, the charm quark dominates the mass and quark mixing terms, i.e. $m_c \lambda$ ($\lambda$ being the Cabibbo angle), with sub-leading term from the top quark $\sim m_t \lambda^5$.

To calculate the contribution of Lagrangian~(\ref{eqn:Leff}) to the $K^0-\overline{K}^0$ mixing, we need to know the hadronic matrix elements when the operators are sandwiched by the $K^0$ states:
\begin{eqnarray}
\langle K^0 | \mathcal{O}_i | \overline{K}^0 \rangle &\ = \ &
N_i m_K f_K^2 B_i (\mu) R_K^2 (\mu) \,, 
\end{eqnarray}
with $i =2,$ 4, and the $K$ decay constant $f_K = 113$ MeV, $N_2 = 5/3$, $N_4 = -2$ and the parameters $B_2 = 0.679$, $B_4 = 0.810$ from lattice calculation~\cite{Babich:2006bh}. The mass ratio factor $R_K = m_K / (m_d + m_s)$ is evaluated at the energy scale $\mu = 2$ GeV. As the strong interaction conserves parity, we have $\langle K^0 | \widetilde{\mathcal{O}}_2 | \overline{K}^0 \rangle = \langle K^0 | \mathcal{O}_2 | \overline{K}^0 \rangle$. Then the $K^0$ mass difference
\begin{eqnarray}
\Delta m_K \ \simeq \ 2 {\rm Re} \, \sum_i \eta_i (\mu)
\langle K^0 | \mathcal{L}_{H_3}^{(i)} | \overline{K}^0 \rangle
\end{eqnarray}
with $\eta_2= 2.052$ and $\eta_4=3.2$ the QCD radiative corrections running from the EW scale down to the scale of $\mu \sim 2$ GeV~\cite{Buras:2001ra}.

On the experimental side, the $K^0 - \overline{K}^0$ mixing has been measured to a high accuracy, i.e.~$\Delta m_K = (3.473 \pm 0.006) \times 10^{-15}$ GeV~\cite{PDG}; on the theoretical side, the short- and long-distance contributions to $\Delta m_K$ are much larger than the experimental errors, up to 20\% of the central value. Conservatively we use 50\% of the experimental central value~\cite{Bertolini:2014sua} to set upper limits on the mixing angles $\sin {\theta}_{1,2}$, as shown in Figure~\ref{fig:limits_K} (blue solid lines).
As expected, in the (narrow) resonance region where $m_{H_3} \simeq m_K$ the limit on the mixing angles could be largely strengthened. When the $H_3$ mass gets lower, the $H_3$ propagator is dominated by the momentum term
\begin{eqnarray}
\frac{1}{q^2 - m_{H_3}^2 + i m_{H_3} \Gamma_{H_3}} \ \to \ \frac{1}{q^2} \ \simeq \ \frac{1}{m_K^2} \,,
\end{eqnarray}
and the limits approach to a constant value. On the other hand, when $m_{H_3} \gg m_K$, the constraints are similar to that for the heavy scalar $H_1$, and scale as $\sin\theta_{1,2}^{\rm limit} \propto m_{H_3}$.



The calculation of flavor constraints from $B_d - \overline{B}_d$ and $B_s - \overline{B}_s$ mixings are quite similar to those from $K^0$, with the QCD correction coefficient $\eta_2 = 1.654$ and $\eta_4 = 2.254$ at the $B$ meson scale~\cite{Buras:2001ra}, and the $B$-parameters for the effective operators with respect to the bottom quark and $d$ quark (or the $s$ quark) are respectively~\cite{Becirevic:2001xt}
\begin{eqnarray}
&& B_2 (B_d) \ = \ 0.82 \,, \quad
B_4 (B_d) \ = \ 1.16 \,, \nonumber \\
&& B_2 (B_s) \ = \ 0.83 \,, \quad
B_4 (B_s) \ = \ 1.17 \,.
\end{eqnarray}
Different from the $K$ meson case, for the $B_{d,s}$ mesons, the top quark contribution dominates in Eq.~\eqref{eqn:Leff}, which largely improves the effective coupling $\sum_i m_i \lambda_i^{LR,\, RL}$ and strengthens the limits on the coupling of $H_3$ to the bottom quark. 
The experimental values of $\Delta m_{B_{d,s}}$ agree well with the SM predictions~\cite{PDG}, allowing new physics contributions of only $9.3 \times 10^{-14}$ GeV and $2.7 \times 10^{-12}$ GeV respectively at the $2\sigma$ level by the current CKM fitter global fit, when CP violation is neglected~\cite{Charles:2015gya}.
The corresponding upper limits on the mixing angles $\sin\theta_{1,2}$
are presented in Figure~\ref{fig:limits_B}.
The $B$ mesons are roughly 10 times heavier than the $K$ meson, and the absolute values of error bars for the $B$ mass differences are much larger than $\Delta m_K$, thus the limits for the case of $m_{H_3} \lesssim m_{\rm meson}$ is weaker for the $B$ mixing case than that from $K$ mixing. However, this could be partially compensated by the large effective coupling $\sum_i m_i \lambda_i^{LR,\,RL}$ when $H_3$ is heavier, especially for the $B_d$ meson. Thus when $m_{H_3} \gtrsim m_b$, the limits on $\sin \tilde{\theta}_2$ from the $B$ meson mixings turn out to be more stringent. 






\subsection{Meson decay} \label{sec:mdecay}

Since $H_3$ acquires flavor-changing couplings to the SM quarks via its mixing with the heavy Higgs $H_1$, it could be produced from the flavor changing decay of $K$ and $B$ mesons, when kinematically allowed. The constraints coming from the up-sector FCNC are very weak and we do not discuss them here. In the down-type quark sector, we have the parton-level processes $b \to d H_3 ,\, s H_3$ and $s \to d H_3$ at the tree level. Depending on the mass $m_{H_3}$ and the mixing angles $\sin\theta_{1,2}$, after being produced in $K$ or $B$ decays, $H_3$ will decay into dileptons $\ell^+ \ell^-$, hadronic states $q \bar q$, $gg$, or two photons $\gamma\gamma$,
with the invariant mass of the final states close to the $H_3$ mass. Thus, we should expect flavor-violating signals of the form
\begin{eqnarray}
d_j \ \to \ d_i H_3^0,\, \qquad {\rm with} \quad H_3 \ \to \ {\rm leptons,~hadrons,~photons} \, .
\label{eq:decay}
\end{eqnarray}
The corresponding SM decay modes, like $B^+ \to K^+ \gamma\gamma$, are generally highly suppressed by the CKM matrix elements and loop factors, thus these rare decay channels are expected to set severe limits on the mixing angles $\sin\theta_{1,2}$ and hence the flavor  changing couplings of $H_3$ to quarks.

The most stringent bounds on the decay branching ratios for the process~\eqref{eq:decay} from various low-energy flavor experiments are collected in Table~\ref{tab:limits}, some of which follow to some extent the discussion of Refs.~\cite{Bezrukov:2009yw,Clarke:2013aya,Dolan:2014ska,Izaguirre:2016dfi}. The first simple but robust limits come from the observed total widths of $K$ and $B$ mesons, which depend only on the flavor-changing couplings of $H_3$ but not on how $H_3$ decays or the details of FCNC data. The lifetimes of $K^\pm$ and $K^0$ are both precisely measured to the level of $10^{-3}$, however the absolute theoretical values are subject to a large uncertainty of the strange quark mass, up to the order of 10\%~\cite{PDG}. Thus, to be conservative, we take 20\% of the experimental values to constrain the light scalar $H_3$, which are respectively $1.33 \times 10^{-17}$ GeV and $3.21 \times 10^{-18}$ GeV when converted to the maximum allowed discrepancy in the total widths of $K^\pm$ and $K^0$. On the other hand, for the $B$  meson, though the lifetime ratios such as $\tau_{B^\pm} / \tau_{B^0}$ can be determined up to the level of a few \%, the absolute values of $\tau_B$ are subject to large uncertainties in the form factors, at the level of 10\%~\cite{PDG}. The $2\sigma$ lifetime uncertainties lead to an allowed discrepancy of up to $1.05 \times 10^{-13}$ GeV in the total decay width of $B$ mesons.
The $K$ and $B$ meson width limits on $\sin\theta_{1,2}$ are shown in Figures~\ref{fig:limits_K} and \ref{fig:limits_B} as functions of $m_{H_3}$. 
All the regions above these lines are excluded, wherein the flavor-changing decays are enhanced by the large mixing angles.

Before going into the details of other meson decay limits, let us make some general comments on the constraints from meson decays. Roughly speaking, regarding the flavor changing couplings mediated by $H_3$, there are essentially two different classes of experiments that are applicable to our case. The first ones are the {\it visible} decays, i.e. those with visible SM particles in the final state such as $K^+ \to \pi^+ \gamma\gamma$ and $B \to K \mu^+ \mu^-$. The SM backgrounds for these rare decays are generally very small, and the tree-level flavor-changing couplings of $H_3$ could be severely constrained. However, the sensitivity depends largely on the selection procedure of signals, e.g. the vetoes, cuts and detector position and energy resolutions etc. The second class of processes are the {\it invisible} decays, i.e. the signal of type $d_j \to d_i + {\rm inv.}$ at the parton level. The invisible part, or missing energy at colliders, could be from the neutrinos, such as $K^+ \to \pi^+ \nu \bar\nu$. We include in this category the null result of dedicated searches for a light neutral particle $X^0$ from meson decay, e.g. $K^+ \to \pi^+ X^0$, with the light particle long-lived enough to escape from the detector without leaving any observable footprints. These invisible decays are expected to be very sensitive to the scalar $H_3$ which is an LLP from the detector perspective, as long as it is light and the mixings $\sin\theta_{1,2}$ are small. In this case, we take the conservative assumption that the sensitivity depends only on the detector size but not too much on the data analysis. For these two distinct categories of searches, we use the following two branching ratios to set limits on the mixing angles $\sin\theta_{1,2}$ and $m_{H_3}$:
\begin{eqnarray}
\label{eqn:BR}
{\rm visible}: &&
{\rm BR} (d_j \to d_i H_3) \,
{\rm BR} (H_3 \to \chi\chi) \,
\left[
\exp\left( -\frac{L \Gamma_{H_3}^{}}{b} \right) -
\exp\left( -\frac{{(L+\Delta L)} \Gamma_{H_3}^{}}{b} \right) \right] ,
\nonumber  \\ && \\
\label{eqn:BR2}
{\rm invisible}: &&
{\rm BR} (d_j \to d_i H_3)
\exp\left( -\frac{R \Gamma_{H_3}^{}}{b} \right) ,
\end{eqnarray}
where $\chi\chi = \ell^+ \ell^-$, hadrons and $\gamma\gamma$ are the visible SM particles, $b$ is the Lorentz boost factor, $L$ and $\Delta L$ denote respectively the distance from the primary production vertex and the decay length when $H_3$ decays into visible particles in the detector, and $R$ denotes the detector size in the invisible final state case. The two different search strategies are largely complementary to each other, when applied to constrain the light scalar $H_3$ in the LR model.

\begin{table}[!t]
  \centering
  \caption[]{Summary of meson decay constraints used to derive current/future limits on the mixing angles in Figures~\ref{fig:limits_K}, \ref{fig:limits_B} and \ref{fig:limits_dump}. The last column gives the upper limit on the BR of the process used in our calculation. The corresponding numbers (in parenthesis) for the beam-dump experiments (last six rows) give the limit on the number of events.}
  \label{tab:limits}
  \small
  \begin{tabular}[t]{rlllll}
  \hline\hline
 Experiment & Meson decay & $H_3$ decay & $E_{H_3}$ & Decay length & Limit on BR ($N_{\rm event}$) \\ \hline
  NA48/2~\cite{Batley:2009aa}
  & $K^+ \to \pi^+ H_3$ & $H_3 \to e^+ e^-$ & $\sim 30$ GeV & $< 0.1$ mm &
  $2.63 \times 10^{-7}$ \\
  NA48/2~\cite{Batley:2011zz}
  & $K^+ \to \pi^+ H_3$ & $H_3 \to \mu^+ \mu^-$ & $\sim 30$ GeV & $< 0.1$ mm & $8.88\times10^{-8}$ \\
  NA62~\cite{Ceccucci:2014oza}
  & $K^+ \to \pi^+ H_3$ & $H_3 \to \gamma\gamma$ & $\sim 37$ GeV & $< 0.1$ mm & $4.70\times10^{-7}$ \\ \hline
  E949~\cite{Anisimovsky:2004hr,Artamonov:2008qb,Artamonov:2009sz,Artamonov:2005cu}
  & $K^+ \to \pi^+ H_3$ & any (inv.) & $\sim 355$ MeV & $> 4$ m & $4\times10^{-10}$ \\
  NA62~\cite{Anelli:2005ju}
  & $K^+ \to \pi^+ H_3$ & any (inv.) & $\sim 37.5$ GeV & $> 2$ m & $2.4\times10^{-11}$ \\ \hline
 KTeV~\cite{AlaviHarati:2003mr}
  & $K_L \to \pi^0 H_3$ & $H_3 \to e^+ e^-$ & $\sim 30$ GeV & $< 0.1$ mm &
  $2.8 \times 10^{-10}$ \\
  KTeV~\cite{AlaviHarati:2000hs}
  & $K_L \to \pi^0 H_3$ & $H_3 \to \mu^+ \mu^-$ & $\sim 30$ GeV & $< 0.1$ mm & $4\times10^{-10}$ \\
  KTeV~\cite{Abouzaid:2008xm,Alexopoulos:2004sx}
  & $K_L \to \pi^0 H_3$ & $H_3 \to \gamma\gamma$ & $\sim 40$ GeV & $< 0.1$ mm & $3.71\times10^{-7}$ \\ \hline
  BaBar~\cite{Aubert:2003cm}
  & $B \to K H_3$ & $H_3 \to \ell^+ \ell^-$ & $\sim m_B/2$ & $< 0.1$ mm & $7.91\times10^{-7}$ \\
  Belle~\cite{Wei:2009zv}
  & $B \to K H_3$ & $H_3 \to \ell^+ \ell^-$ & $\sim m_B/2$ & $< 0.1$ mm & $4.87\times10^{-7}$ \\
  LHCb~\cite{Aaij:2012vr}
  & $B^+ \to K^+ H_3$ & $H_3 \to \mu^+\mu^-$ & $\sim 150$ GeV & $< 0.1$ mm & $4.61\times10^{-7}$ \\ \hline
  BaBar~\cite{Lees:2013kla}
  & $B \to K H_3$ & any (inv.) & $\sim m_B/2$ & $> 3.5$ m & $3.2\times10^{-5}$ \\
  Belle II~\cite{Abe:2010gxa}
  & $B \to K H_3$ & any (inv.) & $\sim m_B/2$ & $> 3$ m & $4.1\times10^{-6}$ \\
  \hline
  LHCb~\cite{CMS:2014xfa} &
  $B_s \to \mu\mu$ & $-$ & $-$ & $-$ & $2.51\times10^{-9}$ \\
  BaBar~\cite{delAmoSanchez:2010bx} &
  $B_d \to \gamma\gamma$ & $-$ & $-$ & $-$ & $3.3\times10^{-7}$ \\
  Belle~\cite{Dutta:2014sxo}  &
  $B_s \to \gamma\gamma$ & $-$ & $-$ & $-$ & $3.1\times10^{-6}$ \\ \hline
  BaBar~\cite{Lees:2011wb}
  & $\Upsilon \to \gamma H_3$ & $H_3 \to qq,\,gg$ & $\sim m_\Upsilon/2$ & $< 3.5$ m & $[1,\,80] \times 10^{-6}$ \\ \hline
  CHARM~\cite{Bergsma:1985qz}
  & $K \to \pi H_3$ & $H_3 \to \gamma\gamma$ & $\sim 10$ GeV & $[480,\,515]$ m & $(< 2.3)$ \\
  CHARM~\cite{Bergsma:1985qz}
  & $B \to X_s H_3$ & $H_3 \to \gamma\gamma$ & $\sim 10$ GeV & $[480,\,515]$ m & $(< 2.3)$ \\
  SHiP~\cite{Alekhin:2015byh}
  & $K \to \pi H_3$ & $H_3 \to \gamma\gamma$ & $\sim 25$ GeV & $[70,\,125]$ m & $(< 3)$ \\
  SHiP~\cite{Alekhin:2015byh}
  & $B \to X_s H_3$ & $H_3 \to \gamma\gamma$ & $\sim 25$ GeV & $[70,\,125]$ m & $(< 3)$ \\
  DUNE~\cite{Adams:2013qkq}
  & $K \to \pi H_3$ & $H_3 \to \gamma\gamma$ & $\sim 12$ GeV & $[500,\,507]$ m & $(< 3)$ \\
  DUNE~\cite{Adams:2013qkq}
  & $B \to X_s H_3$ & $H_3 \to \gamma\gamma$ & $\sim 12$ GeV & $[500,\,507]$ m & $(< 3)$ \\
  \hline\hline
  \end{tabular}
\end{table}


\subsubsection{$K$ meson decay}\label{sec:kdecay}
The partial width for the charged $K$ meson decay is given by~\cite{Dolan:2014ska,Izaguirre:2016dfi}
\begin{eqnarray}
      \label{eqn:Kdecay}
      \Gamma (K^\pm \to \pi^\pm H_3) &\ = \ &
      \frac{G_F m_{K^\pm} \sin^2\tilde{\theta}_2}{8\sqrt2 \pi}
      \left| \sum_i m_i \lambda_{i,21}^{RL} \right|^2
      \left( 1 - \frac{m_{\pi^\pm}^2}{m_{K^\pm}^2} \right)^2 \, ,
\end{eqnarray}
where the kinetic function $\beta_2$ is defined in Eq.~\eqref{eqn:beta2}.
For a CP-even scalar, the decay width for the neutral $K$ meson, i.e. $K_L\to \pi^0H_3$ is related to the charged counterpart by taking the real part of the amplitude in Eq.~\eqref{eqn:Kdecay}~\cite{He:2006uu}.

The ${\rm BR} (K^+ \to \pi^+ e^+ e^-)$ and ${\rm BR} (K^+ \to \pi^+ \mu^+ \mu^-)$ are predicted to be respectively $(3.9 \pm 0.8) \times 10^{-7}$ and $(1.2 \pm 0.3) \times 10^{-7}$ in the SM~\cite{Dubnickova:2006mk}. Taking the largest discrepancy of the theoretical and experimental values from NA48/2~\cite{Batley:2009aa,Batley:2011zz} at the $2\sigma$ C.L., we obtain the maximum allowed contribution from potential beyond SM physics, which are listed in the last column of Table~\ref{tab:limits}.
Regarding the rare kaon decay with two photons in the final state, i.e. $K^+ \to \pi^+ \gamma\gamma$, we adopt the SM prediction of $(9.66 \pm 3.43) \times 10^{-7}$ with the invariant mass of diphoton $(m_{\gamma\gamma} / m_K)^2 >0.2$~\cite{Gerard:2005yk}. Comparing it to the measurement at NA62~\cite{Ceccucci:2014oza}, we arrive at the BR limit of $4.70 \times 10^{-7}$ at $2\sigma$ C.L.

\begin{figure}[!t]
  \centering
  \includegraphics[width=0.49\textwidth]{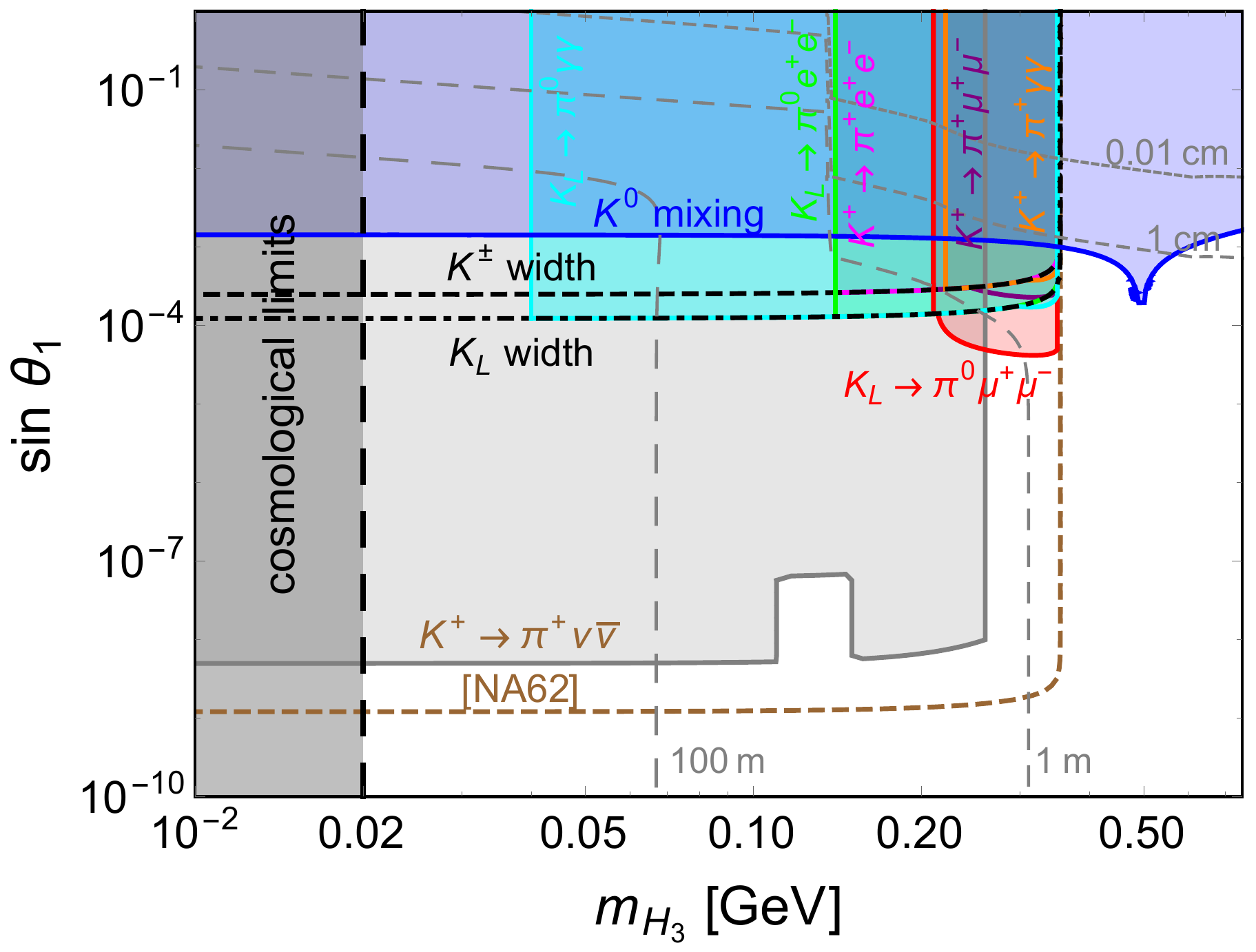}
  \includegraphics[width=0.49\textwidth]{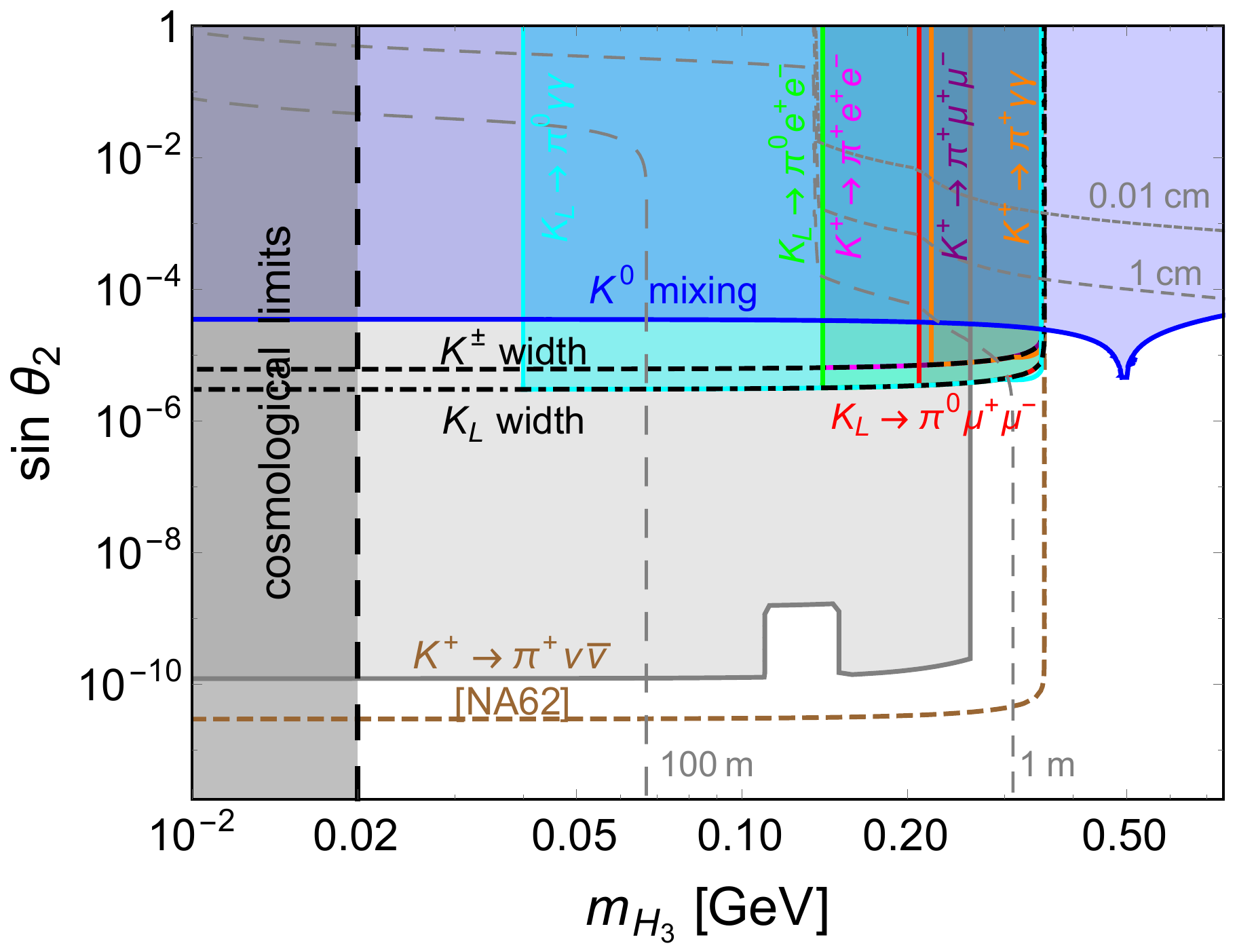}
  \caption{Flavor changing limits on the mixing angles $\sin\theta_{1,2}$ and $m_{H_3}$ from $K-\bar{K}$ mixing (blue) and various charged and neutral kaon decay modes: $K^\pm \to \pi^\pm \chi\chi$, in the final state of leptons $\chi\chi = e^+ e^-$~\cite{Batley:2009aa}, $\mu^+ \mu^-$~\cite{Batley:2011zz} or $\gamma\gamma$~\cite{Ceccucci:2014oza} in the NA48/2 and NA62 experiments, and $K_L \to \pi^0 \chi\chi$, in the final state of leptons $\chi\chi = e^+ e^-$~\cite{AlaviHarati:2003mr}, $\mu^+ \mu^-$~\cite{AlaviHarati:2000hs} or $\gamma\gamma$~\cite{Abouzaid:2008xm,Alexopoulos:2004sx} in the KTeV experiment. The invisible limits from $K^\pm \to \pi^\pm \nu \bar\nu$ is from the E949 experiment~\cite{Anisimovsky:2004hr,
  Artamonov:2008qb,Artamonov:2009sz,Artamonov:2005cu}.  The dashed brown curve is the expected sensitivity from NA62~\cite{Anelli:2005ju}. The dashed (dot-dashed) black curve is the limit from the total $K^\pm$ ($K_L$) width, whereas $m_{H_3}\lesssim 20$ MeV (vertical black-shaded) is disfavored from BBN considerations (cf.~Section~\ref{sec:cosmo}). The dashed gray lines are the {\it proper} lifetime of $H_3$ with values of 0.01 cm, 1 cm, 1 m, and 100 m. See text for more details.}
  \label{fig:limits_K}
\end{figure}

The kaon beam energy at NA48/2 is around $E_{\rm NA} \simeq 60$ GeV or slightly higher, which leads to a large boost factor of $b \simeq E_{\rm NA} / 2m_{H_3}$ in Eq.~(\ref{eqn:BR}). The position resolution of the detector could reach up to $\lesssim 1$ mm; if the LLP $H_3$ were produced in these experiments, the signal would be very different from the SM processes: i.e. there would be displaced $ee$, $\mu\mu$ or $\gamma\gamma$ tracks from the primary kaon vertex, which could be easily identified in the detector layers. To be concrete, we adopt a smaller decay length of $\Delta L = 0.1$ mm, which is more conservative than in Refs.~\cite{Clarke:2013aya,Izaguirre:2016dfi}. Compared to the invisible decays with $H_3$ leaving no trace in the detector, the visible searches are more sensitive to shorter-lived $H_3$ with larger mixing angles $\sin\theta_{1,2}$. The excluded regions from $K^+ \to \pi^+ \ell\ell$ are presented in the plots of Figure~\ref{fig:limits_K}.




In the invisible searches, the most stringent bounds are from the process $K^+ \to \pi^+ \nu\bar\nu$ in the experiment E949~\cite{Anisimovsky:2004hr,Artamonov:2008qb,Artamonov:2009sz,Artamonov:2005cu}, with neutrinos in the final state. In calculation of the limits using Eq.~(\ref{eqn:BR2}),
we set $E_{K} \simeq 710$ MeV~\cite{Adler:2008zza}, adopt conservatively the decay length $L = 4$ m~\cite{Dolan:2014ska}\footnote{A smaller decay length will make the constraints more stringent, see Ref.~\cite{Clarke:2013aya}.}, and use the BR limits of $4 \times 10^{-10}$~\cite{Anisimovsky:2004hr,Artamonov:2008qb,Artamonov:2009sz} to set separate limits on $\sin\theta_1$ and $\sin\theta_2$ by setting the other to be zero, as functions of $H_3$ mass, as shown in Figure~\ref{fig:limits_K}. When $m_{H_3}$ is close to $m_{\pi}$, we adopt the limit from $K^+ \to \pi^+ \pi^0$ with $\pi^0 \to \nu\bar\nu$~\cite{Artamonov:2005cu}, which is less constraining, with the BR up to $6 \times 10^{-8}$. 

The limits from $K^+ \to \pi^+ \nu \bar\nu$ is expected to be more stringent at the proposed running of NA62~\cite{Anelli:2005ju}, with a precision up to
10\% of the SM value~\cite{Anelli:2005ju}.
The theoretical uncertainty is below 4\%~\cite{Buras:2015qea}, thus we take the expected largest $2\sigma$ total uncertainties $2 \times (0.10+0.04) \times 8.4 \times 10^{-11} = 2.35 \times 10^{-11}$ as the expected NA62 limit to constrain the couplings of $H_3$, assuming the measurement of NA62 agrees well with the SM prediction.





The rare decays of neutral kaon $K_L \to \pi^0 \chi\chi$ have been searched for at the KTeV experiment, with lepton pairs $\chi\chi = e^+ e^-,\,\mu^+ \mu^-$~\cite{AlaviHarati:2003mr,AlaviHarati:2000hs} or diphoton $\gamma \gamma$~\cite{Abouzaid:2008xm} in the final state.       The BRs are constrained to be very small, especially for the dileptons. Similar to the visible $K^\pm$ decay at NA48/2 and NA62, the kaon system is highly boosted, with a total energy ranging from 20 GeV to 220 GeV for the dilepton searches, and 40 GeV to 160 GeV for the diphoton decay. Examining the energy distribution, we take the mean energy to be 30 and 40 GeV for the kaon system.  Assuming a decay length of $\Delta L = 1$ mm, we get the visible decay limits as shown in Figure~\ref{fig:limits_K}. Around the $\pi^0$ mass, due to the resonance effect, the SM production rate of $K_L \to \pi^0 \pi^0$ is much larger than elsewhere, and we use the SM BR$(K_L \to \pi^0 \pi^0) = 9 \times 10^{-4}$ to set limits on $H_3$~\cite{Alexopoulos:2004sx}.

\subsubsection{$B$ meson decay}\label{sec:bdecay}
The partial width for the exclusive $B$ meson decay is very similar to that for $K$ meson [cf.~Eq.~(\ref{eqn:Kdecay})]:
\begin{eqnarray}
      \label{eqn:Bdecay}
      \Gamma (B \to K H_3) & \ = \ &
      \frac{G_F m_{B} \sin^2\tilde{\theta}_2}{8\sqrt2 \pi}
      \left| \sum_i m_i \lambda_{i,32}^{RL} \right|^2
      \left( 1 - \frac{m_K^2}{m_B^2} \right)^2
      \left[ f_0^{(K)} (m_{H_3}^2) \right]^2 \nonumber \\
      && \times \beta_2 (m_B, m_K, m_{H_3}) \,,
\end{eqnarray}
where $f_0 (q^2)$ is a form factor of the form~\cite{Ball:2004ye}
\begin{eqnarray}
      f_0 (q^2) \ =  \ \frac{r^2}{1-q^2/m_{\rm fit}^2} \,,
\end{eqnarray}
where for the $K$ meson final state, the parameters are $r_2 = 0.330$ and $m_{\rm fit}^2 = 37.46 \, {\rm GeV}^2$.
For the inclusive $B$ decays, we have
\begin{eqnarray}
      \label{eqn:Bdecay2}
      \Gamma (B \to X_s H_3) & \ = \ &
      \frac{G_F m_{B} \sin^2\tilde{\theta}_2}{4\sqrt2 \pi}
      \left| \sum_i m_i \lambda_{i,32}^{RL} \right|^2
      \left( 1 - \frac{m_{H_3}^2}{m_B^2} \right)^2 \,,
\end{eqnarray}
with $X_s$ standing for any strange-flavored meson.

\begin{figure}[!t]
  \centering
  \includegraphics[width=0.49\textwidth]{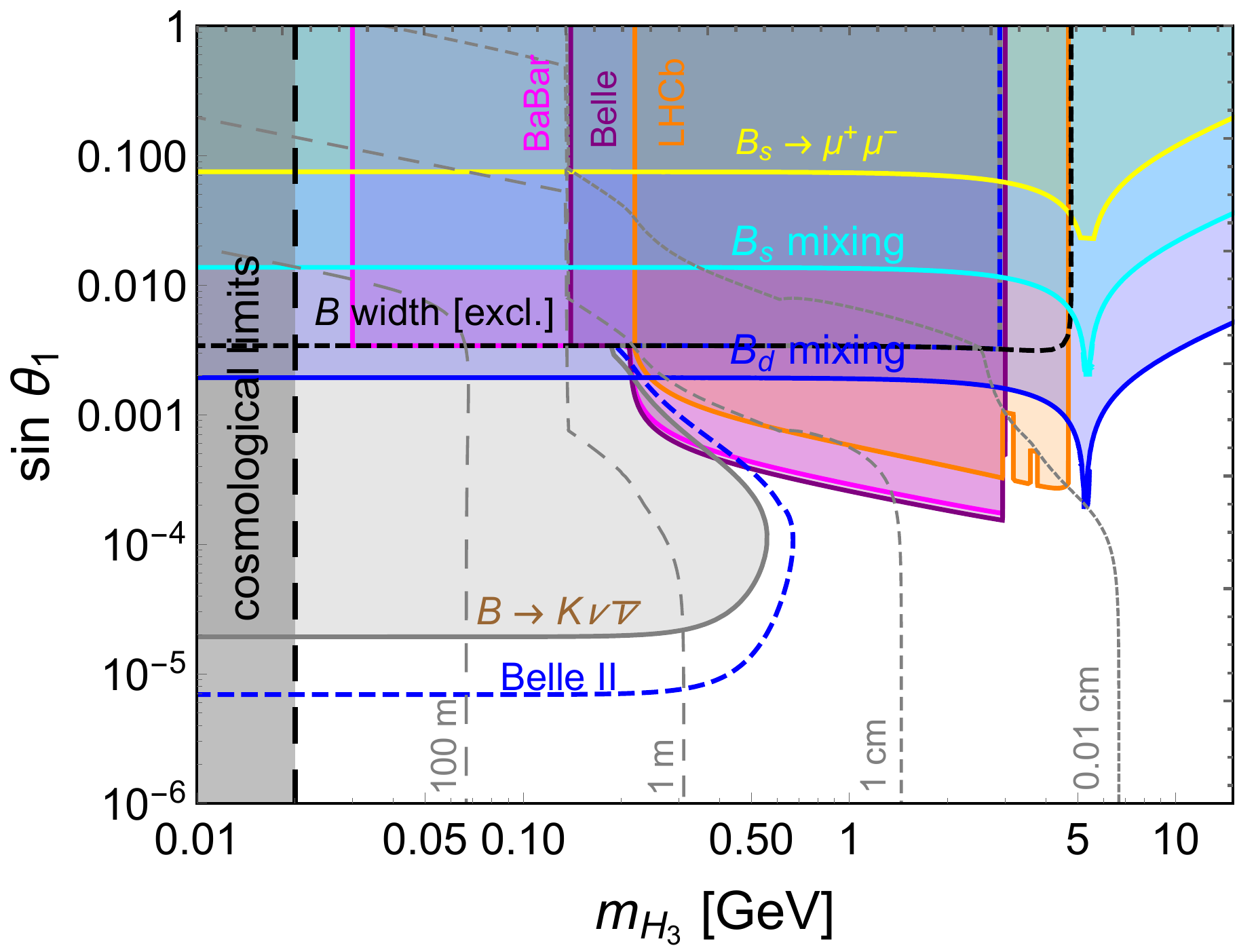}
  \includegraphics[width=0.49\textwidth]{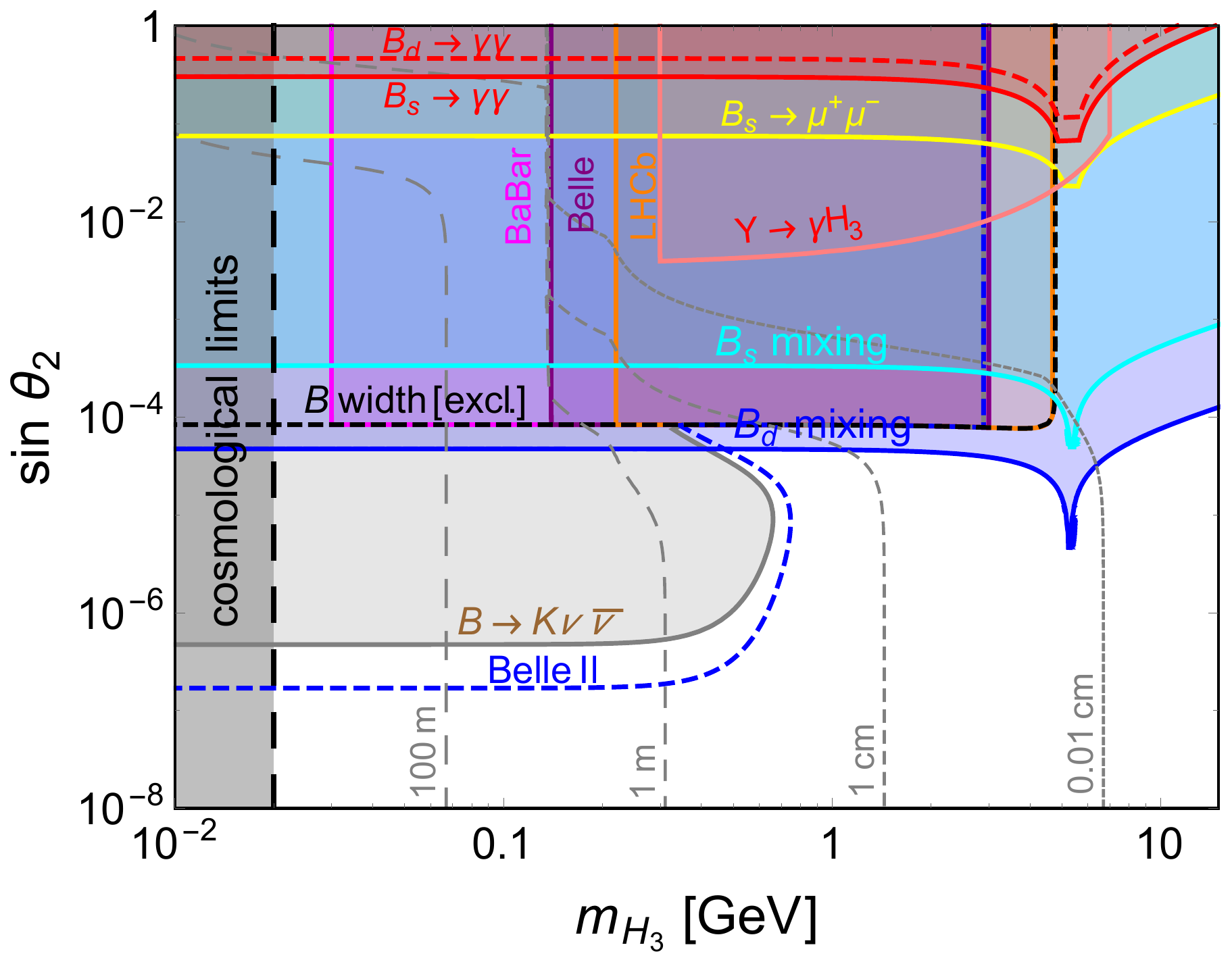}
  \caption{Flavor changing limits on the mixing angles $\sin\theta_{1,2}$ and $m_{H_3}$ from $B_d - \overline{B}_d$ (blue) and $B_s - \overline{B}_s$ (cyan) mixings~\cite{PDG}, $B$ meson decay $B \to K \ell^+ \ell^-$ at BaBar~\cite{Aubert:2003cm} and Belle~\cite{Wei:2009zv} and $B \to K \mu^+ \mu^-$ at LHCb~\cite{Aaij:2012vr}. The invisible limits from $B \to K \nu \bar\nu$ are from BaBar~\cite{Lees:2013kla} and future prospects at Belle II~\cite{Abe:2010gxa}.  The yellow lines are from the decay $B_s \to \mu^+ \mu^-$. In the right panel, there are additional limits from the null result of searches of $B_{d,s}\to \gamma\gamma$ (red)~\cite{delAmoSanchez:2010bx,Dutta:2014sxo}, as well as $\Upsilon \to \gamma H_3$ (pink) at BaBar~\cite{Lees:2011wb}. The dashed black curve is the limit from the exclusive $B$ decay width in Eq.~(\ref{eqn:Bdecay}). The future sensitivity of Belle II~\cite{Abe:2010gxa} is shown by the blue dashed curve. The vertical dashed (black) line is the cosmological limit. The dashed gray lines are the {\it proper} lifetime of $H_3$ with values of 0.01 cm, 1 cm, 1 m, and 100 m. See text for more details.}
  \label{fig:limits_B}
\end{figure}

In the SM, the BR of flavor changing decay $B \to K \ell^+ \ell^-$ ($\ell = e,\, \mu$) is predicted to be $5.7 \times 10^{-7}$, with large uncertainties from the form factor, top quark mass etc, summing up to 35\%~\cite{Ali:1999mm}. Comparing the theoretical prediction to the measurements at BaBar ($(6.5 \pm 1.5) \times 10^{-7}$)~\cite{Aubert:2003cm}, Belle ($(4.8 \pm 0.58) \times 10^{-7}$)~\cite{Wei:2009zv}, and LHCb ($(4.36 \pm 0.23) \times 10^{-7}$)~\cite{Aaij:2012vr} and taking the largest $2\sigma$ discrepancies of theoretical and experimental values, we collect the BR limits in Table~\ref{tab:limits}.
The three detectors all have very good spatial resolutions~\cite{Aubert:2001tu,Abe:2010gxa,Alves:2008zz}; to be concrete we take $\Delta L \sim 0.1$ mm, and setting $L =0$ in Eq.~(\ref{eqn:BR}) for all of them, we obtain the excluded regions shown in in Figure~\ref{fig:limits_B}. Again, for the $B$ mesons at LHCb, the average energy of the $B$ meson is $E_{B}^{(\rm LHCb)} \sim 300$ GeV, so we have a large boost factor. Compared to the $K$ decays, the flavor changing coupling to $b$ quark is largely enhanced by the factor $\sum_i m_i \lambda_{32}^{RL}$. When $m_{H_3} \sim m_{J/\psi}$ or $m_{\psi(2S)}$, we use the SM BRs to set limits on $H_3$~\cite{Dolan:2014ska,PDG}:
\begin{eqnarray}
      {\rm BR} (B \to K J/\psi) & \ = \ &
      {\rm BR} (B \to K \ell^+ \ell^-) \ = \ 5 \times 10^{-5} \,, \\
      {\rm BR} (B \to K \psi (2S)) & \ = \ &
      {\rm BR} (B \to K \ell^+ \ell^-) \ = \ 5 \times 10^{-6} \,.
\end{eqnarray}
With more $B$ mesons collected at Belle II~\cite{Abe:2010gxa}, the constraints in the visible modes could be further strengthened.

A light neutral scalar has been searched for in the decay $B \to K X^0$ in the CLEO experiment~\cite{Ammar:2001gi}, with an upper limit at the level of $5.3 \times 10^{-5}$. The rare decays of $B \to K \nu \bar\nu$ at BaBar leads to a more stringent limit: combining both the channels of $B^+ \to K^+ \nu \bar\nu$ and $B^0 \to K^0 \nu \bar\nu$, the BR is less than $3.2 \times 10^{-5}$ at the 90\% C.L.~\cite{Lees:2013kla},\footnote{There are also searches of $B \to K^\ast \nu \bar\nu$~\cite{Lees:2013kla,Abe:2010gxa}, but the limits are comparatively less constraining, thus we consider here only the $K$ mesons in the final state.} with the SM prediction of ${\rm BR} (B \to K \nu \bar\nu) = (4.5 \pm 0.7) \times 10^{-6}$~\cite{Altmannshofer:2009ma}. With a detector size of $L = 3.5$ m~\cite{Aubert:2001tu}, we can exclude large region in the plane of $m_{H_3}-\sin\theta_{1,2}$, as shown in Figure~\ref{fig:limits_B}. As for the visible $B$ decays, though the absolute value of the limit on ${\rm BR} (B \to K \nu \bar\nu)$ is much smaller than that from ${\rm BR}(K \to \pi \nu \bar\nu)$, the $B$ meson decay is comparatively enhanced by the larger flavor changing coupling of $\sum_i m_i \lambda_{i,32}$, with respect to the coupling of $\sum_i m_i \lambda_{i,21}$ for the $K$ mesons. With a total luminosity of 50 ab$^{-1}$ at SuperKEKB, the $B^+ \to K^+ \nu \bar\nu$ could be measured up to 30\% of the SM BR at Belle II~\cite{Abe:2010gxa}, i.e. $1.5 \times 10^{-6}$. Applied to the LR model, this means the flavor changing couplings to $b$ quark could be more severely constrained, as demonstrated by the dashed blue lines in Figure~\ref{fig:limits_B}.

With the flavor-changing couplings to the quarks in $B$ mesons, $H_3$ could also be produced off-shell from $B$ meson decay and then decays into light SM particles, such as the rare processes $B \to \gamma\gamma$ and $B_s \to \mu^+ \mu^-$. The BR of the latter process is given by~\cite{Altmannshofer:2011gn}
\begin{eqnarray}
\frac{{\rm BR} (B_s \to \mu^+ \mu^-)}{{\rm BR} (B_s \to \mu^+ \mu^-)_{\rm SM}} \ \simeq \
\frac{m_{B_s}^4}{4 m_\mu^2} \frac{|C^{S}|^2}{|C_{10}^{\rm SM}|^2}
\left( 1 - \frac{4m_\mu^2}{m_{B_s}^2} \right) \,,
\end{eqnarray}
where $C_{10}^{\rm SM} = -4.103$ and
\begin{eqnarray}
C^S = \frac{4\pi v_{\rm EW} {\cal Y}_{E, \mu\mu} \sin\tilde\theta_2}{\alpha V_{tb} V_{ts}^\ast m_b (m_{B_s}^2 - m_{H_3}^2 + i \Gamma_{H_3} m_{H_3})}
\left( \sum_i m_i \lambda_{i,32}^{RL} \right)
\end{eqnarray}
which is proportional to the flavor-changing coupling of $H_3$ to strange and bottom quarks and also the Yukawa coupling to muon.

The recent measurement of ${\rm BR} (B_s \to \mu^+ \mu^-)$ by LHCb is $(2.8 \pm 0.6) \times 10^{-9}$, compatible with the SM prediction of $(3.65 \pm 0.23) \times 10^{-9}$~\cite{CMS:2014xfa}, allowing a contribution of $2.51 \times 10^{-9}$ at the $2\sigma$ C.L. from the $H_3$ mediated processes.\footnote{There are also searches of $B^0 \to \mu^+ \mu^-$~\cite{CMS:2014xfa, Aaij:2017vad}, which is however suppressed by the small flavor-changing coupling $\sum_i m_i \lambda_{i, 31}$, as compared to the $B_s$ decay, in our model.} The $2\sigma$ uncertainties are used to set limits on the mixing angles $\sin\theta_{1,2}$ as functions of $m_{H_3}$, as depicted by the yellow lines in Figure~\ref{fig:limits_B}. Analogous to the case of $K$ and $B$ meson oscillations, when $m_{H_3} \sim m_{B_s}$ the constraints are enhanced by the resonance effect.


The rare decay $B_q \to \gamma\gamma$ is closely related to the parton-level flavor-changing process $b \to q \gamma\gamma$ ($q = d,\,s$), both of which are mediated by the  FCNC couplings of $H_3$ to the quarks. There are two distinct sets of Feynman diagrams~\cite{Aranda:2009kz,Aranda:2010qc}: the first one consists of the box diagrams with $H_3$ propagators for the couplings $bq\gamma\gamma$, the triangular diagrams for the trilinear coupling $bq\gamma$ and the $H_3$ mediated $bq$ bilinear coupling; the second class are those with an $s$-channel $H_3$ coupled to the two photons via the fermion, scalar and gauge boson loops. It is expected that the second class of diagrams dominates, producing the partial width~\cite{Aranda:2009kz,Aranda:2010qc}
\begin{eqnarray}
\Gamma (B_q \to \gamma\gamma) & \ = \ &
\frac{\alpha^3 m_{B_q}^5 f_{B_q}^2}{128\pi^2 s_W^2 m_W^2}
\frac{|{\cal Y}_{d,qb}|^2}{|m_{B_q}^2 - m_{H_3}^2 + im_{H_3}\Gamma_{H_3}|^2} \nonumber \\
&& \times \left| \sum_{f} Q_f^2 N_{C}^f A_{1/2}(\tau_f) + \frac{v_{\rm EW}}{v_R} \left(\frac{1}{3} + \frac43  -7\right) \right|^2 \,.
\end{eqnarray}
where the factors of $1/3$, $4/3$ and $-7$ are respectively from the heavy $H_1^\pm$, $H_2^{\pm\pm}$ and $W_R$ loops in the limit of $m_{H_3} \to 0$ [cf. Eq.~(\ref{eqn:loopfunction})]. The current most stringent upper limits on ${\rm BR} (B_d \to \gamma\gamma)$ and ${\rm BR} (B_s \to \gamma\gamma)$ are respectively $3.3\times 10^{-7}$ from BaBar~\cite{delAmoSanchez:2010bx} and $3.1\times10^{-6}$ from Belle~\cite{Dutta:2014sxo}.

Suppressed by the loop-induced $H_3\gamma\gamma$ coupling, the limits from ${\rm BR} (B_q \to \gamma\gamma)$ are less stringent than the tree-level processes $d_j \to d_i H_3$ discussed above. Furthermore, the FCNC effects are dominated by the $\sin\theta_2$ couplings, thus the limits on $\sin\theta_1$ could hardly be constrained by the diphoton decays of $B$ mesons. The limits on $\sin\theta_2$ are presented in Figure~\ref{fig:limits_B}. Though ${\rm BR} (B_s \to \gamma\gamma)$ is less constrained than that of $B_d$ meson, it is comparatively enhanced by the larger coupling of $H_3 \bar{s} b$ than $H_3 \bar{d}b$, thus the former could exclude a larger region in Figure~\ref{fig:limits_B}.

The bottomonium mesons $\Upsilon$ could decays into $\gamma H_3$ at the tree level, triggered by the flavor-conserving coupling of $H_3$ to $b$ quark. A light scalar has been searched for in the final state of $\mu^+ \mu^-$~\cite{Lees:2012iw}, $\tau^+ \tau^-$~\cite{Lees:2012te} and hadrons~\cite{Lees:2011wb}. The BR can be normalized to BR$(\Upsilon \to \mu^+ \mu^-)$, with
      \begin{eqnarray}
      \label{eqn:upsilon}
      \frac{{\rm BR} (\Upsilon \to \gamma H_3)}
      {{\rm BR} (\Upsilon \to \mu^+ \mu^-)_{\rm SM}}
     \ = \ \frac{|{\cal Y}_{D,bb}|^2}{4\pi\alpha}
      \left( 1 - \frac{m_{H_3}^2}{m_{\Upsilon}^2} \right)
      {\cal F} (m_{H_3}) \,,
      \end{eqnarray}
where ${\cal Y}_{D,bb}$ is the Yukawa coupling of $H_3$ to $b$ quark, from both the $\sin\tilde\theta_1$ and $\sin\tilde\theta_2$ terms, and ${\cal F}(m_{H_3})$ is the QCD form factor, including relativistic corrections~\cite{Wilczek:1977zn,Dolan:2014ska}. As shown in Table~\ref{tab:coupling}, there is a relative minus sign between the $\sin\tilde\theta_1$ and $\sin\tilde\theta_2$ terms, thus for the specific well-motivated VEV ratio $\xi = \kappa'/\kappa = m_b/m_t$ adopted throughout this paper, the two terms proportional to $\sin\theta_1$ almost cancel with each other coincidentally,
      \begin{eqnarray}
      \label{eqn:H3bb1}
      \hat{Y}_{D,bb} \sin\tilde\theta_1 &\ = \ & y_b (\sin\theta_1 + \xi \sin\theta_2)
     \ = \ y_b \sin\theta_1 + y_t \xi^2 \sin\theta_2 \,, \\
      \label{eqn:H3bb2}
      \left( V_L^\dagger \hat{Y}_{U} V_R \right)_{33} \sin\tilde\theta_2 & \ \simeq \ & y_t (\xi \sin\theta_1 + \sin\theta_2)
     \ = \ y_b \sin\theta_1 + y_t \sin\theta_2 ,,
      \end{eqnarray}
with only the contribution from mixing with the first two generations, which is suppressed by the small quark masses and CKM mixings to the 3rd generation. The $\sin\theta_2$ terms, however, are not canceled, and the couplings of $H_3$ to bottom quarks are dominated by the $y_t \sin\theta_2$ term from the flavor changing part [cf.~Eq.~(\ref{eqn:H3bb2})]. For other  small values of $\xi = \kappa'/\kappa$ not necessarily equal to $m_b/m_t$, we will ``recover'' the $y_b \sin\theta_1$ in Eqs.~(\ref{eqn:H3bb1}) and (\ref{eqn:H3bb2}) with an ${\cal O} (1)$ coefficient, while the $y_t \sin\theta_2$ term will be affected only at the $\xi^2$ level. Consequently in this case the constraints on $\sin\theta_1$ is still weaker than $\sin\theta_2$. In this paper, we will not scan the full range of the small parameters $\xi$ and $\epsilon$.

Gathering both the contributions from $gg$ and $q\bar q$, the BR of the hadronic channel $H_3 \to {\rm hadrons}$ is generally larger than the leptonic modes, thus in Figure~\ref{fig:limits_B} we show only the constraint on $\sin\theta_2$ from the hadronic decay of $H_3$, with the BR limits from $1 \times 10^{-6}$ to $8 \times 10^{-5}$ for $H_3$ mass ranging from $\sim 300$ MeV to $\sim 8$ GeV. Requiring that $H_3$ decays inside the detector with a radius of $3.5$ m~\cite{Aubert:2001tu}, we obtain the limits on $\sin\theta_2$ as shown on the right panel of Figure~\ref{fig:limits_B}. Note that the form factor in Eq.~(\ref{eqn:upsilon}) becomes smaller when $H_3$ is heavier, and the phase space also shrinks, thus the limit becomes less stringent for heavier $H_3$.





\subsubsection{Beam-dump experiments}\label{sec:beam}
\begin{figure}[!t]
  \centering
  \includegraphics[width=0.49\textwidth]{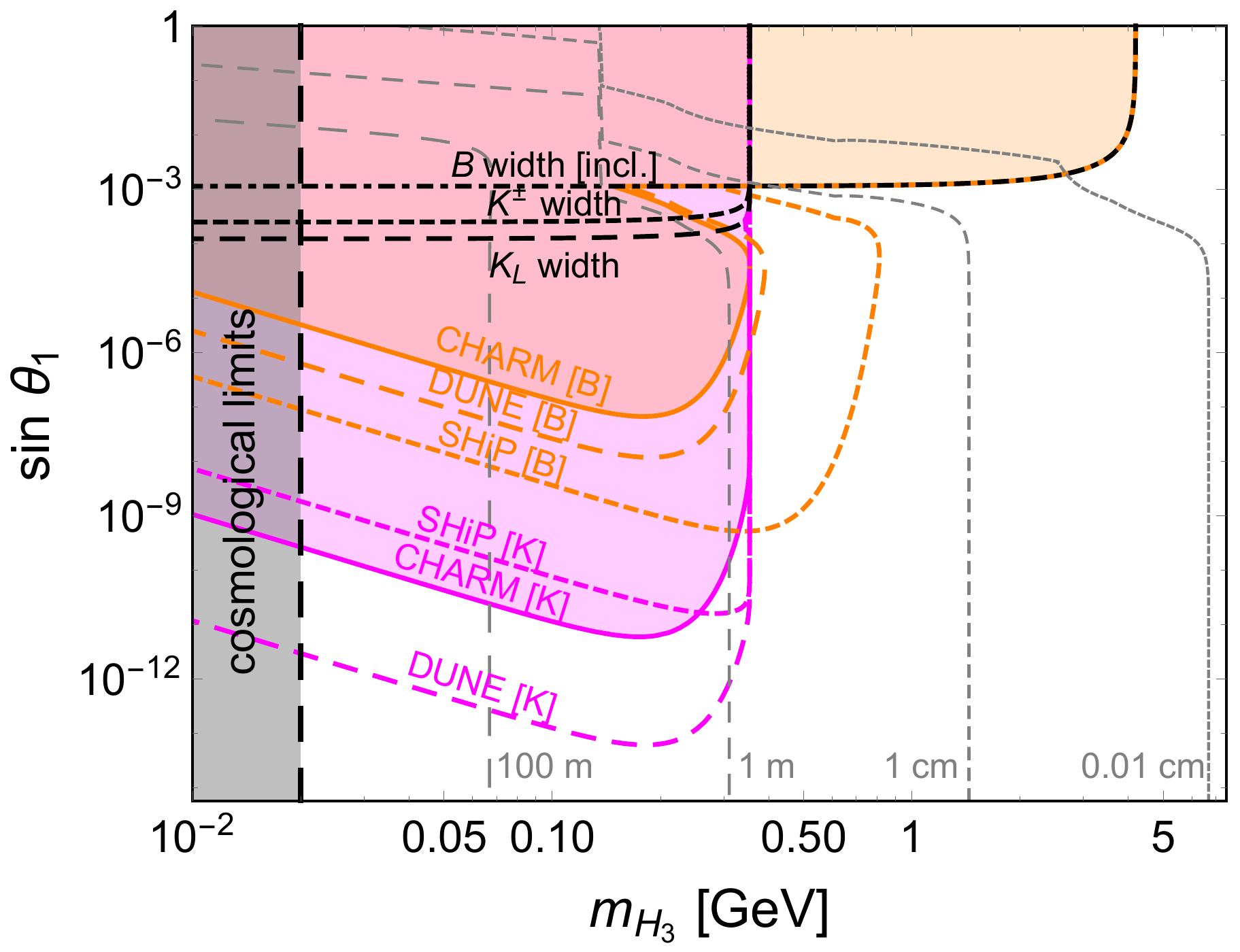}
  \includegraphics[width=0.49\textwidth]{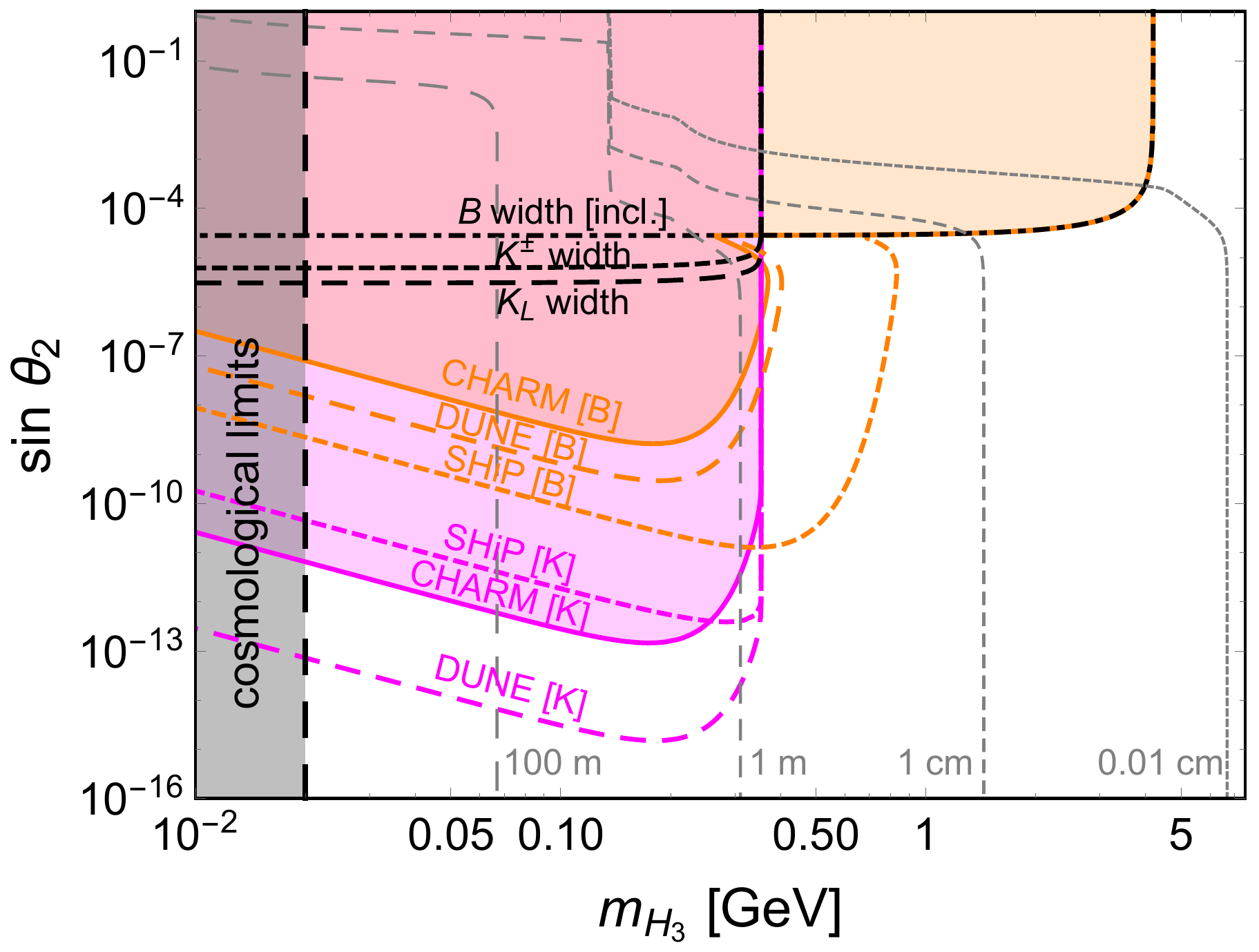}
  \caption{Limits on the mixing angles $\sin\theta_{1,2}$ and $m_{H_3}$ from the proton beam dump experiment CHARM~\cite{Bergsma:1985qz} and the future prospects at SHiP~\cite{Alekhin:2015byh} and DUNE~\cite{Adams:2013qkq}, in the flavor-changing decays of $K \to \pi \gamma\gamma$ and $B \to X_s \gamma\gamma$. For comparison, we also show the limits from the total width of $K$ and inclusive $B$ decays. The vertical black dashed line shows the cosmological limit. The dashed gray lines are the {\it proper} lifetime of $H_3$ with values of 0.01 cm, 1 cm, 1 m, and 100 m. See text for more details.}
  \label{fig:limits_dump}
\end{figure}

With a huge number of protons on target (PoT), the proton fixed target experiments, like CHARM~\cite{Bergsma:1985qz}, SHiP~\cite{Alekhin:2015byh} and DUNE~\cite{Adams:2013qkq}, provide a unique opportunity to generate a large number of LLPs, and thus, complementary constraints to the collision experiments discussed above.\footnote{The muon beam dump experiment could in principle be used to produce $H_3$ from bremsstrahlung processes~\cite{Chen:2017awl}, however, this is suppressed by the small Yukawa couplings of muon in the SM.} In the beam dump experiments, a light $H_3$ could be produced from $K$ and $B$ meson decay via $K^+ \to \pi^+ H_3$, $K_L \to \pi^0 H_3$ and $B \to X_s H_3$. The searches for $e^+ e^-$, $\mu^+ \mu^-$ and $\gamma\gamma$  final states have been carried out at CHARM~\cite{Bergsma:1985qz}, but no signal event was found, which sets an upper limit of $N_{\rm event} < 2.3$ at the 90\% C.L. on the contribution from beyond SM physics, as shown in Table~\ref{tab:limits}. Following Refs.~\cite{Bezrukov:2009yw,Clarke:2013aya,Dolan:2014ska}, the $H_3$ production cross section is given by
      \begin{eqnarray}
      \sigma_{H_3} \ \simeq \ \sigma_{pp} M_{pp} \left[
      \frac12 \chi_s {\rm BR} (K^+ \to \pi^+ H_3) +
      \frac14 \chi_s {\rm BR} (K^0 \to \pi^0 H_3) +
      \chi_b {\rm BR} (B \to X_s H_3) \right] \,, \nonumber \\ &&
      \end{eqnarray}
with $\chi_s = 1/7$ and $\chi_b = 3 \times 10^{-8}$ the fractions of charm and bottom pair-production rates respectively, $\sigma_{pp}$ the proton-proton cross section and $M_{pp}$ the average hadron multiplicity. Normalized to the neutral pion yield $\sigma_{\pi^0} \simeq \sigma_{pp} M_{pp} /3$, we can predict the total number of $N_{H_3} \simeq 2.9 \times 10^{17} \sigma_{H_3} / \sigma_{\pi^0}$. Then the number of events collected by the detector turns out to be
       \begin{align}
       N_{\rm event} \ = \ N_{H_3}
       \left( \sum_{\chi = e,\mu,\gamma} {\rm BR} (H_3 \to \chi\chi) \right)
       \left[
       \exp\left( -\frac{L \Gamma_{H_3}}{b} \right) -
       \exp\left( -\frac{(L+\Delta L) \Gamma_{H_3}}{b} \right)
       \right] \,, 
       \end{align}
with $L = 480$ m, $\Delta L = 35$ m, $b = E_{H_3} / m_{H_3}$ the boost factor where $E_{H_3} \sim 10$ GeV~\cite{Bergsma:1985qz}. Due to the huge number of events $N_{H_3}$, the mixing angles $\sin\theta_{1,2}$ are expected to be severely constrained, which implies that the most stringent limits are from the $\gamma\gamma$ channel, since this is the dominant decay mode of $H_3$ for small mixing [cf.~Figure~\ref{fig:BR}]. Indeed, the $\gamma\gamma$ limits from CHARM are much stronger than the meson decay limits discussed above, especially those from the kaon decays, and could reach $\sim 10^{-11}$ for both $\sin\theta_1$ and $\sin\theta_2$, as shown in Figure~\ref{fig:limits_dump}. For lighter $H_3$, the boost factor $b$ becomes larger, and fewer $H_3$ decays inside the detector, thus the constraints get much weaker.

Regarding the future SHiP experiment~\cite{Alekhin:2015byh}, it is quite analogous to CHARM but with a PoT number of $2\times 10^{20}$. There, we could collect $8 \times 10^{18}$ kaon and $7 \times 10^{13}$ $B$ meson events. With $E_{H_3} \sim 25$ GeV, $L = 70$ m, $\Delta L = 55$ m and $N_{\rm event} < 3$, the most stringent constraints possible are also from the $H_3 \to \gamma\gamma$ decay mode. As shown in Figure~\ref{fig:limits_dump}, the $K$ decay limits overlap largely with those from CHARM, while the $B$ limits could be largely improved and broadened.

As for the DUNE experiment~\cite{Adams:2013qkq}, with an even larger PoT of $5 \times 10^{21}$, we can collect more kaons at the near detector upstream 500 m away from the source. The total kaon number can be estimated as $N_K\simeq N_{\rm PoT} M_{pp} \chi_s\sim 8 \times 10^{21}$~\cite{Gorbunov:2007ak} with
the multiplicity $M_{pp} = 11$ and $\chi_s = 1/7$ for DUNE. This could largely improve the CHARM limits by about two orders of magnitude; see Figure~\ref{fig:limits_dump}. With a small $\chi_b = 10^{-10}$~\cite{Gorbunov:2007ak}, the number of $B$ mesons is much less and the expected limits from this are much weaker in Figure~\ref{fig:limits_dump}. The limits from $D$ meson decays will be somewhat intermediate and we do not show them here.

\subsection{SM Higgs, $Z$ and top decays} \label{sec:SM}

\begin{figure}[!t]
  \centering
  \includegraphics[width=0.52\textwidth]{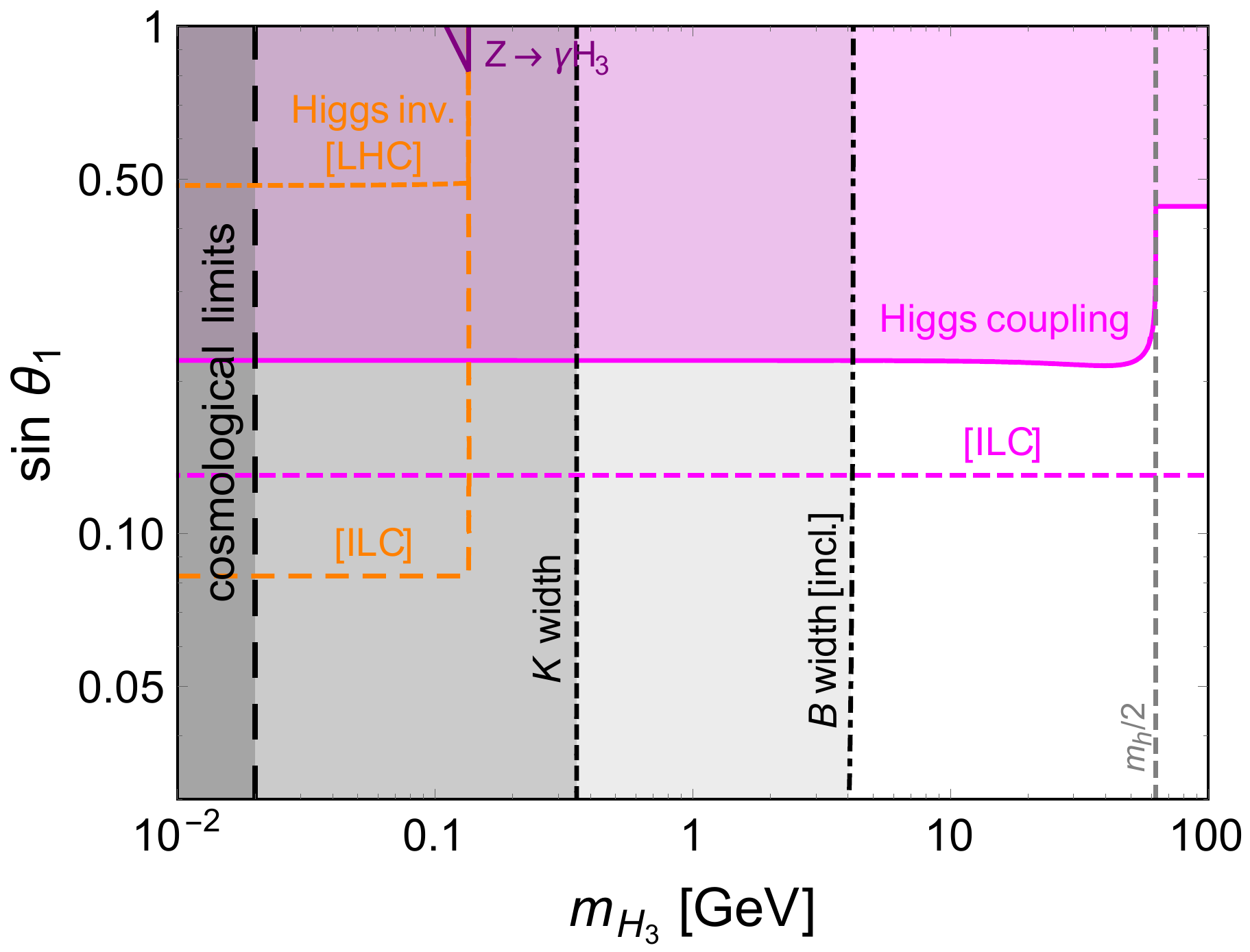}
  \caption{Limits on the mixing angle $\sin\theta_{1}$ as a function of $m_{H_3}$ from precision measurements at the LHC (magenta solid)~\cite{Falkowski:2015iwa} and future prospects at ILC (magenta dashed)~\cite{Profumo:2014opa}, as well as the limits from the invisible decay of SM Higgs by the $\sqrt s=14$ TeV LHC (orange, small-dashed)~\cite{Peskin:2012we} and $\sqrt s=1$ TeV ILC (orange, dashed)~\cite{Baer:2013cma} data, when $m_{H_3} < m_h/2$. The small purple region is excluded by searches of rare $Z \to \gamma H_3$ decay~\cite{Jaeckel:2015jla}.
  The dashed and dot-dashed black curves are the total width limit from the $K$ and inclusive $B$ decays. The region below 20 MeV is cosmologically disfavored. See text for more details. }
  \label{fig:limits_higgs}
\end{figure}

The existence of a light $H_3$ could induce some rare or unusual decay modes for the heavier SM particles, e.g. the $t$ quark, the Higgs and EW gauge bosons. Thus the couplings of $H_3$ could be limited from the relevant observations of these rare decays.
Firstly, the $h - H_3$ mixing could rescale all the SM Higgs couplings universally. The current precision Higgs measurements at the LHC constrain a generic scalar mixing $\sin\theta_1 < 0.22$~\cite{Falkowski:2015iwa}, which is almost a constant for $m_{H_3} < m_h/2$. Future more precise measurements could  significantly improve this up to 0.13~\cite{Profumo:2014opa}. When $m_{H_3} < m_h/2$, we have the extra scalar decay mode for the SM Higgs,
\begin{eqnarray}
\Gamma (h \to H_3 H_3) \ = \ \frac{m_h^3 \sin^2 \theta_1}{16\pi v_R^2}
\sqrt{1-\frac{4m_{H_3}^2}{m_h^2}} \,.
\end{eqnarray}
If $H_3$ is long-lived enough to escape the detector without leaving any signal, then it contributes to the invisible decay width of the SM Higgs. At the $\sqrt s=14$ TeV LHC, with an integrated luminosity of 300 fb$^{-1}$, the Higgs invisible BR can be constrained to be smaller than $9\%$ at the 95\% C.L.~\cite{Peskin:2012we}, while at $\sqrt s=1$ TeV ILC with a luminosity of 1000 fb$^{-1}$, the BR limit can reach up to 0.26\%~\cite{Baer:2013cma}. The corresponding limits on $\sin\theta_1$ are respectively $0.49$ and $0.083$. The Higgs coupling and invisible decay constraints on $\sin\theta_1$ are summarized in Figure~\ref{fig:limits_higgs}, as solid/dashed magenta and orange lines respectively.

When $m_{H_3} < m_Z$, we have the rare $Z$ decay $Z \to \gamma H_3$ at one-loop level mediated by mixing with the SM Higgs, with the partial width $\Gamma (Z \to \gamma H_3)$ given in~\ref{app:Zdecay}. With $H_3$ decaying into two photons, we would have the three-photon final states $Z \to \gamma H_3 \to 3\gamma$. However, if $m_{H_3} \ll m_Z$, the two photons from $H_3$ decay are highly collimated and they can not be separated experimentally. For instance, an angular separation of $20^\circ$ requires that $H_3$ must be above the GeV scale~\cite{Jaeckel:2015jla}. At LEP, the rare decay $Z \to \gamma \pi^0$ has been performed, with an upper bound of $5.2 \times 10^{-5}$ on the BR~\cite{Acciarri:1995gy}. But this helps to constrain the mixing $\sin\theta_1$ only by a marginal amount, as shown close to the upper border of Figure~\ref{fig:limits_higgs}, because this decay into $H_3$ arises at loop level, and we have set the decay length at 10 cm. With a huge number $\gtrsim 10^9$ of $Z$ events to be collected at FCC-ee~\cite{Gomez-Ceballos:2013zzn}, the limit on $\sin\theta_1$ could be improved significantly, but it may not be able to compete with the Higgs constraints.

The flavor-changing decay of top quark into up and charm quarks, i.e. $t \to u H_3, c H_3$ with $H_3 \to \gamma\gamma$ could also be used to constrain the mixing angle $\theta_1$. Again for $m_{H_3} \lesssim {\rm GeV}$, the photon pair can not be separated apart at the LHC, and we expect to see the signals $t \to u \gamma, c\gamma$ with collimated photon jets. The current limit of $1.3 \times 10^{-4}$ ($1.7 \times 10^{-3}$) for $u\gamma$ ($c\gamma$)~\cite{Khachatryan:2015att}, can not provide any competent limits on the mixing angles $\sin\theta_{1,2}$, which is largely due to the small CKM mixing of the third generation with the first two in the SM.

Finally, we also note that constraints from flavor changing leptonic processes such as $\mu\to 3 e$, are not more stringent than the  hadronic decays considered above since they necessarily involve $H_3-H_1$ mixing as well as electron and muon Yukawa couplings, which are very small.

The most important laboratory constraints discussed in this section (i.e. those ruling out some part of the parameter space not already ruled out by others) are summarized in Figure~\ref{fig:limits_all}, together with the collider sensitivity curves to be discussed in the next section. Here the shaded regions are all excluded. The bottom line of this summary plot is that for a GeV-scale $H_3$ boson in the minimal LR model, the FCNC constraints necessarily imply small $h-H_3$ and $H_1-H_3$ mixing angles $\sin\theta_{1,2}\lesssim 10^{-4}$. This naturally makes the $H_3$ a good LLP candidate, with distinct displaced vertex signatures of collimated diphotons, as discussed below. This is a unique feature of the minimal LR model, not shared by either a generic $U(1)_{B-L}$ model, as we will show explicitly in Section~\ref{sec:U1}, or by other new physics scenarios with a light scalar, such as NMSSM~\cite{Ellwanger:2015uaz}.

\section{Production and displaced vertex searches at colliders}
\label{sec:collider}
In this section, we discuss the production of light $H_3$ in high-energy proton-proton collisions, and its subsequent decay to displaced photon signatures.

\subsection{Production cross section}\label{sec:prod}

In the minimal LR model, the scalar $H_3^0$ can be produced from its coupling to the heavy RH gauge bosons $W_R$ and $Z_R$, as well as through its coupling to the SM Higgs~\cite{Dev:2016dja}.\footnote{There is also the production of $H_3$ from photon fusion $\gamma\gamma \to H_3$, mediated by the $W_R$ and scalar loops, analogous to the diagrams in Figure~2 of Ref.~\cite{Babu:2016rcr}. However, these loop-level processes turn out to be much smaller than the direct fusion of $W_R$ and $Z_R$ bosons in our case.} For a small mixing $\sin\theta_1 \lesssim 10^{-4}$ of our interest, which implies the scalar quartic coupling $\alpha_1 \simeq \lambda_1 \sin\theta_1 (v_{\rm EW} / v_R) \lesssim 10^{-6}$ [cf.~Eq.~\eqref{eqn:theta1}], the Higgs portal can be neglected, and we focus here only on the gauge portal production, which is through the associated production with a heavy $W_R$ boson which decays predominantly into the SM quark jets ($J = u,\, d,\, s,\, c,\, b,\, t$):
\begin{eqnarray}
pp \ \to \ W_R^\ast \ \to \ W_R H_3 \,, \quad
W_R \to JJ \,.
\end{eqnarray}
Here for simplicty we have assumed that the decay mode into on-shell heavy RHNs $W_R \to \ell N$ is kinematically forbidden. If it is open, then we could have the smoking-gun $\ell^\pm \ell^\pm jj$ signal of the $W_R$ boson, in association with the the unique displaced photon jet from the light scalar $H_3$. Other decay modes such as $W_R \to WZ$, $Wh$ are heavily suppressed by the small $W - W_R$ mixing angle in the minimal model we are considering~\cite{Dev:2015pga}. One should note that the $H_3 jj$ processes (with $j = u,\, d,\, s,\, c$) also receive (small) contributions from the heavy vector boson fusion (VBF) $pp \ \to\  W_R^{\ast} W_R^\ast jj \ \to \ H_3 jj$, which is however suppressed by the three-body phase space and the off-shell $W_R$. At the LHC Run II, limited by the total center-of-mass energy, the associated production with the $Z_R$ boson is always highly suppressed, as it is heavier than the $W_R$ boson in the minimal LR scenario.

Dictated by the gauge interaction in the RH sector, the production of $H_3$ is only sensitive to the value of $g_R$,\footnote{In the limit of $v_R \gg \sqrt{s}$, the dependence of the gauge couplings and $W_R$ mass on the gauge coupling $g_R$ completely cancels out, leaving only the effective dimension-seven interaction of $H_3$ with the SM quarks $H_3 (\bar{q}_R \gamma^\mu q'_R) (\bar{q}^{\prime\prime}_R \gamma_\mu q^{\prime\prime\prime}_R) / v_R^3$ (note that the vertex $H_3 W_R^+ W_R^-$ is proportional to $v_R$, see Table~\ref{tab:coupling}), with the production cross section suppressed by $v_R^{-6}$. Thich is rather analogous to the $W_R$ mediated loop contribution to the $K$ and $B$ meson mixing, where $m_{K,\,B} \ll m_{W_R}$.} as it determines not only the $W_R$ mass for fixed $v_R$ but also the magnitudes of couplings of $W_R$ to the initial partons and $H_3$. The leading order production cross sections at the $\sqrt s=14$ TeV LHC for different values of $g_R/g_L = 0.6$, 1, and 1.5 are presented in the left panel of Figure~\ref{fig:H3production}, where we have set the RH scale $v_R = 5$ TeV and adopted the basic trigger cuts for the jets $p_T (J) > 25$ GeV and $\Delta\phi (JJ) > 0.4$ in a {\tt MadGraph5} set up~\cite{Alwall:2014hca}.\footnote{Here for simplicity we do not distinguish the heavy flavor jets from the light quark jets from $W_R$ decay, which are all expected to be highly boosted. The bottom and top jet tags might help to further suppress the SM background.} For a smaller $g_R < g_L$, the $W_R$ boson is lighter and the production of $H_3$ can be significantly enhanced. When $m_{H_3} \lesssim 10$ GeV, the production rates are almost constant for a given $v_R$, and is sensitive only to the gauge coupling $g_R$.

\begin{figure}[t!]
  \def\topdiff{0.25}
  \def\toppos{-1.5}
  \def\vertexstart{-4}
  \def\vertex{\vertexstart+1.5}
  \def\length{1.4}
  \def\size{0.75}
  \def\topoffset{0.75}

  \centering

  \begin{tabular}{cc}
  \begin{tikzpicture}[]
  \draw[dashed,thick](\vertexstart+0.5,0)--(\vertex,-1)node[right]{{\footnotesize$H_3$}};
  \draw[ZZ,thick](\vertexstart+0.5,0)--(\vertex,1)node[right]{{\footnotesize$V_R$}};
  \draw[ZZ,thick](\vertexstart-0.5,0)--(\vertexstart,0)node[above]{{\footnotesize$V_R$}} -- (\vertexstart+0.5,0);
  \draw[quark,thick] (\vertexstart-1.5,1)node[left]{{\footnotesize$q$}} -- (\vertexstart-0.5,0);
  \draw[antiquark,thick] (\vertexstart-1.5,-1)node[left]{{\footnotesize$\bar{q}$}} -- (\vertexstart-0.5,0);
  \end{tikzpicture} 
  \end{tabular}
  \caption{Production of $H_3$ at hadron colliders in associated production with a heavy $W_R/Z_R$ boson.}
  \label{fig:prod}
\end{figure}
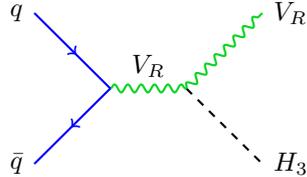

\begin{figure}[!t]
  \centering
  \includegraphics[width=0.48\textwidth]{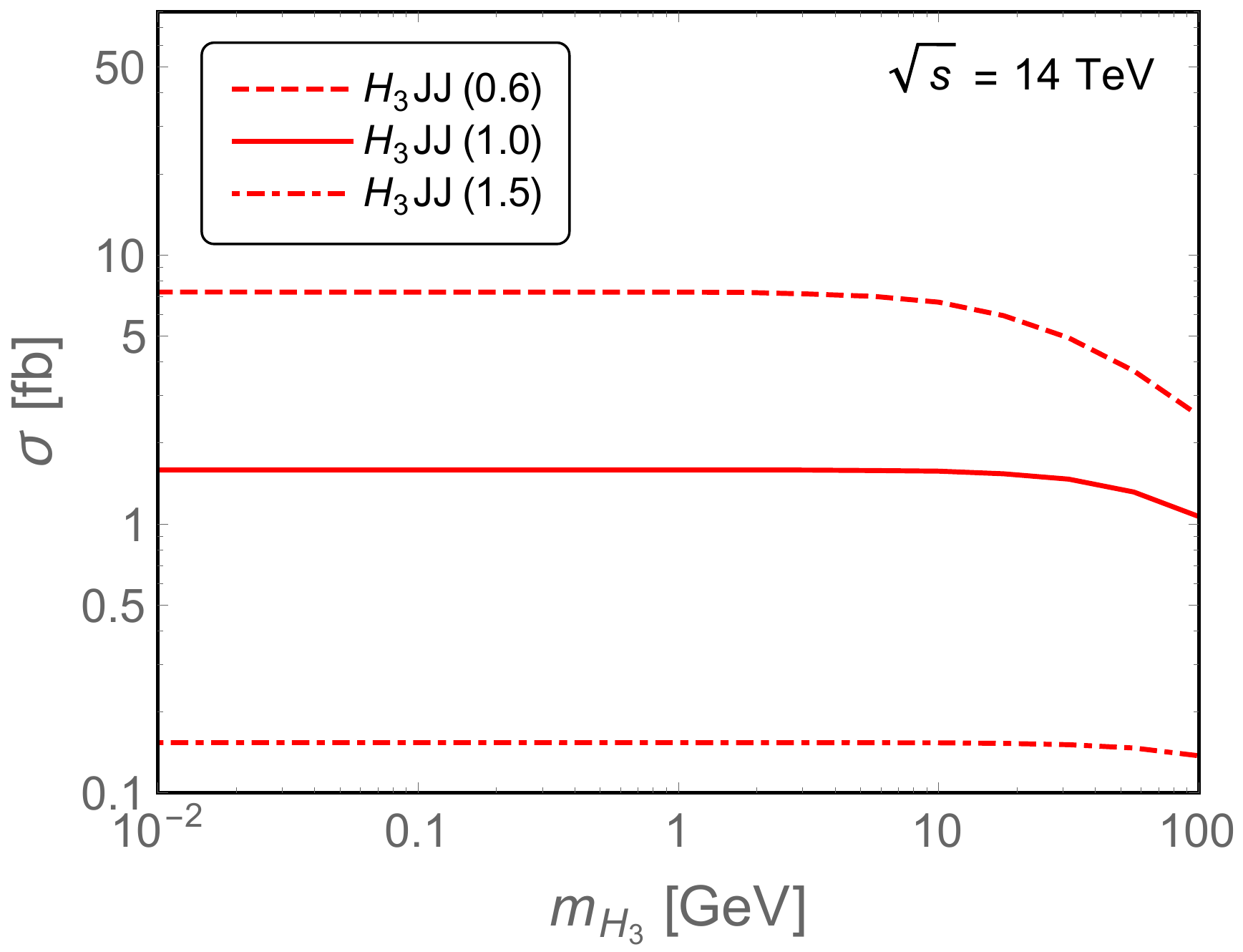}
  \includegraphics[width=0.48\textwidth]{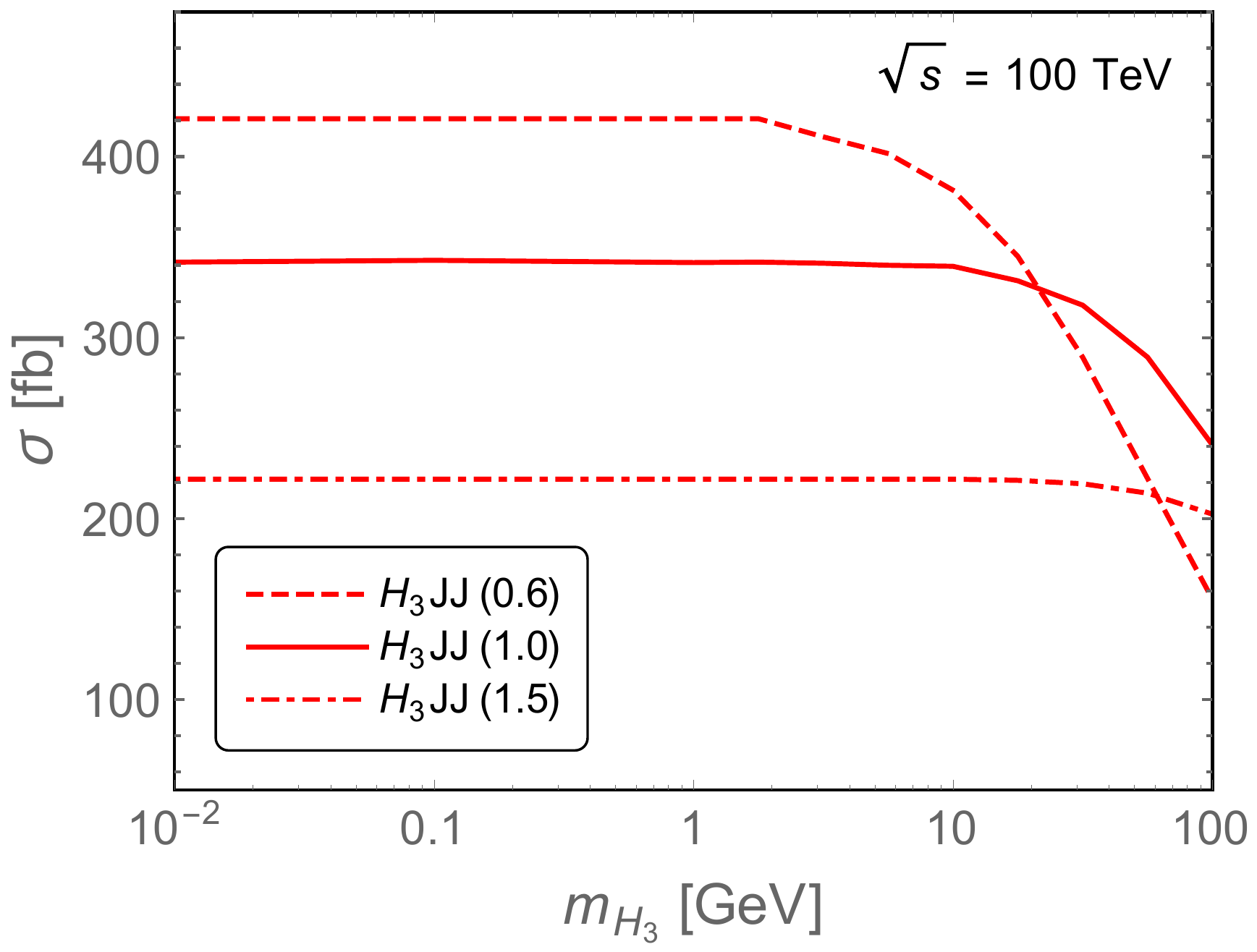}
  \caption{Production cross section of $H_3$ at $\sqrt s=14$ TeV LHC (left) and future 100 TeV FCC-hh collider (right) in association with two quark jets, as function of its mass. The numbers in parentheses are the values of $g_R/g_L$. In both the plots we have set the RH scale $v_R = 5$ TeV. }
  \label{fig:H3production}
\end{figure}

\subsection{Prospects at the LHC} \label{sec:LHC}
\begin{figure}[!t]
  \centering
  \includegraphics[width=0.51\textwidth]{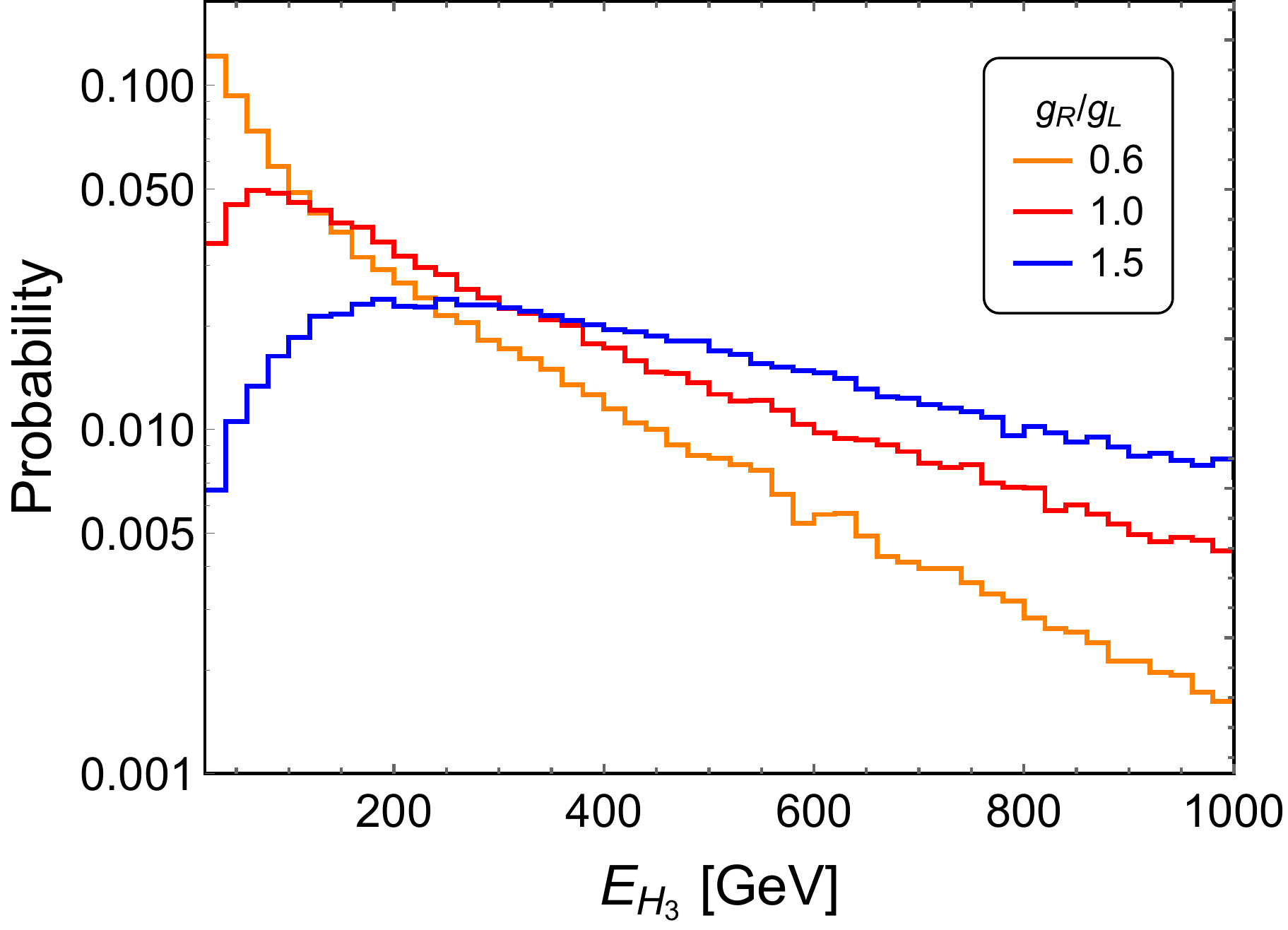}
  \caption{Energy distributions of the simulated events of $H_3$ production from $W_R$ VBF, with the probability $(\frac{1}{\sigma} \frac{d\sigma}{dE_{H_3}}) \Delta E_{H_3}$ where $\sigma$ is the production cross section and $\Delta E_{H_3} = 20$ GeV is the size of the energy bins considered.}
  \label{fig:distribution}
\end{figure}

Limited by the flavor constraints in Section~\ref{sec:lab}, a light $H_3$ decays mostly into two photons at the LHC after being produced and flying over a distance of $L = b L_0$. For a GeV mass, the decay-at-rest length $L_0$ is of order of cm. The boost factor $b = E_{H_3} / m_{H_3}$ depends on the distribution of energy $E_{H_3}$ at the LHC, which is different for different values of $g_R$. When the gauge coupling $g_R$ is smaller, the $W_R$ boson is lighter and has a larger momentum, so the scalar $H_3$ tends to be more highly boosted, with respect to the case with a heavier $W_R$. This effect can be seen from the energy distributions in Figure~\ref{fig:distribution} from a parton-level simulation. Roughly speaking, the energy $E_{H_3}$ has a peak at the hundred GeV scale, with a long tail up to few TeV. For our rough sensitivity estimates, we use a boost factor of order $\sim 100$. Then the actual decay length is expected to be of order of meter, comparable to the radius of the Electromagnetic Calorimeter (ECAL) of ATLAS and CMS detectors, which are respectively 1.5 m~\cite{Aad:2009wy} and 1.3 m~\cite{Ball:2007zza,Chatrchyan:2008aa}.

\begin{figure}[!t]
  \centering
  \includegraphics[width=0.49\textwidth]{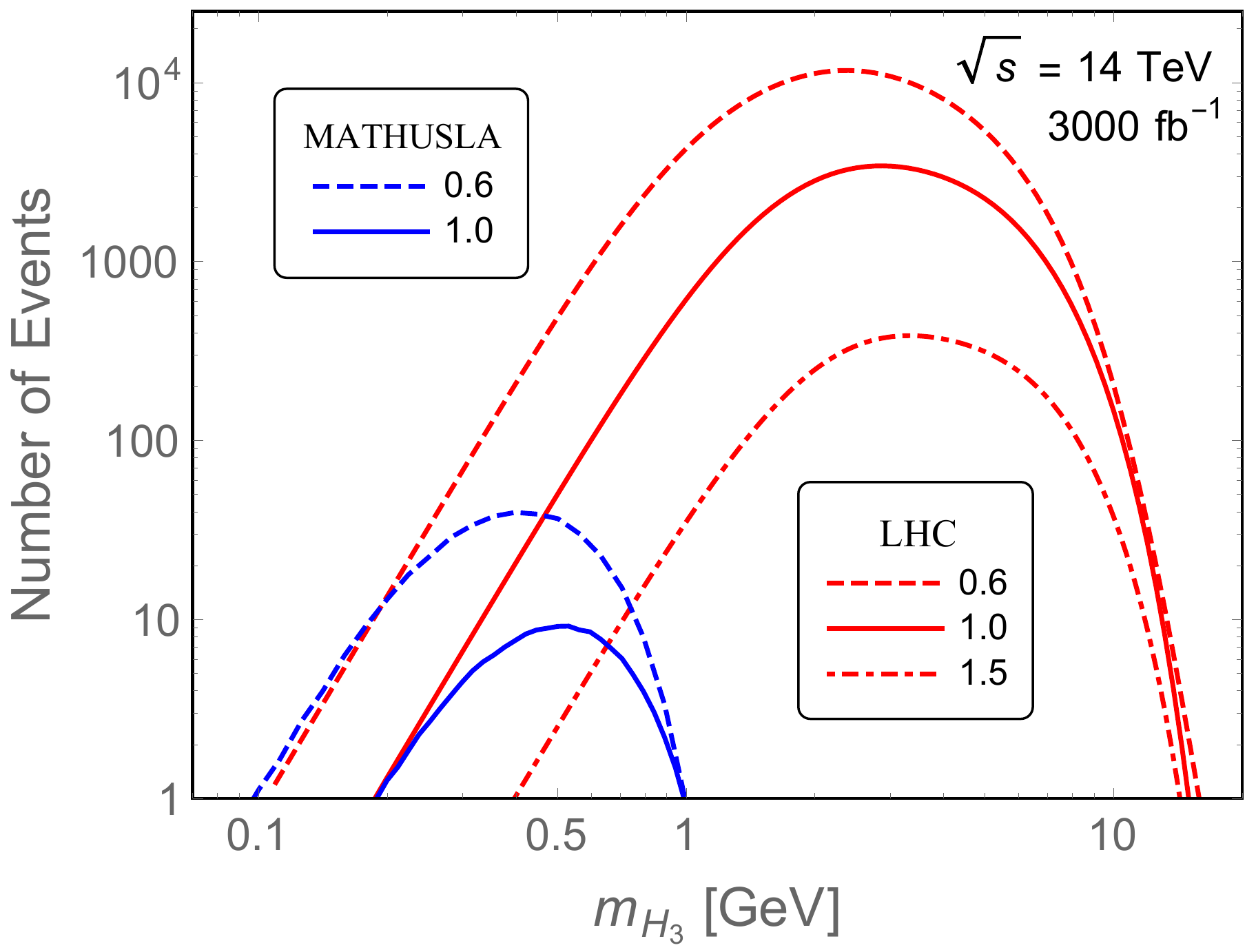}
  \includegraphics[width=0.49\textwidth]{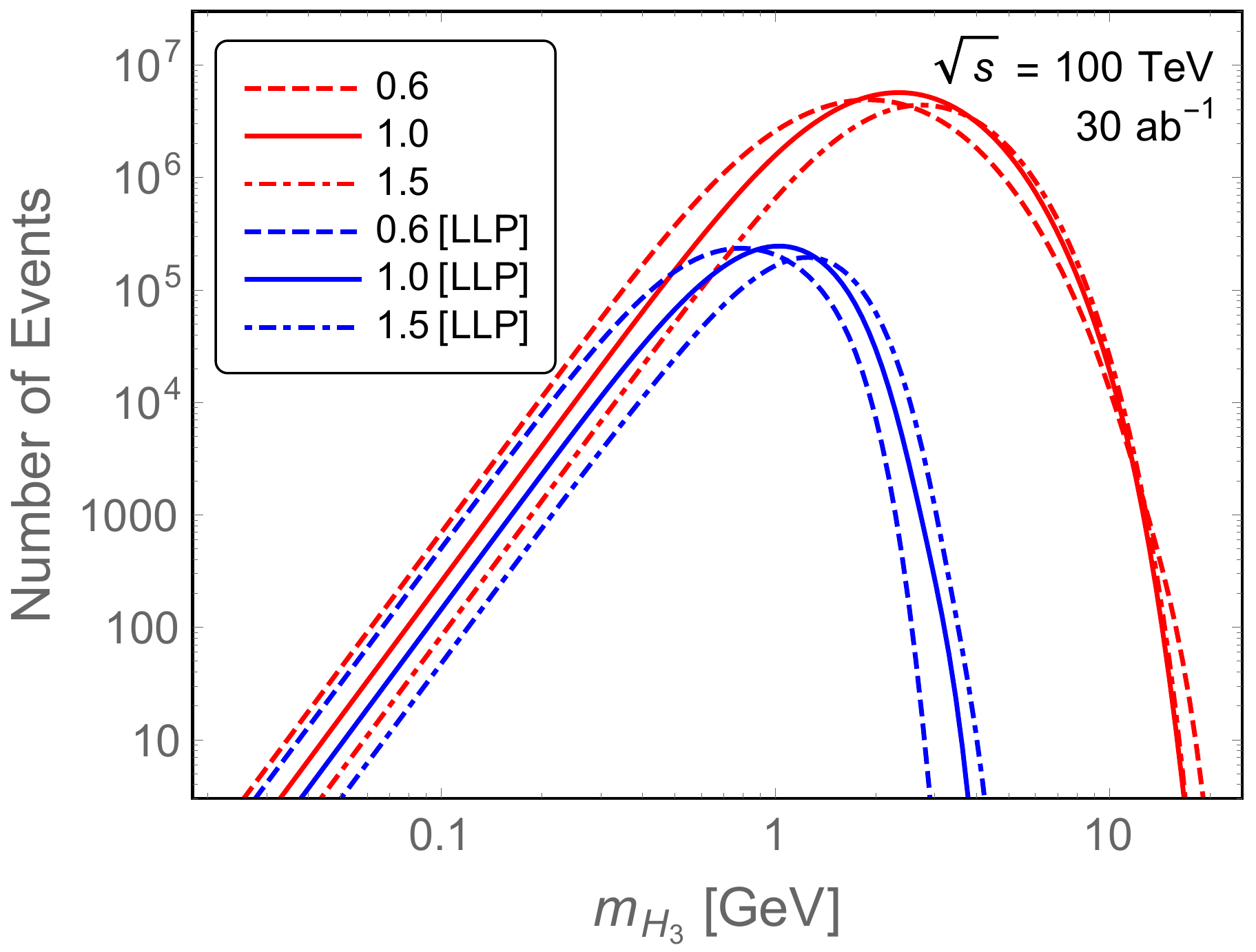}
  \caption{{\it Left:} Predicted numbers of displaced photon events from $H_3$ decay within the ECAL of ATLAS/CMS (red) and at the proposed surface detector MATHUSLA (blue), with an integrated luminosity of 3000 fb$^{-1}$ at $\sqrt{s} = 14$ TeV, for $g_R / g_L = 0.6$, 1 and 1.5. {\it Right:} The corresponding numbers of displaced photon signals at FCC-hh (red) and a forward LLP detector (blue), based on a luminosity of 30 ab$^{-1}$ at $\sqrt s=100$ TeV.}
  \label{fig:signal}
\end{figure}

The final-state photons from $H_3$ decay are highly collimated with a separation of $\Delta R \sim m_{H_3} / E_{H_3}$. Thus, in a large range of parameter space, most of the photon pairs can not be separated with the angular resolution of $\Delta \eta \times \Delta \phi = 0.025 \times 0.025$ (ATLAS) and $0.0174 \times 0.0174$ (CMS)~\cite{Aad:2009wy,Ball:2007zza,Chatrchyan:2008aa}, and would be identified as a high-energy single-photon jet. Counting conservatively these single photon jets within $1\, {\rm cm} < L < R_{\rm ECAL}$, we predict the numbers of displaced diphoton events from $H_3$ decay in ATLAS/CMS for an integrated luminosity of 3000 fb$^{-1}$ at $\sqrt{s} = 14$ TeV LHC -- the ultimate high-luminosity phase of LHC (HL-LHC). Our results are shown in Figure~\ref{fig:signal} for three benchmark values of $g_R/g_L = 0.6$, 1.0 and 1.5 with $v_R=5$ TeV. Here we have applied the basic trigger cuts $p_T (J) > 25$ GeV and $\Delta\phi (JJ) > 0.4$ on the jets and have assumed the SM fake rate for the displaced diphotons to be small~\cite{Dasgupta:2016wxw,Tsai:2016lfg,Fukuda:2016qah}. We find it promising that for a GeV-scale $H_3$, one could find up to ${\cal O}(10^4)$ displaced photon events at the LHC, which would constitute a ``smoking gun'' signature of the $H_3$ decays as predicted by the minimal LR model.

If the scalar is lighter, i.e.
$m_{H_3}\lesssim 1$ GeV, the decay length would exceed the size of LHC detectors, and could be suitable for future dedicated ultra LLP (ULLP) search experiments, such as MATHUSLA~\cite{Chou:2016lxi}. Although the surface detector MATHUSLA is much farther away from the collision point, at the 100 m scale, which provides better sensitivity for low-mass LLPs, the effective solid angle of the detector being very small, at the order of $0.1 \times 4\pi$, the number of events turns out to be much smaller than those at ATLAS/CMS, as shown in Figure~\ref{fig:signal} with a high luminosity of 3000 fb$^{-1}$.
However, the background at MATHUSLA is rather low or almost negligible~\cite{Chou:2016lxi, Coccaro:2016lnz}, whereas the displaced photon signals at ATLAS/CMS could potentially suffer from a non-negligible background, mostly from $\pi^0\to \gamma\gamma$, which has not been considered in our preliminary analysis. Thus, we believe MATHUSLA is largely complementary to the LLP searches at ATLAS/CMS, and could extend to lower mass range of $H_3$ in the minimal LR model.

\subsection{Prospects at future 100 TeV collider}\label{sec:fcc}

As the physics case~\cite{Arkani-Hamed:2015vfh, Golling:2016gvc, Contino:2016spe} for a future high-energy collider , such as FCC-hh or SPPC, with the center-of-mass energy $\sqrt{s} = 80-100$ TeV, is growing rapidly, we find it worthwhile analyzing the detectable parameter space of $H_3$ in this scenario. The production cross section are collected in the right panel of Figure~\ref{fig:H3production}, where we include also the $Z_R$ mediated processes:
\begin{eqnarray}
pp \to Z_R H_3 \,, \quad
Z_R \to q \bar{q},\, \ell^+ \ell^-  \,.
\end{eqnarray}
We do not include the decays $Z_R \to NN$ which, depending on the RH neutrino mass, could give rise to distinct LNV signatures. Similarly, we do not consider $Z_R\to \nu\nu$, which is suppressed by the $Z-Z_R$ mixing. 
For the sake of comparison we retain $v_R = 5$ TeV and change only the trigger cut to $p_T (J) > 50$ GeV for the SM quark and lepton jets. For a light $H_3$ with mass $\lesssim 10$ GeV, the cross sections are much larger than at LHC, at the level of few 100 fb, and less sensitive to the gauge coupling $g_R$, as in this case the center-of-mass energy of initial partons is much larger than the $W_R$ mass, i.e. $\hat{s} \gg M_{W_R}$, and the differences of cross section in Figure~\ref{fig:H3production} is mainly due to the changes of couplings as we change $g_R$.

Given a ATLAS-like detector at FCC-hh, we predict the numbers of signal events at FCC-hh, with an integrated luminosity of 30 ab$^{-1}$ of running at 100 TeV, which are shown by the red lines in the right panel of Figure~\ref{fig:signal}. For concreteness, we assume the future detector has the same angular resolution as ATLAS, and count again only the highly collimated photon events, with a larger decay length ranging from $L = 10$ cm to $3$ m. With a larger cross section and higher luminosity, we can collect up to $100$ times more events than at the LHC. For the LR models with a larger $v_R$ which is beyond the scope of LHC detectability, the future higher energy colliders are the only facility to study the properties of $H_3$.

With a dedicated forward LLP detector at FCC-hh, similar to the one proposed in Ref.~\cite{Chou:2016lxi}, the background can be significantly reduced to almost zero. With the MATHUSLA detector geometry as in Ref.~\cite{Chou:2016lxi}, we obtain the numbers of signal events at the forward LLP detector, depicted as the blue lines in the right panel of Figure~\ref{fig:signal}, for different values of $g_R/g_L = 0.6$, 1, 1.5.


\subsection{Probing the LR seesaw model}\label{sec:sens}
\begin{figure}[!t]
  \centering
  \includegraphics[width=0.49\textwidth]{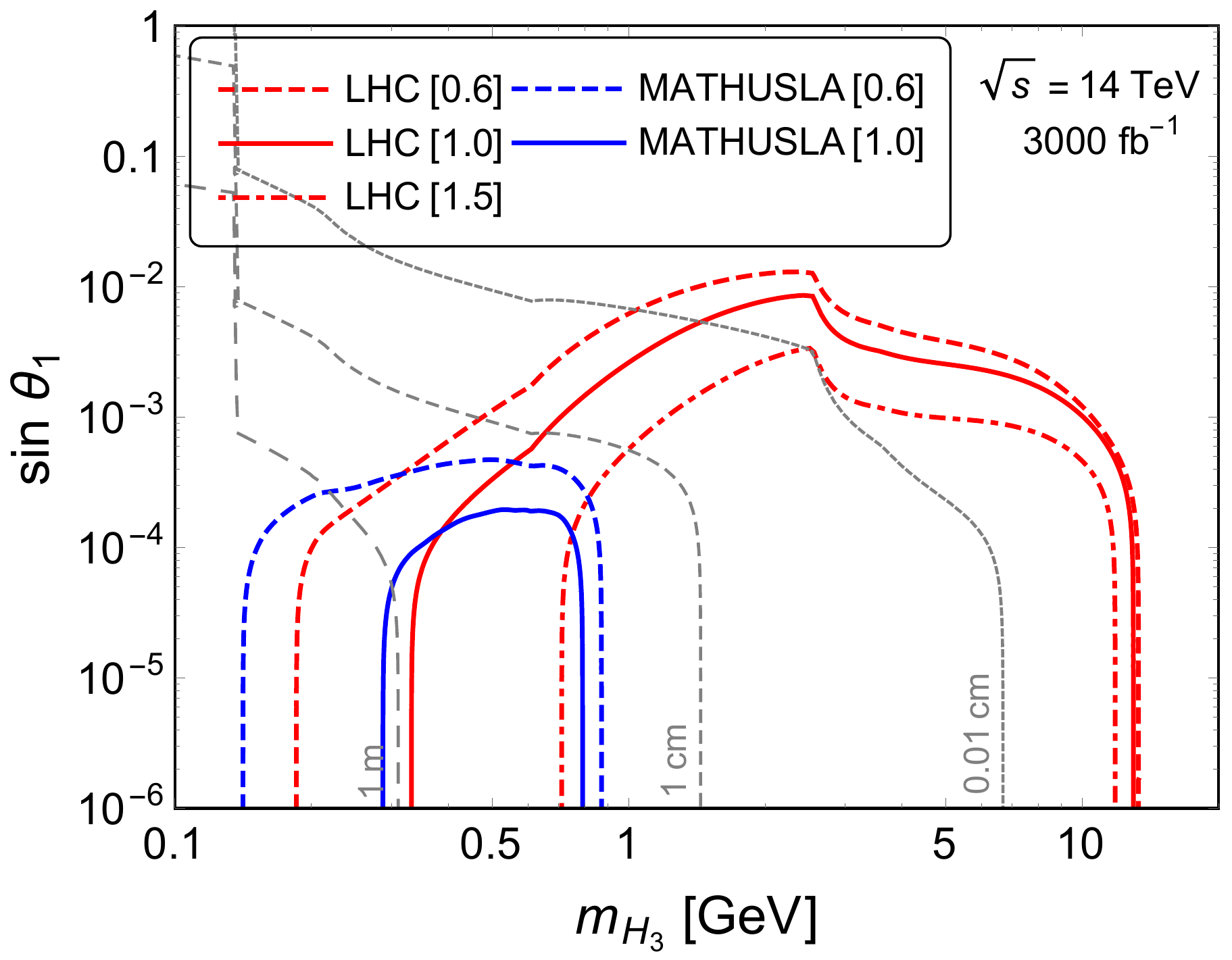}
  \includegraphics[width=0.49\textwidth]{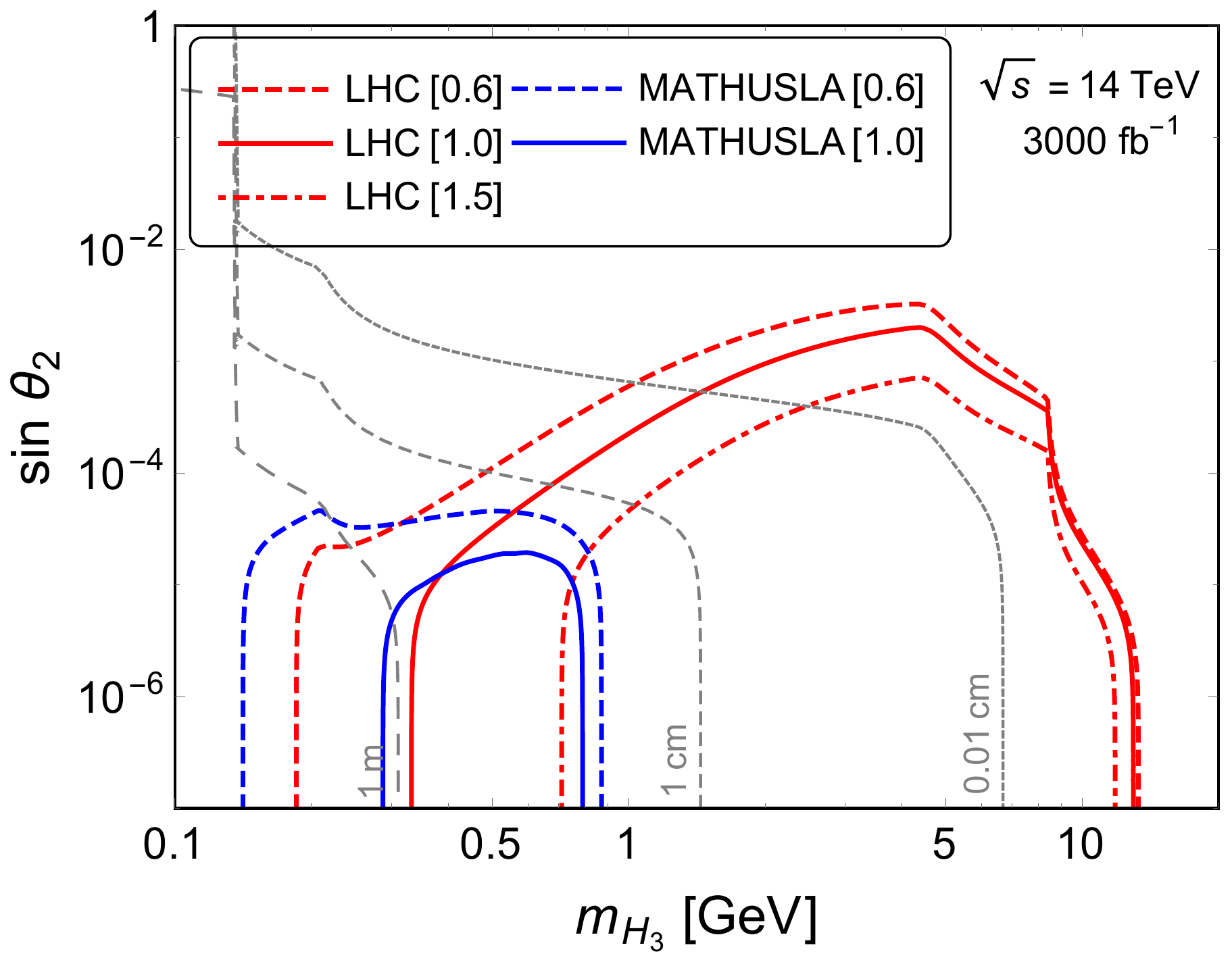}
  \caption{Sensitivity contours in the mass-mixing plane from future LLP searches at LHC and MATHUSLA, with an integrated luminosity of 3000 fb$^{-1}$ running at $\sqrt{s} = 14$ TeV, for $g_R/g_L = 0.6$, 1 and 1.5. The regions below these lines are probable with 10 signal events at LHC and 4 at MATHUSLA. The dashed gray lines are the {\it proper} lifetime of $H_3$ with values of 0.01 cm, 1 cm, 1 m, and 100 m.}
  \label{fig:limits_LLP}
\end{figure}

\begin{figure}[!t]
  \centering
  \includegraphics[width=0.49\textwidth]{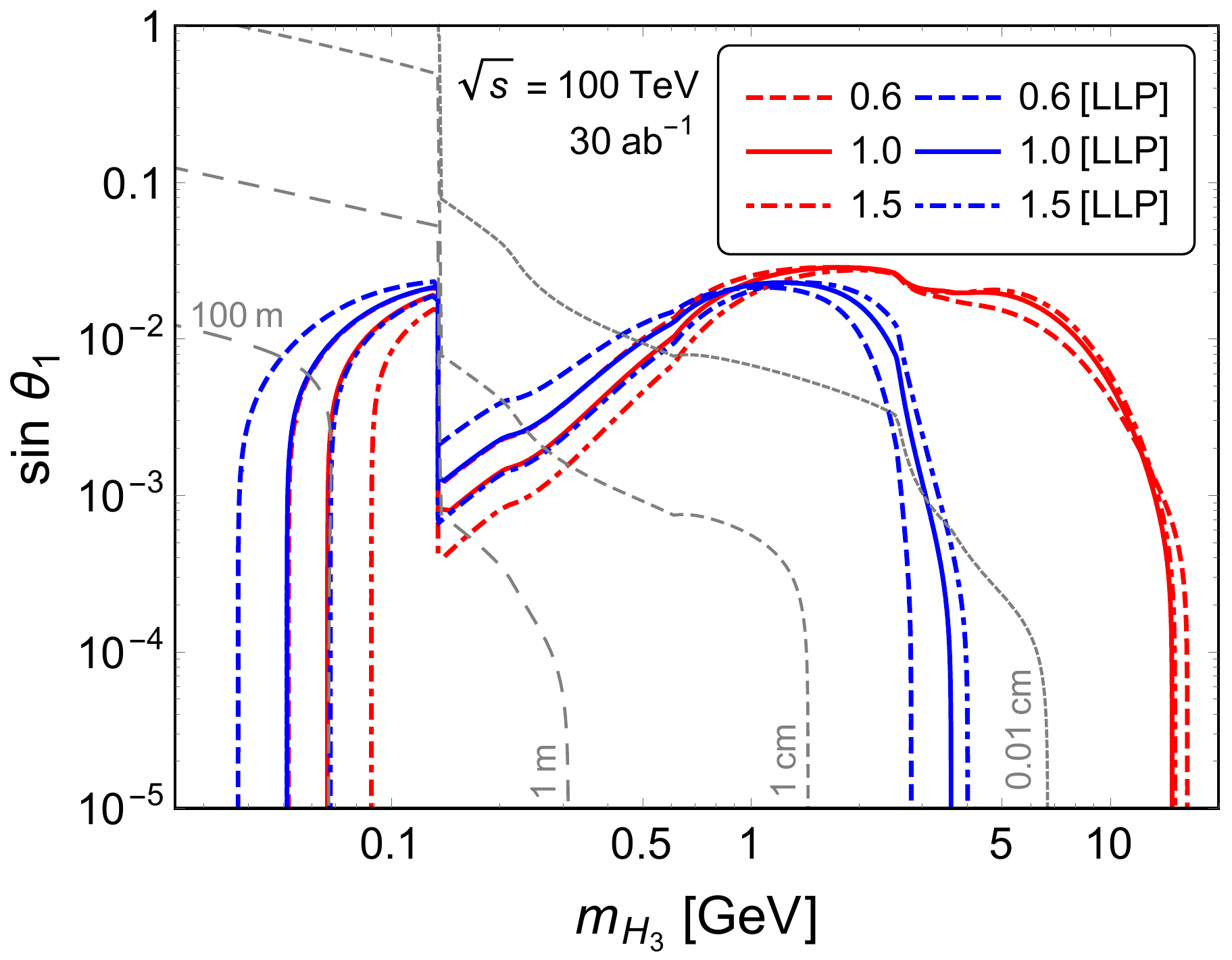}
  \includegraphics[width=0.49\textwidth]{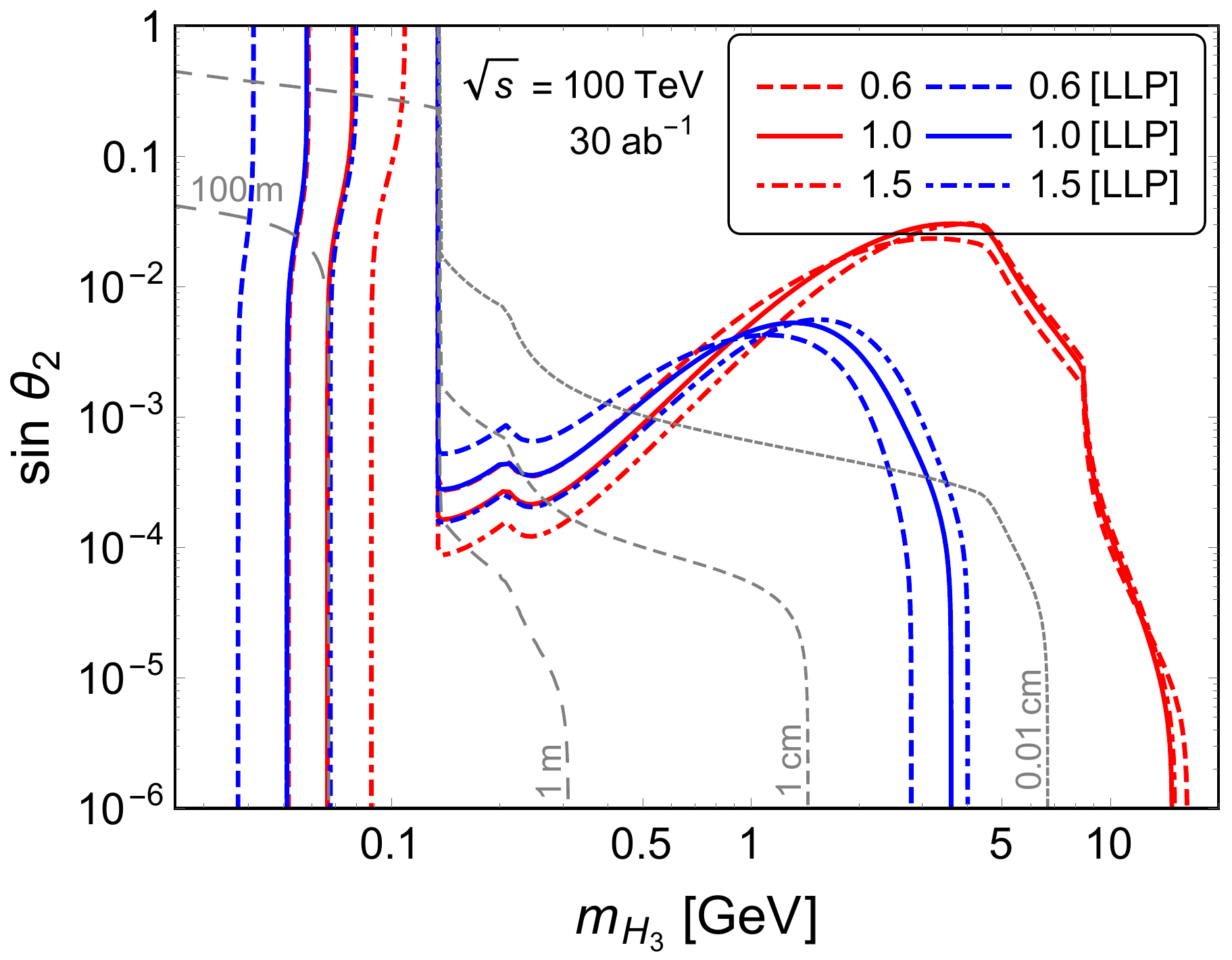}
  \caption{Sensitivity contours in the mass-mixing plane from future LLP searches at FCC-hh and forward LLP detector therein, with an integrated luminosity of 30 ab$^{-1}$ running at $\sqrt{s} = 100$ TeV, for $g_R/g_L = 0.6$, 1 and 1.5. The regions below these lines are probable with 50 signal events at FCC-hh and 10 at the forward detector. The dashed gray lines are the {\it proper} lifetime of $H_3$ with values of 0.01 cm, 1 cm, 1 m, and 100 m.}
  \label{fig:limits_LLP2}
\end{figure}

With the expected numbers of events at LHC and MATHUSLA in Figure~\ref{fig:signal}, we can easily translate them to the sensitivity regions in the plane of $m_{H_3}$ and $\sin\theta_1$ (or $\sin\theta_2$) in the LR model, assuming that the SM background for the (ultra) LLP signal is under control. Instead of embarking on a full-fledged simulation of the detector noise for the high energy displaced photon signals, we just assume a signal number of 10 (4) at LHC (MATHUSLA) to set limits on the mass $m_{H_3}$ and mixing angles $\sin\theta_{1,2}$ for illustration purposes. The expected sensitivity regions are shown in Figure~\ref{fig:limits_LLP}, where the regions below the lines can be probed.
As expected, complementary to the limits from FCNC and Higgs data in Section~\ref{sec:lab}, the (U)LLP searches are sensitive to small values of mixing angles $\sin\theta_{1,2}$, as a small mixing angle would suppress the fermionic decays $H_3 \to f \bar{f}$ and ensure the dominance of $H_3 \to \gamma\gamma$ for the displaced vertex signal.

\begin{figure}[!t]
  \centering
  \includegraphics[width=0.47\textwidth]{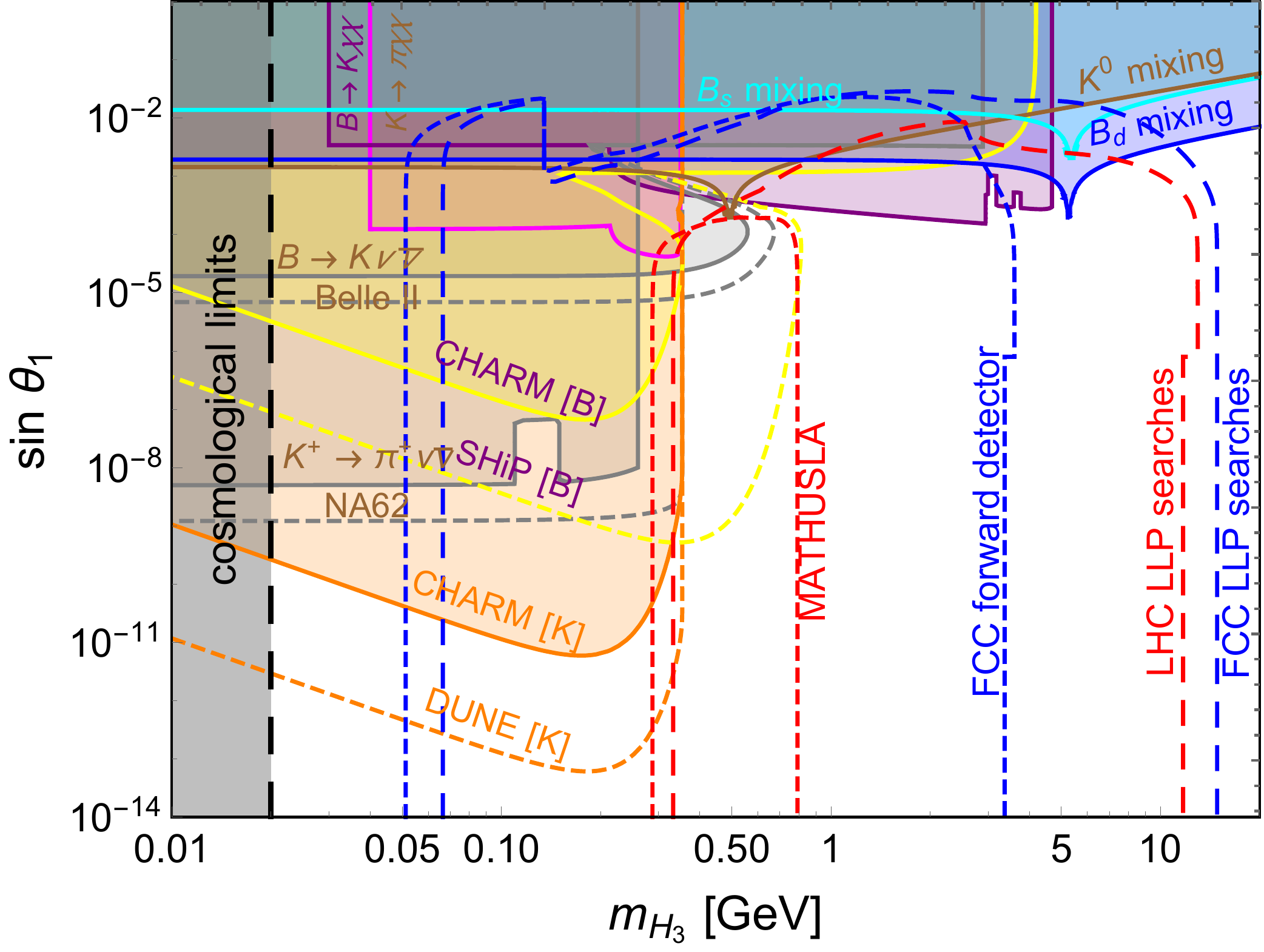} 
  \includegraphics[width=0.47\textwidth]{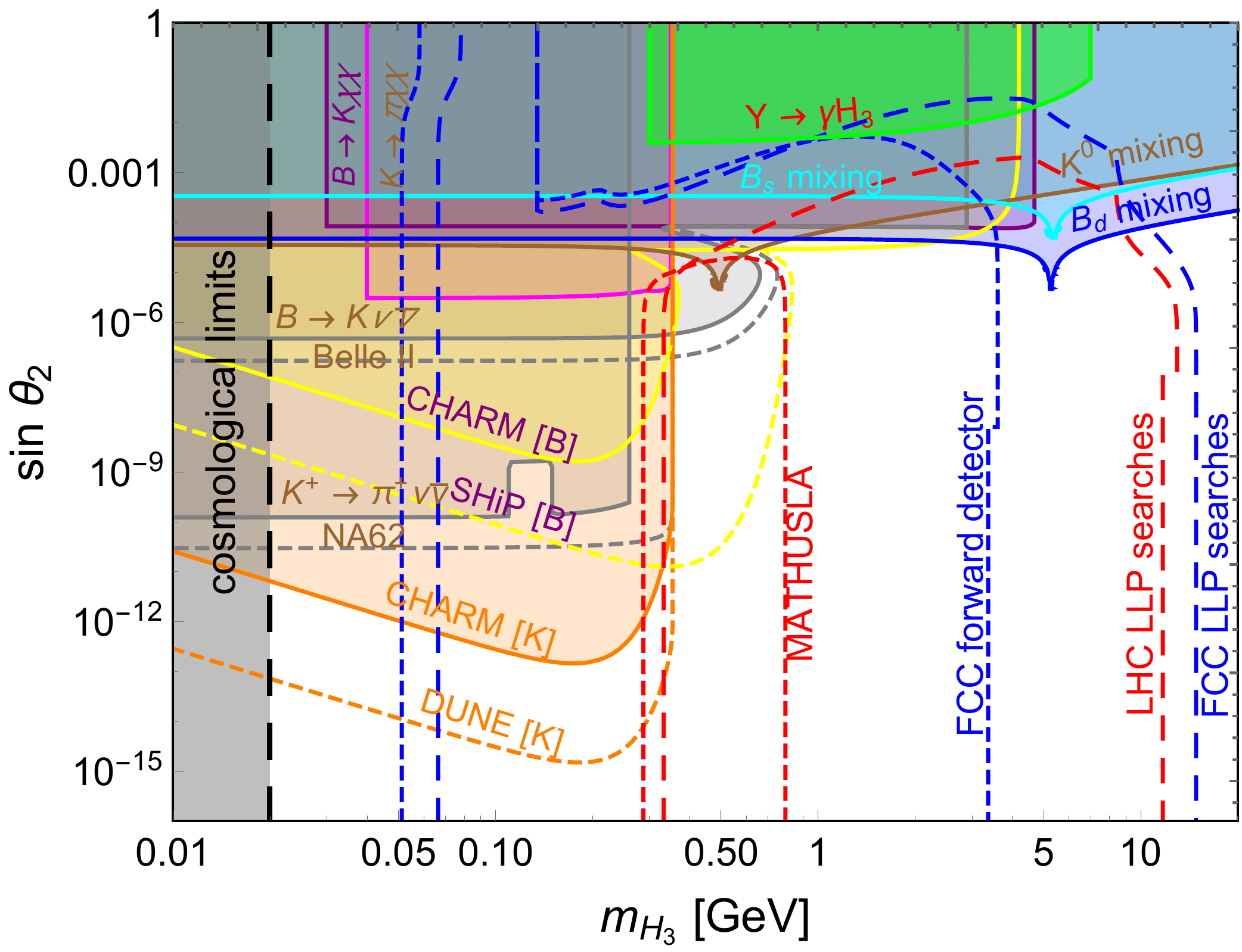}
  \caption{A summary of the important limits and sensitivity curves in the $m_{H_3}-\sin\theta_{1,2}$ plane, extracted from Figures~\ref{fig:limits_K}, \ref{fig:limits_B}, \ref{fig:limits_dump}, \ref{fig:limits_LLP}, \ref{fig:limits_LLP2}. The shaded regions are excluded. For the sensitivity contours of LLP searches at LHC, MATHUSLA, FCC and the forward detector, the gauge coupling $g_R = g_L$. For details, see Sections~\ref{sec:lab} and \ref{sec:collider}.
}
  \label{fig:limits_all}
\end{figure}

Analogously, the sensitivity regions for the LLP searches at FCC-hh and the forward detector are collected in Figure~\ref{fig:limits_LLP2}, where we have assumed the signal numbers to be respectively 50 and 10 for FCC-hh and the LLP forward detector. It is clearly obvious that compared to the regions to be probed at LHC and MATHUSLA in Figure~\ref{fig:limits_LLP}, the future higher energy colliders could probe larger regions of $H_3$ parameter space in the minimal LR model, as well as a larger range of the gauge coupling $g_R$, as the $H_3$ production at LHC is largely limited by kinematics due to the heaviness of $W_R$. In the right panel of Figure~\ref{fig:limits_LLP2}, when the mixing angle $\sin\theta_2$ is large, e.g. $\gtrsim 10^{-2}$, and $m_{H_3} < m_\pi$ (the hadronic decays are kinematically forbidden), though $H_3 \to \gamma\gamma$ is sub-leading to the leptonic decays $H_3 \to \ell^+ \ell^-$ (see Figure~\ref{fig:BR}), the $\gamma\gamma$ channel is yet the dominant channel, here mediated mainly by the SM fermion loops. With the huge number of signal events at both FCC-hh and the forward detector shown in Figure~\ref{fig:signal}, the LLP searches can probe the mixing angle $\sin\theta_2$ up to order one, which is truly complementary to the {\it indirect} limits from lower energy flavor and Higgs data. This is further depicted by the summary plots in Figure~\ref{fig:limits_all}.

\begin{figure}[!t]
  \centering
  \includegraphics[width=0.48\textwidth]{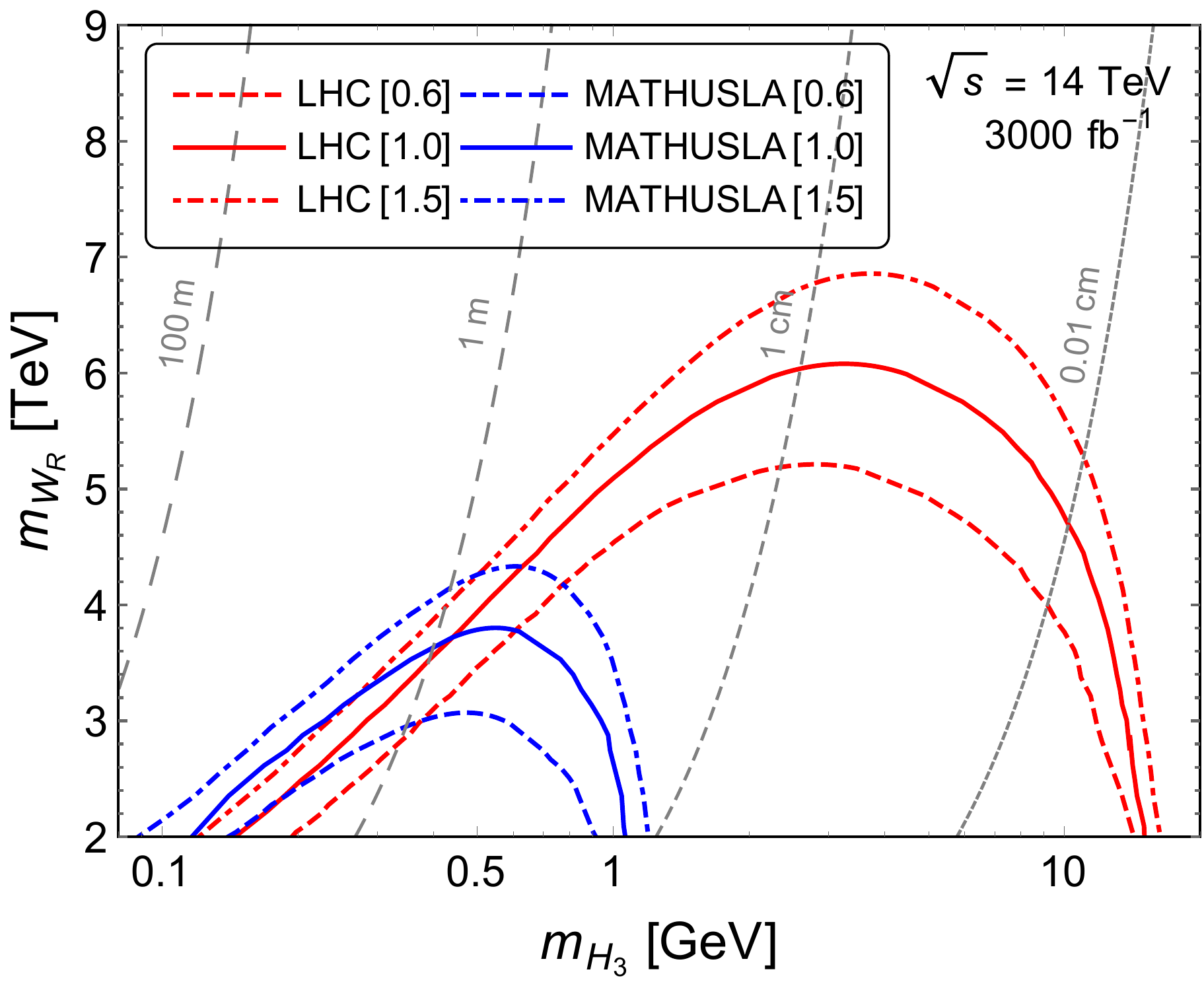}
  \includegraphics[width=0.5\textwidth]{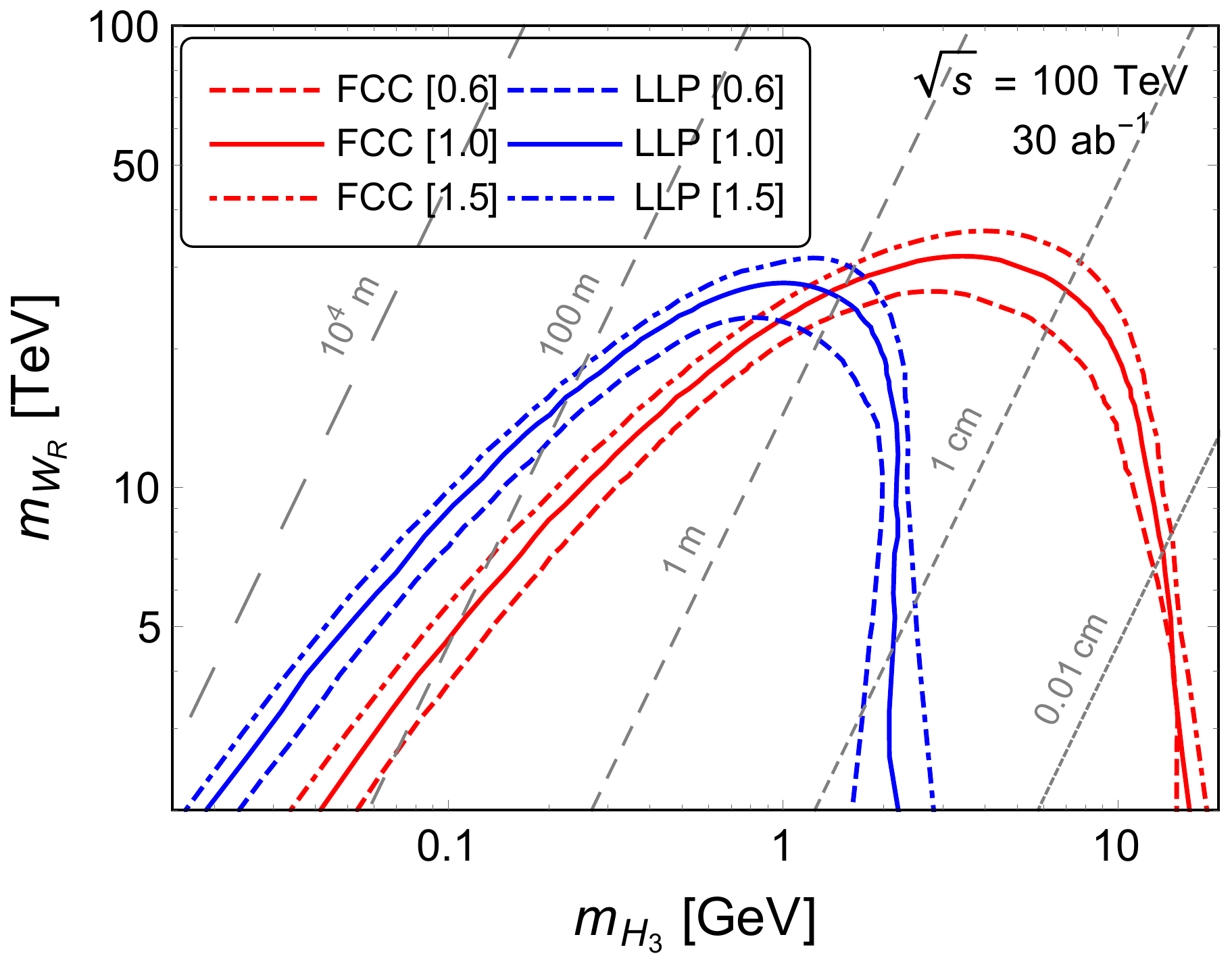}
  \caption{Collider sensitivity contours in the $m_{H_3}$-$m_{W_R}$ plane from future LLP searches at LHC and FCC-hh. The grey contours indicate the proper lifetime of $H_3$ with $g_R = g_L$; for $g_R \neq g_L$, the lifetime has to be rescaled by the factor of $(g_R/g_L)^{-2}$.}
  \label{fig:limits_LLP3}
\end{figure}

From Figures~\ref{fig:limits_LLP} and \ref{fig:limits_LLP2}, one might have noticed that the sensitivities become independent of $\sin\theta_{1,2}$ for very small mixing values, as the dominant contribution to the $H_3 \to \gamma\gamma$ mode only depends on the heavy gauge boson loops, and therefore, on the RH gauge coupling $g_R$ and the RH scale $v_R$. Hence, it is instructive to translate the collider sensitivity regions in the $m_{H_3}-m_{W_R}$ plane by varying $g_R$ and $v_R$, and assuming very small values of $\sin\theta_{1,2}$ to ensure that the $H_3\to \gamma\gamma$ BR is almost 100\%. This is shown in Figure~\ref{fig:limits_LLP3} for different values of $g_R/g_L$. For $g_R=g_L$, we can probe $m_{W_R}$ values up to 6 TeV or so at the LHC, which is complementary to the conventional collider searches of LR models through the same-sign dilepton plus multi-jet signal~\cite{Keung:1983uu,Ferrari:2000,Nemevsek:2011hz,Das:2012ii,
AguilarSaavedra:2012gf,Chen:2013fna,Deppisch:2015qwa,Gluza:2015goa,Ng:2015hba,
Dev:2015kca,Lindner:2016lxq,Mitra:2016kov, Biswal:2017nfl, Gluza:2016qqv}, or other collider signals in the heavy Higgs boson sector~\cite{Gunion:1986im, Dev:2016dja, Maiezza:2015lza, CMS:2017pet, Azuelos:2004mwa, Perez:2008ha, Jung:2008pz, Bambhaniya:2013wza, Dutta:2014dba, Bambhaniya:2014cia, Bambhaniya:2015wna, Nemevsek:2016enw}.

For completeness, we also present in~\ref{app:light_RHN} an updated sensitivity study for the displaced vertex signal in the fermion sector of the LR model, namely, from light RHN decays. Again, this probes a region complementary to those being probed by the traditional collider searches~\cite{Khachatryan:2014dka, Aad:2015xaa, Khachatryan:2016jqo}.

\section{Light neutral scalar in $U(1)_{B-L}$ model}\label{sec:U1}
\label{sec:U1}

In this section, we discuss the light neutral scalar phenomenology in a simpler model based on $SU(2)_L\times U(1)_{I_{3R}}\times U(1)_{B-L}$ local symmetry. This $U(1)_{B-L}$ model can be viewed in some sense as the ``effective'' theory of LR model at TeV scale with the $SU(2)_R$ breaking scale and the mass of the heavy $W_R$ bosons much higher than the TeV scale.  The SM fermions are assigned to the gauge group $SU(2)_L\times U(1)_{I_{3R}}\times U(1)_{B-L}$ as
\begin{align}
& Q=(u_L, \: d_L)^{\sf T}: \left({\bf 2}, 0, \frac13\right);  \quad
L=(\nu, \:  e_L)^{\sf T}: \left({\bf 2}, 0, -1 \right); \nonumber \\
& u_R: \left({\bf 1}, \frac12, \frac13 \right); \quad
d_R: \left({\bf 1}, -\frac12, \frac13 \right); \quad
e_R: \left({\bf 1},-\frac12, -1 \right).
\end{align}
Anomaly freedom requires that this model has three RHNs with gauge quantum numbers $N_a: ({\bf 1}, 1/2, -1)$. The minimal Higgs fields in the model include $H({\bf 2},-1/2, 0)$ and $\Delta({\bf 1},-1,2)$ with the following Yukawa couplings:
\begin{eqnarray}
{\cal L}_Y \ = \ h_u\overline{Q}Hu_R+h_d\overline{Q}\widetilde{H}d_R+h_e\overline{L}\widetilde{H}e_R
+h_\nu\overline{L}{H}N+f\overline{N}^c\Delta N+ {\rm H.c.} \, .
\label{eq:1}
\end{eqnarray}
Note that $\langle\Delta^0\rangle=v_R$ breaks the gauge symmetry down to the SM gauge group which is  further broken by  $\langle H^0\rangle=v_{\rm EW}$ to $U(1)_{\rm em}$. From the Yukawa interactions in Eq.~\eqref{eq:1} it is clear that after symmetry breaking this leads to the type I seesaw formula for neutrino masses. In this model, $H_3 = {\rm Re}(\Delta^0)$, which mixes with the SM Higgs, governed by the angle $\sin\theta$. Different from the LR model, in the $U(1)_{B-L}$ model, we do not have the extra heavy gauge bosons, as well as the heavy doublet, which change essentially the production and decay properties of the light scalar $H_3$.

\subsection{Couplings and decay}\label{sec:decayU1}

The couplings of $H_3$ to the SM fermions are proportional to the SM Yukawa couplings, rescaled by the mixing angle $\sin\theta$, all of which are flavor conserving. However, flavor-changing coupling $H_3 \bar{s} b$ can arise at one-loop level, through the $W - {\rm top}$ loop~\cite{Batell:2009jf}:
\begin{eqnarray}
{\cal L}_{\rm eff} \ = \ \frac{3\sqrt2 G_F m_t^2 V_{ts}^\ast V_{tb} \sin\theta}{16\pi^2} \,
\frac{m_b H_3 \bar{s}_L b_R}{\sqrt2 v_{\rm EW}}  + {\rm H.c.} \,.
\end{eqnarray}
Similarly, we can have the loop-induced flavor-changing couplings to $ds$ and $db$, which are all dominated by the top-quark loops.

If $m_{H_3} \lesssim {\rm GeV}$, it decays predominantly into the SM fermions at tree level, and into $\gamma\gamma$ and $gg$ at one-loop level, with the proper lifetime and branching ratios into leptons, hadrons and photons shown in Figure~\ref{fig:lifetime}. For the hadronic modes, we combine the decays into quarks and gluons. The BRs do not depend on the mixing angle but only on $H_3$ mass, as all the couplings are universally proportional to the mixing angle.

\begin{figure}[!t]
  \centering
  \includegraphics[width=0.455\textwidth]{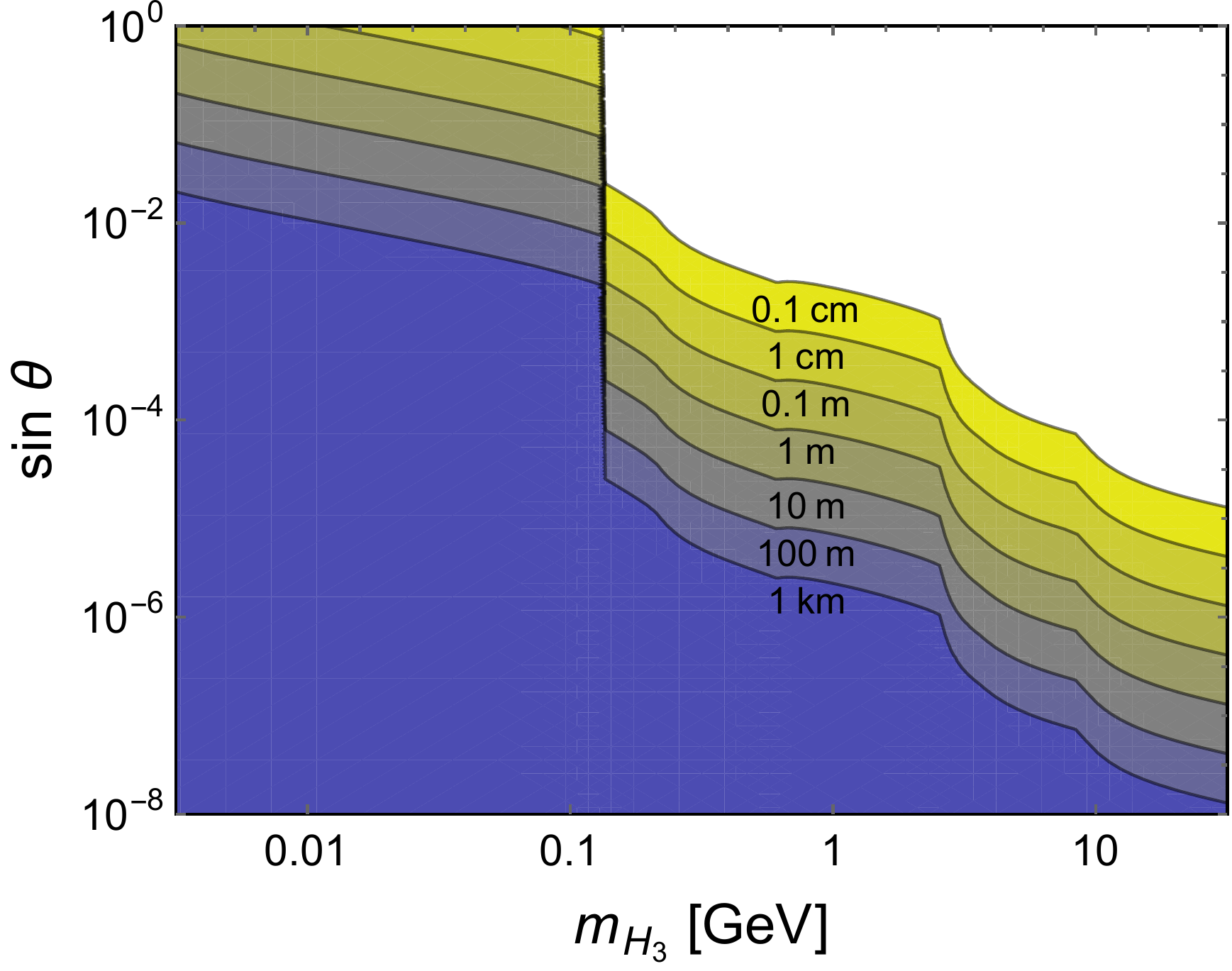}
  \includegraphics[width=0.48\textwidth]{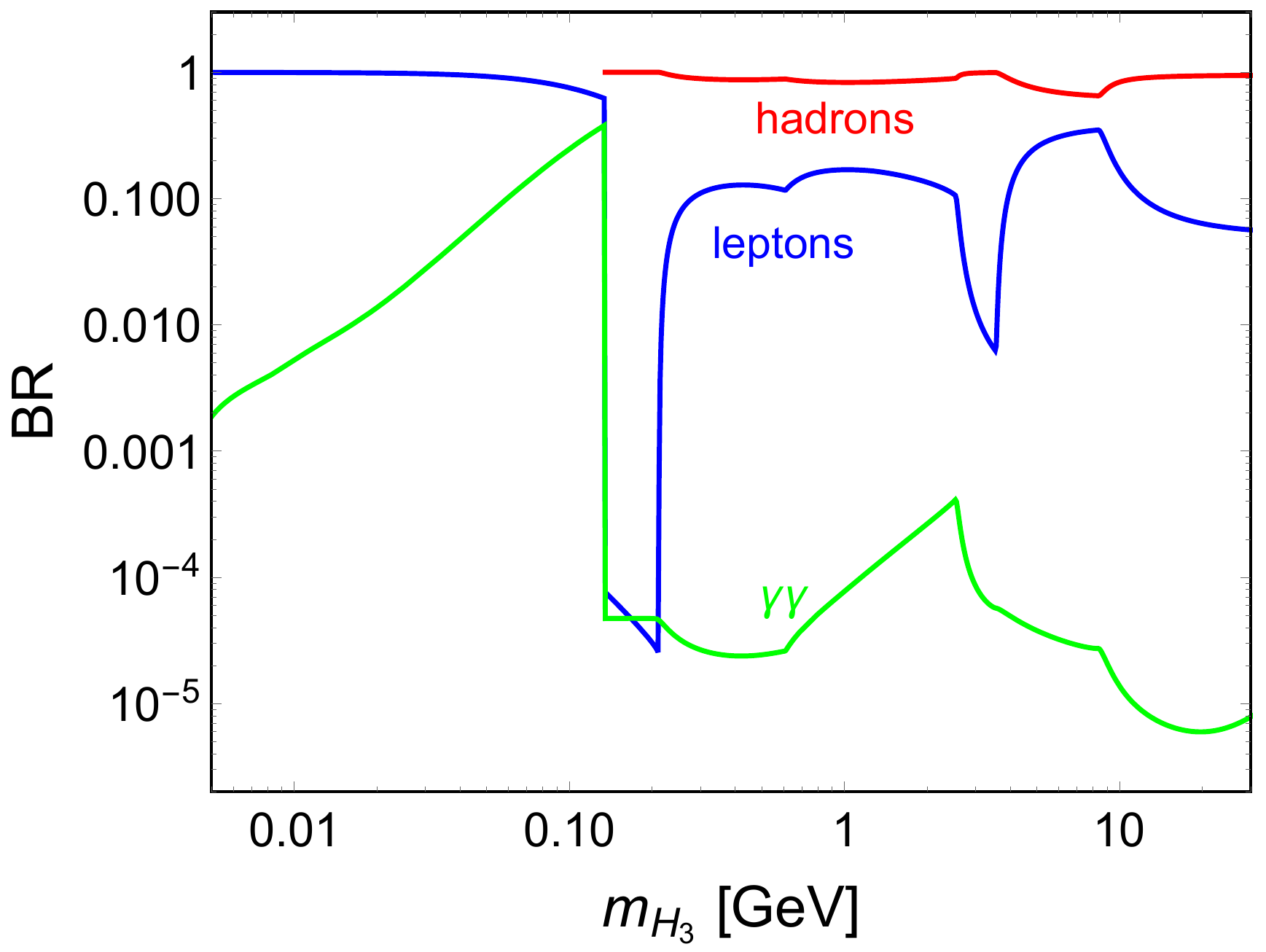}
  \caption{{\it Proper} lifetime and branching ratios of $H_3$ in the $U(1)_{B-L}$ model. }
  \label{fig:lifetime_U(1)}
\end{figure}

\subsection{Meson limits}\label{sec:mesonU1}

\begin{figure}[!t]
  \centering
  \includegraphics[width=0.48\textwidth]{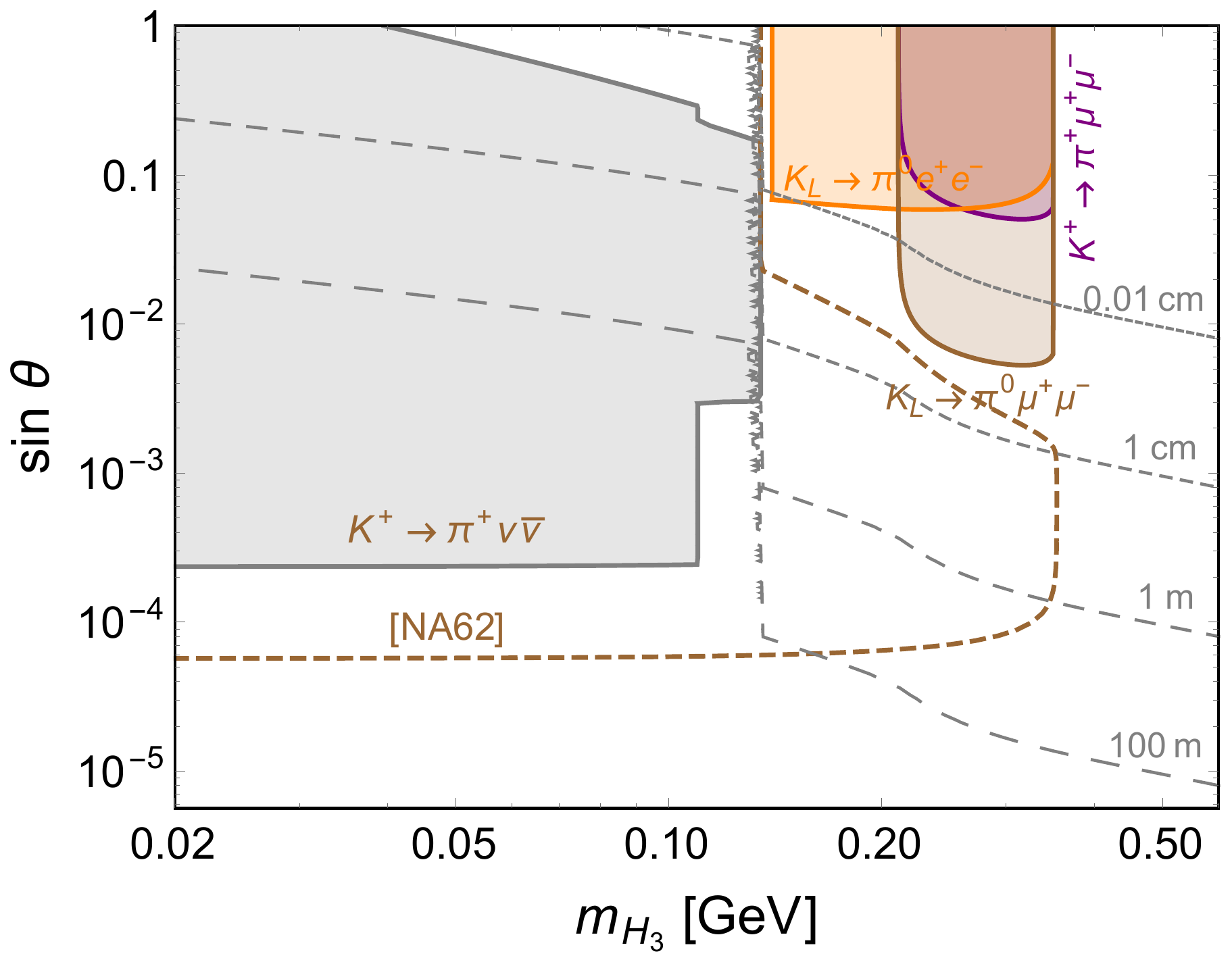}
  \includegraphics[width=0.48\textwidth]{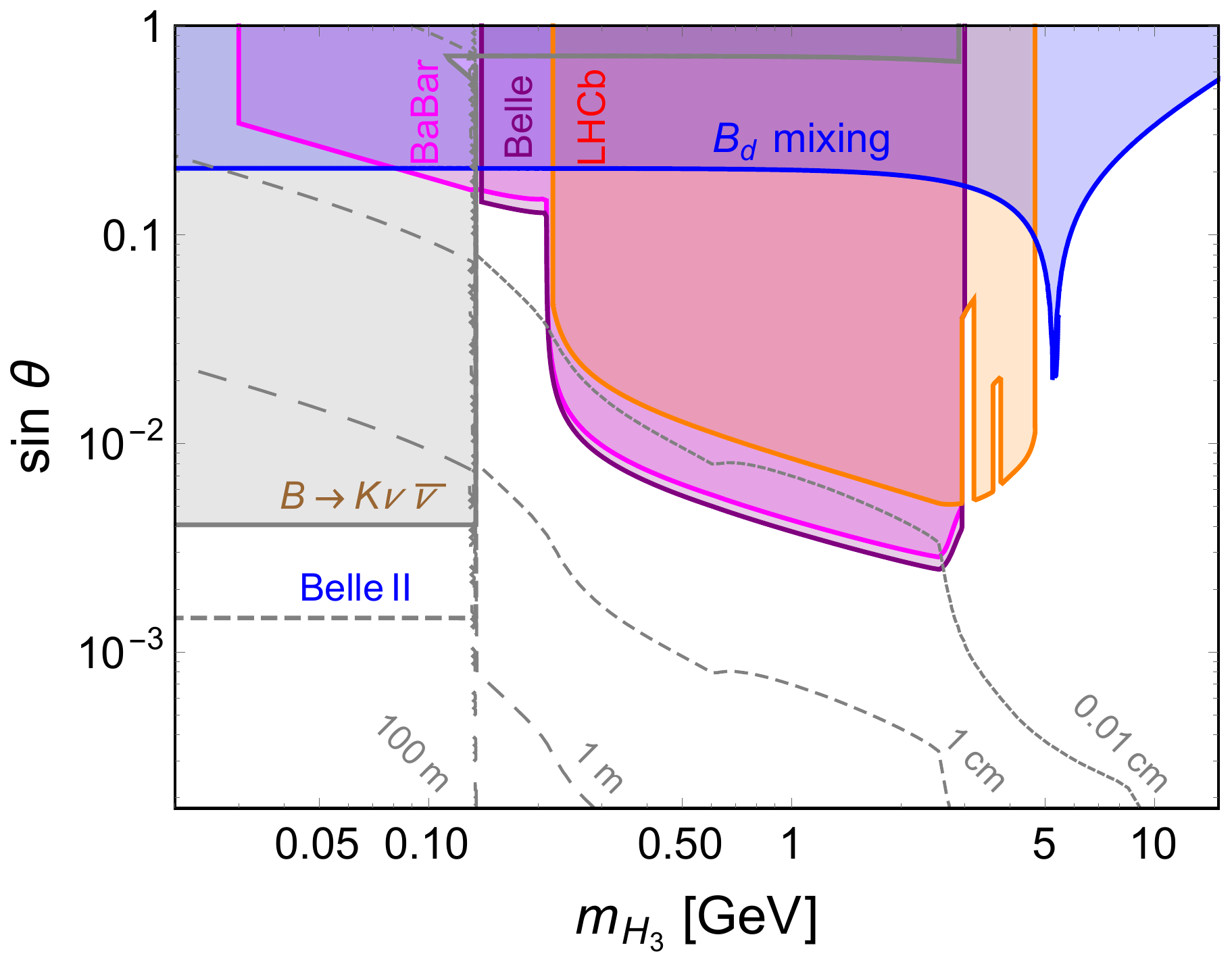} \\
  \includegraphics[width=0.48\textwidth]{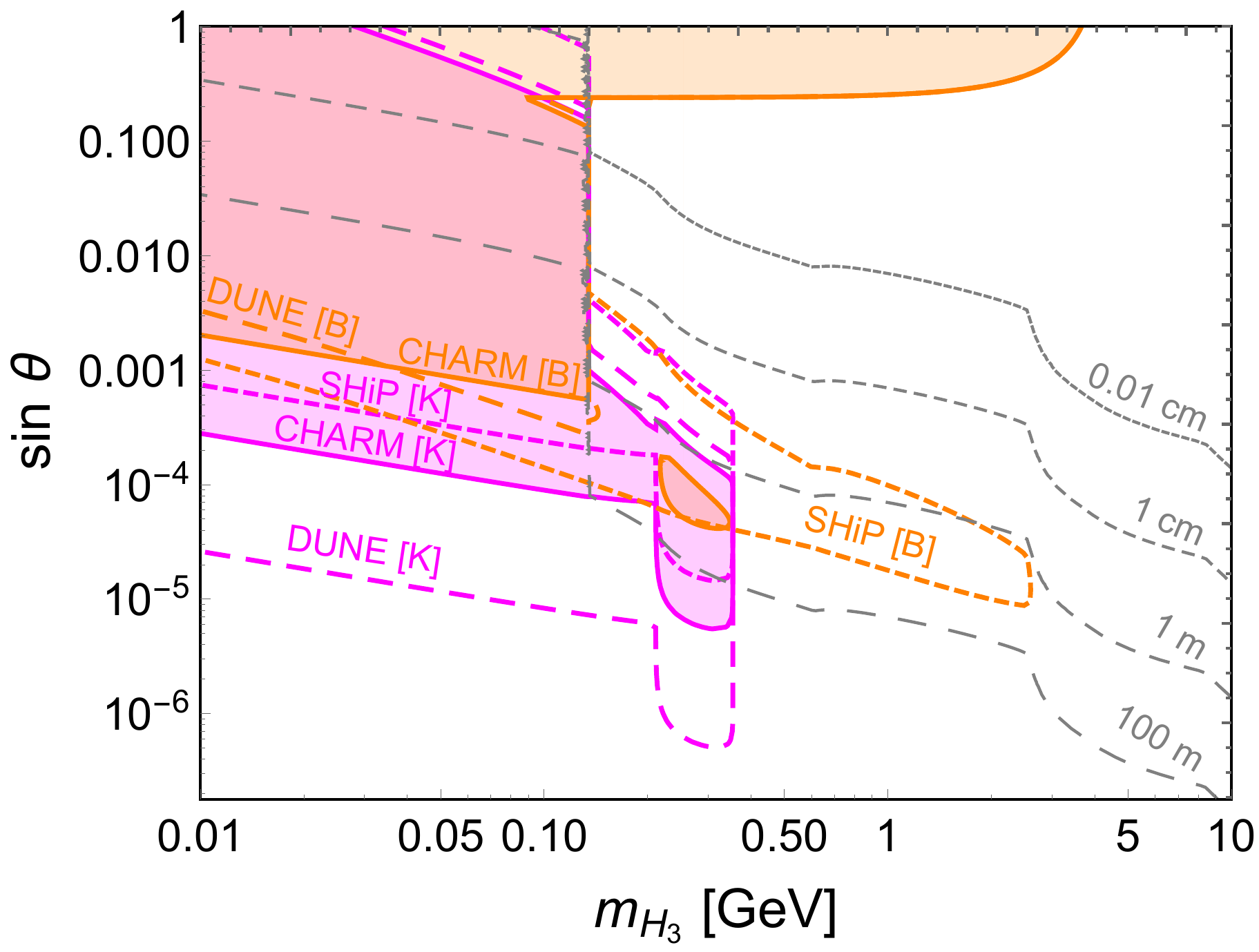}
  \caption{Limits on the light scalar mass $m_{H_3}$ and its mixing with the SM Higgs $\sin\theta$ from the $K$ meson decay (top, left), $B$ meson decay and oscillation (top, right) and beam-dump experiments (bottom). The $K$ and $B_s$ mixing limits are very weak and not shown here.}
  \label{fig:limits_U(1)}
\end{figure}

With the loop-level flavor-changing couplings (and tree level flavor conserving couplings), we can apply the limits from $K$ and $B$ meson oscillations and rare decays, as for the light scalar in the LR model (see Sections~\ref{sec:osc} and \ref{sec:mdecay}). Compared to the LR case, the constraints are much weaker, and the most promising limits are from the lepton ($ee$ and $\mu\mu$) and hadron decays but not the diphoton channel, which can be understood from the BR plot (Figure~\ref{fig:lifetime_U(1)}). All the limits on the mixing angle as a function of the $H_3$ mass are collected in Figure~\ref{fig:limits_U(1)}. Note that some of the limits are very weak and not shown in the plots, such as those from the total width of $B$ mesons, the decays $B_s \to \mu\mu$ and $\Upsilon \to \gamma H_3$, and those from Higgs measurements and SM Higgs invisible decay. Comparing Figures~\ref{fig:limits_all} and \ref{fig:limits_U(1)}, we find a new key feature that distinguishes the LR model from the $U(1)_{B-L}$ model, namely, in the former case, the meson oscillation and decay constraints rule out larger mixing angles, thus naturally ensuring the long-lived nature and diphoton decay of the light scalar, whereas in the latter case, the FCNC constraints are not so stringent, and moreover, the diphoton mode is not the dominant one.

\subsection{Production and LLP searches}\label{sec:colU1}

In the $U(1)$ model, the FCNC couplings of $H_3$ to the SM quarks arise at loop level; therefore, compared to the LR scenarios, the flavor limits on the mixing angle $\sin\theta$ is much weaker, as shown in Figure~\ref{fig:limits_U(1)}. Consequently, the light scalar $H_3$ could be produced either from mixing with the SM Higgs or through the gauge interaction with the heavy $Z_R$ boson. For a large mixing angle $\sin\theta$, the scalar $H_3$ could couple to the top quark, whch induces an effective $H_3 gg$ coupling at the one-loop level, like the SM Higgs case. Then the dominant production mode in the scalar portal is
\begin{eqnarray}
gg \to H_3 g \,,
\end{eqnarray}
at the parton-level,\footnote{Without the associated hard jet(s) in the final state, the light scalar $H_3$ is very likely to go in the beam pipe direction, which makes it hard to write such events to tape, nor could the decay products of $H_3$ generate any signal in the surface detector MATHUSLA.} with subleading contributions from the quark parton processes. When $H_3$ is light as we are considering, the production cross section is almost a constant,
\begin{eqnarray}
\sigma (pp \to H_3 j + X) \simeq (25 \, {\rm pb}) \times \sin^2\theta \,,
\end{eqnarray}
with a conservative $k$-factor of 1.5, and a simple cut on the jet $p_T (j) > 50$ GeV. $H_3$ could also produced via other modes, e.g. heavy VBF, whose cross sections are, however, much smaller. 

\begin{figure}[!t]
  \centering
  \includegraphics[width=0.48\textwidth]{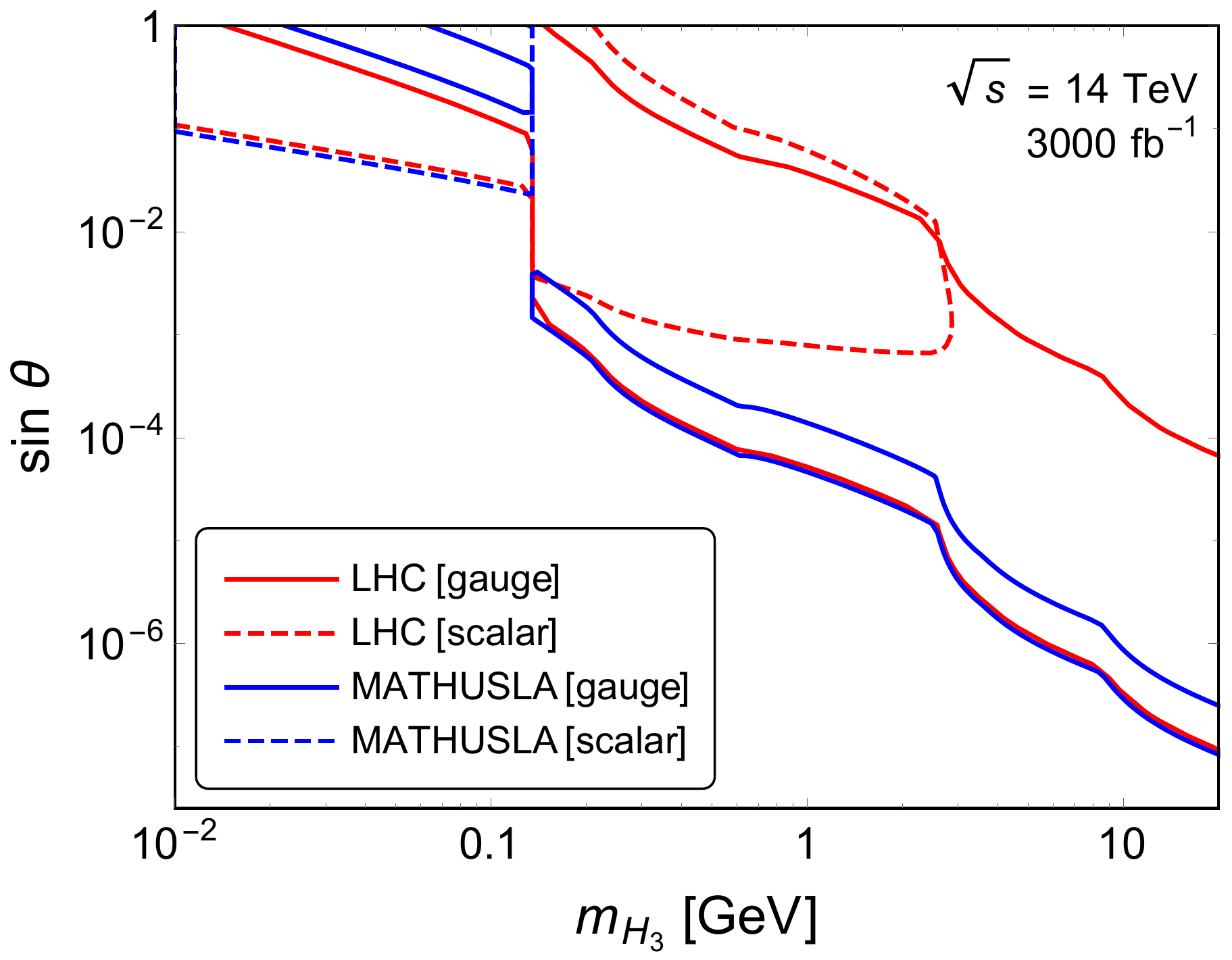}
  \includegraphics[width=0.48\textwidth]{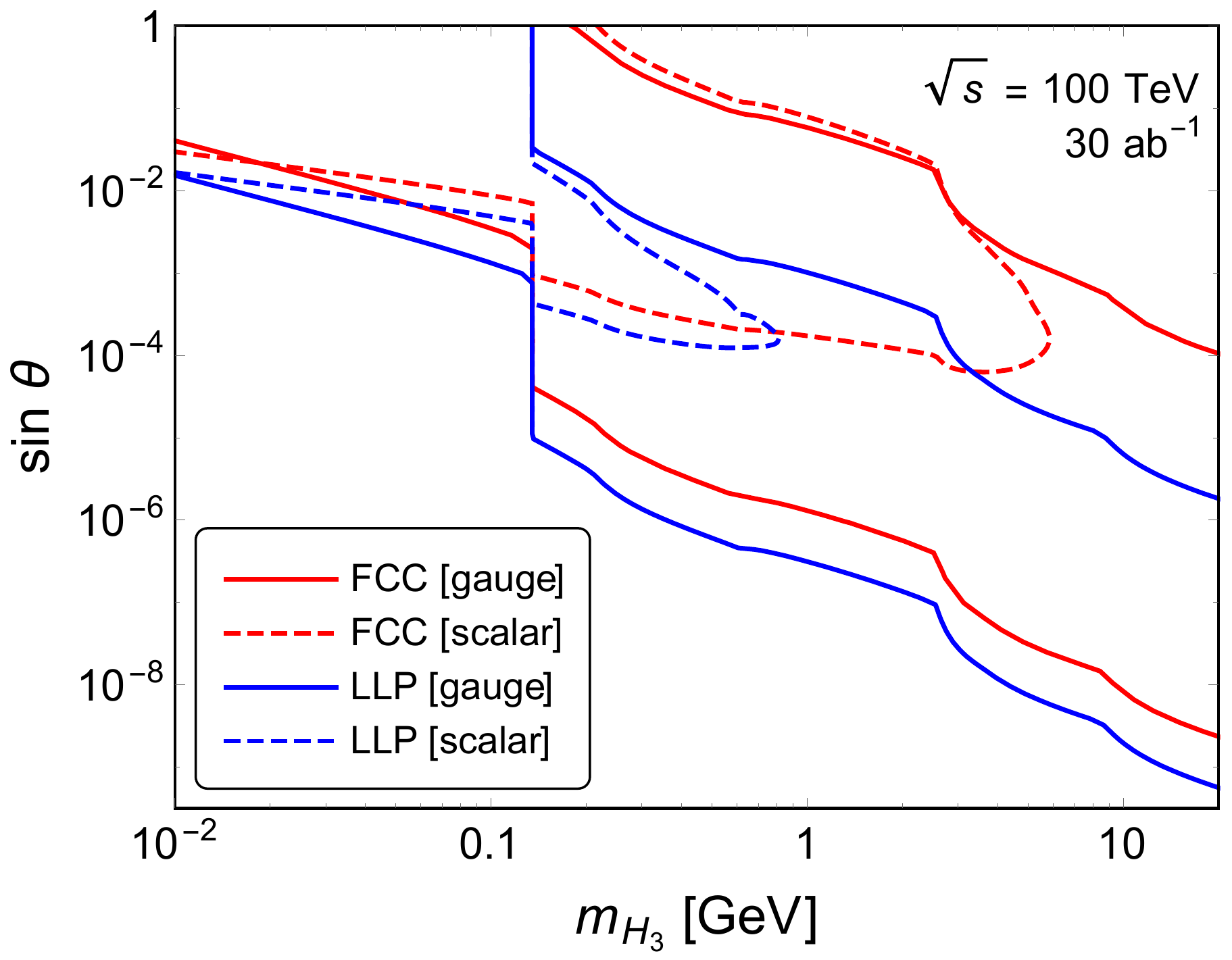}
  \caption{LLP search sensitivities at LHC, MATHUSLA (left), FCC-hh and the forward detector therein (right) in the $U(1)_{B-L}$ model, through both the gauge and scalar portals by coupling to the $Z_R$ boson or mixing with the SM Higgs, with $g_R = 0.835 g_L$. For the LHC (FCC-hh) we assume 30 (50) signal events for the hadronic decays and 10 (20) for the lepton final state, while for MATHUSLA (forward detector) we assume 4 (10) signal events.}
  \label{fig:LLP_U1}
\end{figure}

With a sufficiently small $m_{H_3}$, i.e. lighter than the pion mass, such that $H_3$ could decay only into $e^+ e^-$ or $\gamma\gamma$ which are suppressed by the tiny Yukawa coupling or the loop factor, $H_3$ could be long-lived enough to generate displaced signals in the ATLAS/CMS detector or even at the surface detector MATHUSLA.
Parton level simulations reveal that the associated jet in the final state tends to be very soft, mostly from the gluon bremsstrahlung processes, and only a small portion of the events could arrive at the MATHUSLA detector. It turns out that the dedicated LLP searches at the surface detector could yet probe a large region in the scalar portal if the mixing angle $\sin\theta > 10^{-2}$ and the scalar is lighter than roughly 100 MeV, assuming 4 signal events, as seen in Fig.~\ref{fig:LLP_U1}. With more events collected by ATLAS/CMS and much shorter decay length, the displaced vertex searches at LHC could probe a much larger region as shown in Fig.~\ref{fig:LLP_U1}, which is largely complementary to the ULLP searches at MATHUSLA and a cross-check in the overlapped regions.


\begin{figure}[!t]
  \centering
  \includegraphics[width=0.48\textwidth]{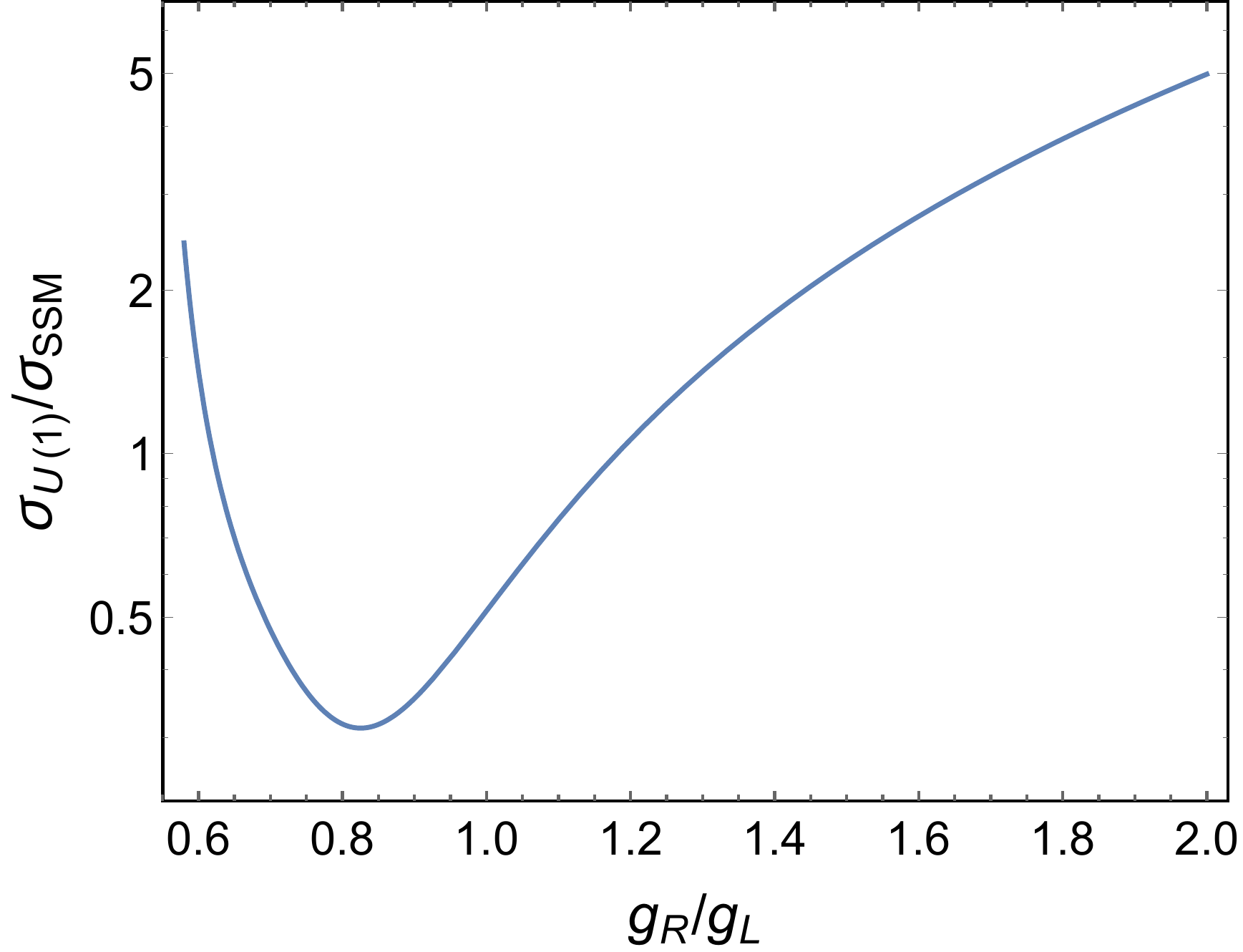}
  \caption{Production cross section of $\sigma (pp \to Z_R)$ times the branching ratio into dileptons $ee$ and $\mu\mu$, in units of the corresponding cross section in the sequential SM.}
  \label{fig:productionU1}
\end{figure}

In the gauge portal, i.e. via interaction with the $Z_R$ boson, the dominant production mode is the associated production with a heavy $Z_R$ boson, as in the LR models, with $Z_R$ decaying further into the SM quarks and charged leptons (here again we do not consider the decays $Z_R \to \nu\bar\nu$, $NN$ as in the LR case):
\begin{eqnarray}
pp \to Z_R^{\ast} \to Z_R H_3 \,, \quad
Z_R \to q\bar{q},\, \ell^+ \ell^- \,,
\end{eqnarray}
with subleading contributions from the VBF of two $Z_R$ bosons $pp \to Z_R^\ast Z_R^\ast jj \to H_3 jj$. With the heavy $Z_R$ boson taking away most of the energy in the final state, the light scalar $H_3$ tends to be very soft, with a transverse momentum typically $\lesssim 100$ GeV for most of the events. Therefore only a small portion of the $H_3$ events could arrive at the surface detector, as in the scalar portal.
With the current LHC dilepton searches~\cite{ATLAS:2016cyf,CMS:2016abv, Patra:2015bga, Lindner:2016lpp}, the $Z_R$ mass is required to be above the TeV scale. Rescaling from the $Z'$ boson in sequential SM, we get the mass limit to be 3.72 TeV with the gauge coupling $g_R = g_L$ in the $U(1)_{B-L}$ model. With a smaller $g_R = 0.835 g_L$, the constraint becomes slightly less stringent, at $3.64$ TeV. This is the optimistic scenario, as when $g_R$ gets smaller, the couplings to quarks and leptons would be larger benefitting from a larger gauge coupling $g_{BL}$. The exact dependence of the production cross section (with respect to that in the sequential SM) with the ratio $g_R/g_L$ is shown in Figure~\ref{fig:productionU1}. Note that the ratio $g_R/g_L$ is bounded from below, otherwise the  $U(1)_{B-L}$ gauge coupling $g_{BL}$ becomes non-perturbative:
\begin{align}
\frac{g_R}{g_L} \ > \  \tan\theta_w\left(1-\frac{e^2}{g_{BL}^2\cos^2\theta_w}\right)^{-1/2} \, .
\end{align}
With $Z_R$ mass set at the optimistic value $g_R = 0.835 g_L$, we obtain the production cross section of 0.97 fb in the gague portal, assuming a $k$-factor of $1.2$.





With the optimistic value $g_R = 0.835 g_L$, which corresponds to the weakest limit on $m_{Z'}$ from Figure~\ref{fig:productionU1},  and therefore, the largest $H_3$ production cross section at the LHC, we estimate the expected signal events at $\sqrt{s} = 14$ TeV LHC with a luminosity of 3000 fb$^{-1}$. Assuming the signal number of 30 for the hadronic decays and 10 for the leptonic decays, we can probe a large region in the plane of $m_{H_3}$ and $\sin\theta$, as shown in Figure~\ref{fig:LLP_U1}. Inside the region surround by the red curve, the signal number in the gauge portal could even reach up to few hundreds. The direct LLP searches at LHC are largely complementary to the indirect limits from meson oscillations and rare decays in Figure~\ref{fig:limits_U(1)}.

Limited by the $Z_R$ mass limits and thus the small production cross section, the ULLP signal number at MATHUSLA could probe a narrow band in the parameter space of $m_{H_3} - \sin\theta$, as shown in Fig.~\ref{fig:LLP_U1}, though the mixing angle $\sin\theta$ could be much smaller than that in the scalar portal, with the detector geometry and efficiency as given in Ref.~\cite{Chou:2016lxi}.
When $g_R/g_L$ is different from the optimistic value of 0.835, the $Z'$ limits from LHC become more stringent, and it is much more difficult to collect ULLP signals at MATHUSLA, irrespective of the final states being leptons or jets. However, the virtually background-free environment in the MATHUSLA detector might make it possible to probe the $U(1)_{B-L}$ model even with such small number of signal events. The prospects at a future 100 TeV collider with a dedicated forward detector is more promising for ULLP searches.

\begin{figure}[!t]
  \centering
  \includegraphics[width=0.58\textwidth]{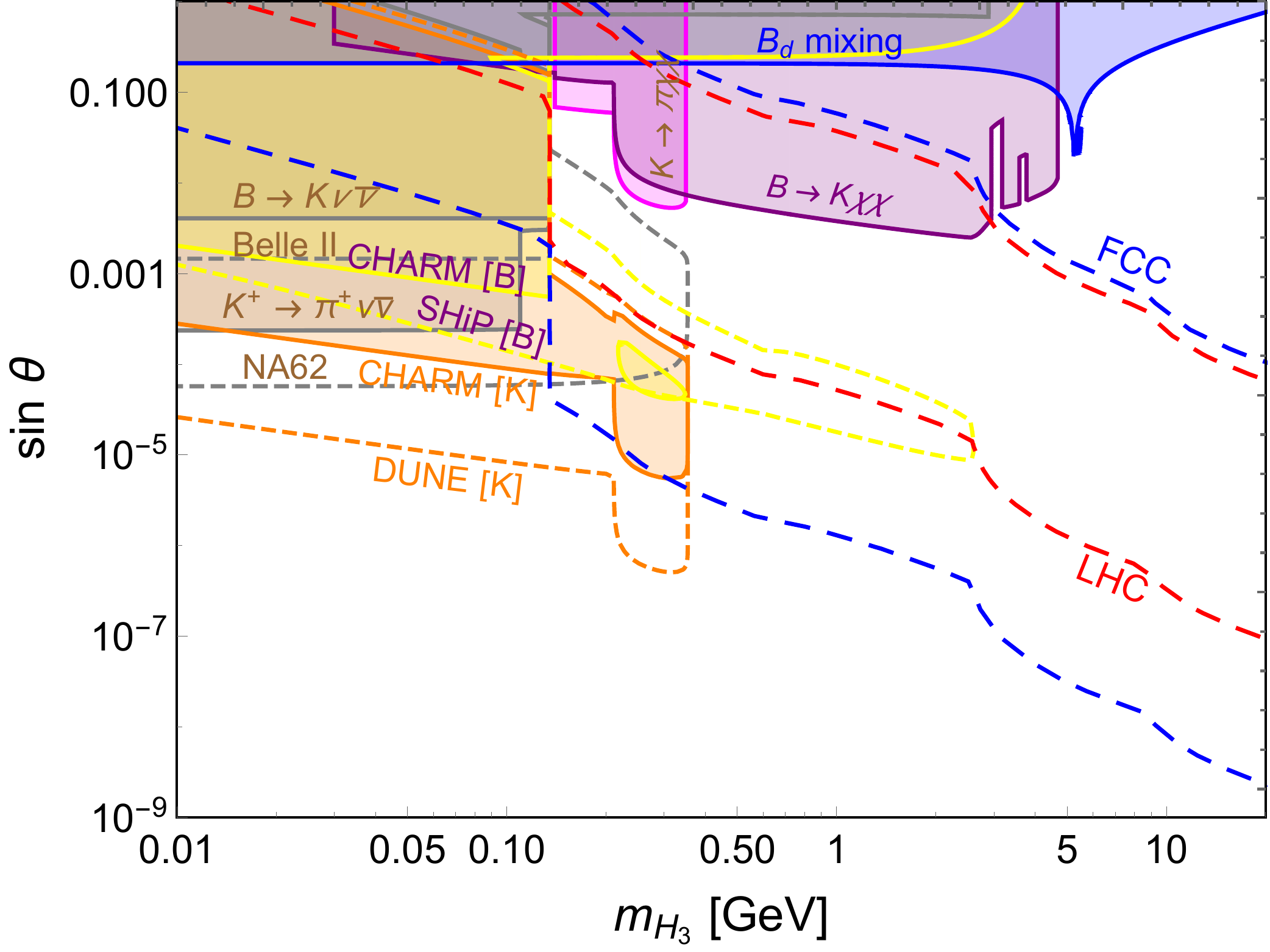}
  \caption{A summary of the important limits and sensitivity curves in the $m_{H_3}-\sin\theta$ plane in the $U(1)_{B-L}$ model, extracted from Figures~\ref{fig:limits_U(1)} and \ref{fig:LLP_U1}. The shaded regions are excluded, while those surrounded by the dashed red and blue lines are the expected sensitivities in the gauge portal from LLP searches at LHC and future 100 TeV collider FCC-hh. For details, see Section~\ref{sec:U1}.}
  \label{fig:limits_all2}
\end{figure}

With the heavy $Z_R$ boson more abundantly produced at the future 100 TeV collider such as FCC-hh, the probable regions of LLP searches could be significantly broadened in both the scalar and gauge portals, as presented in the right panel of Figure~\ref{fig:LLP_U1}, where we have assumed 50 LLP events in the hadronic channel and 20 in the leptonic channel, and 10 events at the forward detector, with the same geometry as for the LR case. With the large production cross section, which could reach about 310 fb,\footnote{Here for the sake of concreteness and comparison with the LHC case we have assumed again the gauge coupling $g_R = 0.835 g_L$; with other values of $g_R$, the production cross section could be even larger at the 100 TeV collider, though $Z_R$ might be heavier, then the sensitivity regions in the right panel of Fig.~\ref{fig:LLP_U1} could even be larger.} and the huge luminosity of 30 ab$^{-1}$ and high center-of-mass energy, the LLP searches at future 100 TeV colliders and forward detector could probe the proper lifetime from 0.01 cm up to $10^4$ m, for a wide range of $H_3$ mass from 10 MeV up to tens of GeV, and could even probe the mixing angle up to $10^{-9}$.
As in the case of LR model, due to the large boost factors in the production of $H_3$ at the high energy colliders, the LLP searches at LHC and future 100 TeV colliders are sensitive to the relatively higher mass range, complementary to the high intensity experiments, which is explicitly shown in Figure~\ref{fig:limits_all2} where we collect all the important limits and prospects.

\section{Conclusion}  \label{sec:con}

We have pointed out that, the real part of Higgs field that breaks local $B-L$ symmetry in low-scale type I seesaw models for neutrino masses can be very light with mass in the GeV to sub-GeV range. When $B-L$ is part of the left-right seesaw model, the light scalar couplings to Standard Model fields are so weak due to FCNC constraints on the model that it necessarily becomes a long-lived particle, leading to high energy displaced photons at the current LHC detectors (see the summary plots in Figure~\ref{fig:limits_all}). Searches for them could therefore provide a new probe of the TeV scale left-right seesaw models. We have also carried out an analogous discussion for the simple $U(1)_{B-L}$ scenario. While the FCNC constraints are not so strong for this case, we show that for small mixings (or smaller $H_3$ masses) there can be observable displaced vertex signals (see Figure~\ref{fig:limits_all2}).  We have also commented (in~\ref{app:light_RHN}) on the possibility of light right-handed neutrinos giving displaced vertices.

As clearly shown in Figure~\ref{fig:limits_all} and \ref{fig:limits_all2}, the displaced vertex searches at LHC and future 100 TeV collider, no matter whether the displaced signals are the collimated diphotons, the hadronic jets or the charged leptons, are largely complementary to the probes of the oscillations and rare decays in the meson sector on the light scalar in the seesaw models. Compared to the comparatively lower-energy high intensity experiments,  the high energy frontier tends to extend the probable scalar mass to larger values, and the mixing angles to smaller values. Furthermore, both these two avenues are also complementary to the direct tests of heavy right-handed (or $U(1)_{B-L}$) sector at the hadron colliders. We believe that our work provides another window using experiments in the lifetime frontier to probe the possibility of TeV scale origin of neutrino masses.

\section*{Acknowledgments} We thank David Curtin and Brian Shuve for useful discussions and correspondence on MATHUSLA. The work of R.N.M. is supported by the US National Science Foundation grant No.~PHY1620074. Y.Z. would like to thank the IISN and Belgian Science Policy (IAP VII/37) for support.

\appendix

\section{Partial decay widths of $H_3$} \label{app:decay}
Here we collect all the partial widths for the dominant decay modes of $H_3$:
\begin{eqnarray}
\Gamma (H_3 \to q\bar{q}) & \ = \ &
\frac{3 m_{H_3}}{16\pi}
\left[ \sum_{i,j} \left| \mathcal{Y}_{u, \, ij} \right|^2
\beta^3_2 (m_{H_3}, m_{u_i}, m_{u_j}) \Theta (m_{H_3}- m_{u_i}- m_{u_j}) \nonumber \right.  \\
&& \qquad\quad + \left. \sum_{i,j} \left| \mathcal{Y}_{d, \, ij} \right|^2
\beta^3_2 (m_{H_3}, m_{d_i}, m_{d_j}) \Theta (m_{H_3}- m_{d_i}- m_{d_j})  \right] , \\
\Gamma (H_3 \to \ell^+ \ell^-) & \ = \ &
\frac{m_{H_3}}{16\pi}
\sum_{i,j} \left| \mathcal{Y}_{e, \, ij} \right|^2
\beta^3_2 (m_{H_3}, m_{e_i}, m_{e_j}) \Theta (m_{H_3}- m_{e_i}- m_{e_j}) \,, \\
\label{eqn:H3diphoton}
\Gamma (H_3 \to \gamma\gamma) & \ = \ &
\frac{\alpha^2 m_{H_3}^3}{1028 \pi^3}
\left| \frac{\sqrt2}{v_R} A_{0}(\tau_{H_1^\pm}) +
\frac{4\sqrt2}{v_R} A_{0}(\tau_{H_2^{\pm\pm}}) \right. \nonumber \\
&& \qquad\qquad +\left. \frac{\sqrt2}{v_{\rm EW}} \sum_{f = q, \ell} f_f N^f_C Q_f A_{1/2} (\tau_f) +
\frac{\sqrt2}{v_R} A_{1} (\tau_{W_R}) \right|^2 \,, \\
\Gamma (H_3 \to gg) & \ = \ &
\frac{G_F \alpha_s^2 m_{H_3}^3}{36 \sqrt2 \pi^3}
\left| \frac34 \sum_{f = q} f_f A_{1/2} (\tau_f) \right|^2 \,,
\end{eqnarray}
with the kinetic function
\begin{eqnarray}
\label{eqn:beta2}
\beta_2 (M,\, m_1,\, m_2) & \ \equiv \ & \left[ 1 - \frac{2(m_1^2 + m_2^2)}{M^2} + \frac{(m_1^2 - m_2^2)^2}{M^4} \right]^{1/2} \,,
\end{eqnarray}
$\mathcal{Y}_{u,d,e}$ the Yukawa couplings given in Table~\ref{tab:coupling},
\begin{align}
\mathcal{Y}_{u} & \ = \ \widehat{Y}_U
  \sin\tilde\theta_1
  -  \left( V_L \widehat{Y}_D V_R^\dagger \right) \sin\tilde\theta_2,\\
\mathcal{Y}_{d} & \ = \ \widehat{Y}_D
  \sin\tilde\theta_1
  -\left( V_L^\dagger \widehat{Y}_U V_R \right) \sin\tilde\theta_2,\\
\mathcal{Y}_{e} & \ = \ \widehat{Y}_E
  \sin\tilde\theta_1 - Y_{\nu N}
  \sin\tilde\theta_2 \, ,
\end{align}
$f_f$ the normalization factor with respect to the SM Yukawa couplings,
\begin{eqnarray}
f_{u,i} & \ = \ & \sin\tilde\theta_1
- \frac{(V_L \widehat{M}_d V_R^\dagger)_{ii}}{m_{u,i}}
\sin\tilde\theta_2 \,, \\
f_{d,i} & \ = \ & \sin\tilde\theta_1
- \frac{(V_L^\dagger \widehat{M}_u V_R)_{ii}}{m_{d,i}}
\sin\tilde\theta_2 \,, \\
f_{e,i} & \ = \ & \sin\tilde\theta_1
- \frac{Y_{\nu N, ii}}{m_{e,i}/v_{\rm EW}}
\sin\tilde\theta_2 \,,
\end{eqnarray}
and the loop functions
\begin{eqnarray}
A_{0} (\tau) & \ \equiv \ & - \left[ \tau - f(\tau) \right] \tau^{-2} \,, \\
A_{1/2} (\tau) & \ \equiv \ & 2 \left[ \tau + (\tau-1) f(\tau) \right] \tau^{-2} \,, \\
A_1 (\tau) &\equiv& - \left[ 2 \tau^2 + 3\tau + 3 (2\tau-1) f(\tau) \right] \tau^{-2} \,,
\end{eqnarray}
with $\tau_X = m_{H_3}^2 / 4m_X^2$ and
\begin{align}
f(\tau) \ \equiv \ \left\{ \begin{array}{cc}
{\rm \arcsin}^2\sqrt{\tau} & ({\rm for}~\tau\leq 1) \\
-{\displaystyle \frac{1}{4}}\left[\log \left( \frac{1+\sqrt{1-1/\tau}}{1-\sqrt{1-1/\tau}}\right)-i\pi  \right]^2 & ({\rm for}~\tau>1) \;.
\end{array}\right.
\label{fx}
\end{align}
For the heavy particle loops, only the limits below are useful for us
\begin{eqnarray}
A_0 (0) = 1/3 \,, \quad
A_{1/2} (0) = 4/3 \,, \quad
A_1 (0) = -7 \,.
\label{eqn:loopfunction}
\end{eqnarray}
Thus in Eq.~(\ref{eqn:H3diphoton}), we have a suppression factor of $5A_0(0)/A_1(0)=-5/21$ for scalar loops, with the factor of 5 coming from the sum of electric charges squared.

\section{Rare $Z$ decay $Z \to \gamma H_3$}
\label{app:Zdecay}

The partial width of rare $Z$ decay reads
\begin{eqnarray}
\Gamma (Z \to \gamma H_3) &\ = \ &
\frac{G_F^2 \alpha m_W^2}{192 \pi^4} m_Z^3
\left( 1-\frac{m_{H_3}^2}{m_Z^2} \right)^3 \nonumber \\
&& \times \left| \sin\theta_1 A_{1} (\tau_W,\, \lambda_W) +
\sum_{f=q,\ell} \frac{f_f N_C^f Q_f \hat{v}_f}{c_W}
A_{1/2} (\tau_f,\, \lambda_f) \right|^2 \,,
\end{eqnarray}
with $\tau_X = 4 m_X^2 / m_{H_3}^2$, $\lambda_X = 4 m_X^2 / m_Z^2$, and the loop functions are defined as
\begin{eqnarray}
A_{1/2}(\tau,\,\lambda) &\ \equiv \ &
I_1 (\tau,\,\lambda) - I_2 (\tau,\,\lambda) \,, \\
A_{1}(\tau,\,\lambda) &\ \equiv \ &
c_W \left[ 4 \left( 3-\frac{s_W^2}{c_W^2} \right) I_2 (\tau,\,\lambda) +
\left( \left( 1+\frac{2}{\tau} \right) \frac{s_W^2}{c_W^2}
- \left( 5+\frac{2}{\tau} \right)
\right) I_1 (\tau,\,\lambda) \right] \,, \nonumber \\
\end{eqnarray}
with
\begin{eqnarray}
I_1 (\tau, \lambda) & \ \equiv \ &
\frac{\tau\lambda}{2(\tau-\lambda)} +
\frac{\tau^2 \lambda^2}{2(\tau-\lambda)^2}
\left[ f(\tau^{-1}) - f(\lambda^{-1}) \right] \nonumber \\
&&+ \frac{\tau^2 \lambda}{(\tau-\lambda)^2}
\left[ g(\tau^{-1}) - g(\lambda^{-1}) \right] \,, \\
I_2 (\tau, \lambda) & \ \equiv \ &
- \frac{\tau \lambda}{2(\tau-\lambda)}
\left[ f(\tau^{-1}) - f(\lambda^{-1}) \right] \,,
\end{eqnarray}
$f (x)$ as defined in Eq.~(\ref{fx}), and
\begin{align}
g(\tau) \ \equiv \ \left\{ \begin{array}{cc}
\sqrt{\tau^{-1}-1} \: {\rm \arcsin}\sqrt{\tau} & ({\rm for}~\tau\leq 1) \\
{\displaystyle \frac{\sqrt{1-\tau^{-1}}}{2}}
\left[\log \left( \frac{1+\sqrt{1-1/\tau}}{1-\sqrt{1-1/\tau}}\right)-i\pi  \right] & ({\rm for}~\tau>1) \;.
\end{array}\right.
\label{gx}
\end{align}

\section{Light RH neutrinos in the LR seesaw model}\label{app:light_RHN}


The RHNs have extra charged current interactions in the LR model mediated by the heavy $W_R$ boson, which are not suppressed by the heavy-light neutrino mixings, but sensitive to the RH scale $v_R$ (or equivalently the $W_R$ mass, up to the gauge coupling $g_R$). Though the ULLP signals in this case are very similar to the light scalar case discussed in Section~\ref{sec:sens}, i.e. highly collimated leptonic and hadronic jets, the production modes and phenomenological implications are very different.

In the LR model, the decay of RHNs are predominantly mediated by an off-shell heavy $W_R$ boson: $N\to W_R^*\ell \to \ell jj$. When $m_N \ll m_{W_R}$ which is the case for a light RHN to be viable ULLP candidate at the LHC, the three-body decay width is given by
\begin{eqnarray}
\Gamma_N \ \simeq \ \frac{3 G_F^2}{32 \pi^3} \, m_N^5
\left( \frac{m_W}{m_{W_R}} \frac{g_R}{g_L} \right)^4 \,.
\end{eqnarray}
For a few-TeV scale $W_R$, if the RHN mass is order 10 GeV, then its {\it proper} lifetime would be at the cm level:
\begin{eqnarray}
\label{eqn:lifetime}
\tau^0_N \ \simeq \ 9.3 \times 10^{-3}
\left( \frac{m_N}{10\, {\rm GeV}} \right)^{-5}
\left( \frac{m_{W_R}}{3\, {\rm TeV}} \right)^{4}
\left( \frac{g_R}{g_L} \right)^{-4} \,
{\rm m} \,.
\end{eqnarray}
Such a light RHN at the GeV scale can be produced from rare meson decays, such as $D_s \to e N$ (here for simplicity we assume the RHN is of electron flavor), with the subsequent decay $N \to e \pi$~\cite{Castillo-Felisola:2015bha}. Both the production of $N$ from mesons and decay into lighter states are mediated by its gauge interaction to the heavy $W_R$ boson. The masses $m_{W_R}$, $m_N$ and the gauge coupling $g_R$ can thus be probed at dedicated beam dump experiments, such as SHiP~\cite{Castillo-Felisola:2015bha, Alekhin:2015byh}. Here we present a sensitivity study for the displaced vertex signal at the LHC.\footnote{In the minimal type I seesaw model without the LR symmetry, the small Yukawa couplings of the RHNs also make them long-lived with displaced vertex signatures at colliders~\cite{Helo:2013esa, Antusch:2016ejd}; however, their production cross section will also be suppressed by the same Yukawa couplings.}

\begin{figure}[!t]
  \centering
  \includegraphics[width=0.55\textwidth]{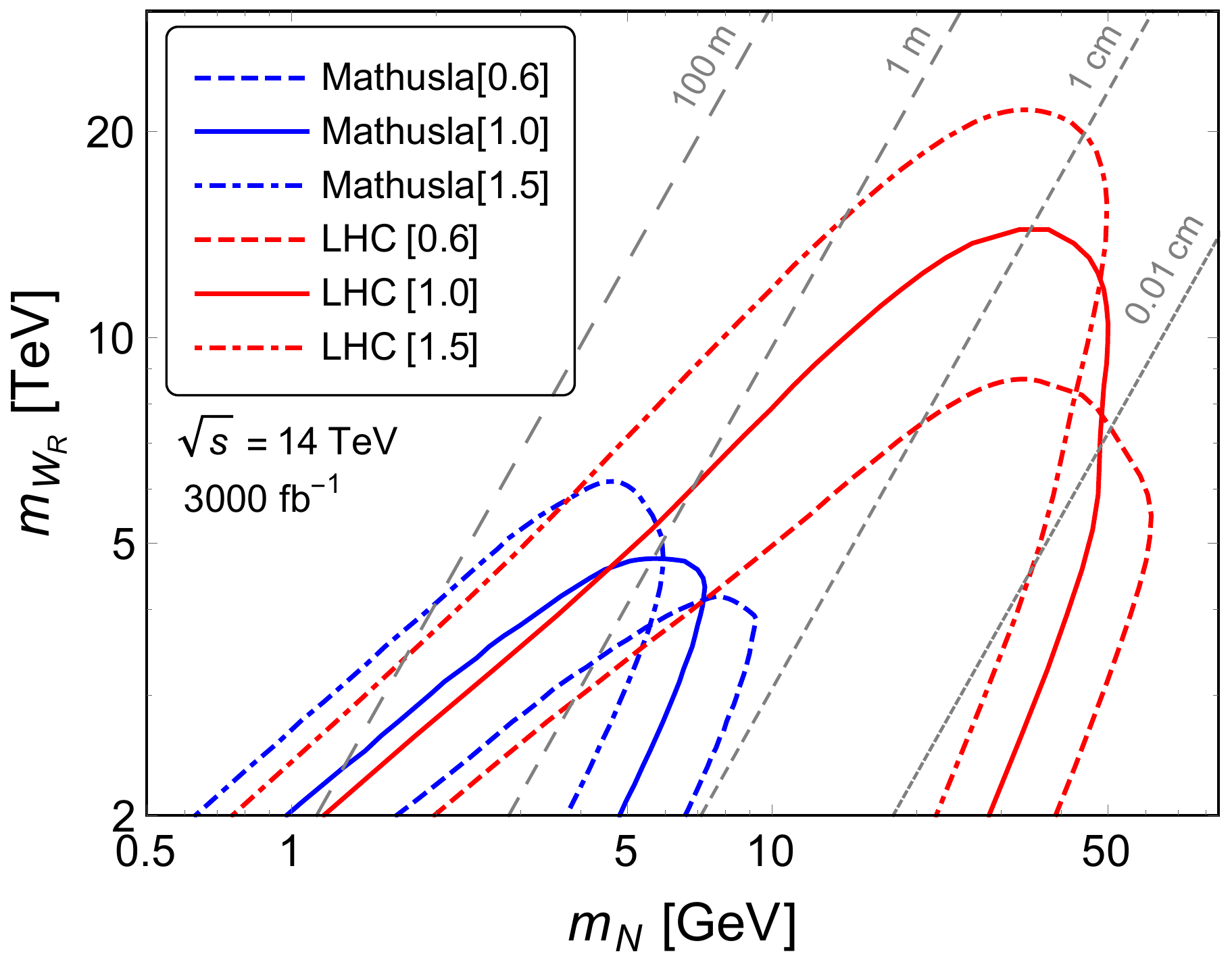}
  \caption{Light RHN sensitivity in the minimal LR model from the ULLP searches at the $\sqrt s=14$ TeV LHC (red) and MATHUSLA (blue) for three different values of $g_R/g_L = 0.6$, 1 and 1.5. The grey contours indicate the proper decay length of RHN with $g_R = g_L$; for $g_R \neq g_L$, the lifetime has to be rescaled by the factor of $(g_R/g_L)^{-4}$. }
  \label{fig:RHN}
\end{figure}

In the minimal LR model with the $SU(2)_R$ gauge symmetry broken by the RH triplet scalar, $m_{Z_R}>m_{W_R}$. Thus the dominant production of RHNs at the LHC is through the $s$-channel $W_R$ exchange: $pp \to W_R^{(*)} \to \ell N$, followed by the three-body decay $N\to W_R^*\ell \to \ell jj$~\cite{Keung:1983uu}.  
With $m_{W_R} \gtrsim 3$ TeV$(g_R/g_L)^4$, as required to satisfy the direct LHC constraints~\cite{Khachatryan:2014dka,Aad:2015xaa,Khachatryan:2016jqo}, as well as the low-energy FCNC constraints~\cite{Bertolini:2014sua}, the production cross section could reach few tens of fb, depending on the $W_R$ mass as well as the gauge coupling $g_R$. Here we have imposed the condition that the associated lepton (jet) must satisfy the basic trigger cuts of $p_T > 25$ GeV and $|\eta| < 2.5$. Requiring that the decay length 1 cm $< L_N <$ 1.5 m with the ECAL size of the ATLAS detector, we obtain the sensitivity reach for light RHN as shown in Figure~\ref{fig:RHN} for three different values of the gauge coupling $g_R/g_L = 0.6$, 1 and 1.5 at the $\sqrt s=14$ TeV HL-LHC with an integrated luminosity of 3000 fb$^{-1}$. For concreteness, we have assumed only the electron flavor $\ell = e$ without RH leptonic mixing, and the number of signal events to be 10. Note that in Figure~\ref{fig:RHN} even when the heavy $W_R$ is off-shell, i.e. $m_{W_R} \gtrsim 5$ TeV, the light RHN could yet be produced abundantly. For the purpose of illustration, we also show the proper lifetime of RHN for $g_R = g_L$, estimated from Eq.~(\ref{eqn:lifetime}); for the values of $g_R \neq g_L$ the lifetime should be rescaled via $(g_R / g_L)^{-4}$ accordingly. Depending on $g_R$, the general-purpose detectors at LHC could probe the lifetime $\tau_N^0$ from $\sim 10$ m to below $0.01$ cm, after  taking into consideration the large Lorentz boost factor $E_{N} / m_N$. The RH sector can be probed up to $m_{W_R} \simeq 20$ TeV for a large $g_R /g_L = 1.5$.

We also show the sensitivity contours for MATHUSLA in Figure~\ref{fig:RHN}, assuming at least 4 signal events. Though the effective cross section is smaller (due to its smaller size), MATHUSLA is sensitive to lighter RHN with mass as low as $\sim 1$ GeV, and could effectively probe a complementary parameter space of LR seesaw with GeV-scale light RHNs.


\end{document}